\documentclass[floatfix,12pt,letter,tightenlines,superscriptaddress,nolinenumbers,notitlepage,longbibliography,nofootinbib]{revtex4-1}
\usepackage{graphicx} 
\usepackage{adjustbox}
\usepackage{amsmath}
\usepackage{amsfonts}
\usepackage{amssymb} 
\usepackage{xcolor}
\usepackage{multirow}
\usepackage{lineno}
\usepackage{url}
\usepackage[utf8]{inputenc}
\usepackage{hyperref}
\usepackage{natbib}
\usepackage{marvosym}

\hypersetup{
     colorlinks=true,
     linkcolor=blue,
     filecolor=blue,
     citecolor = black,      
     urlcolor=cyan,
     }
\graphicspath{ {./Figures/} }

\newcommand{\orcidicon}[1]{\href{https://orcid.org/#1}{\includegraphics[height=\fontcharht\font`\B]{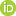}}}

\begin{document}

\title{CORE - a COmpact detectoR for the EIC}
\author{R.~Alarcon}
\affiliation{Arizona State University, Tempe Arizona 85287}
\author{M.~Baker}
\affiliation{Thomas Jefferson National Accelerator Laboratory, Newport News VA 23606}
\author{V.~Baturin}
\affiliation{Old Dominion University, Norfolk Virginia 23529}
\author{P.~Brindza}
\affiliation{Old Dominion University, Norfolk Virginia 23529}
\author{S.~Bueltmann}
\affiliation{Old Dominion University, Norfolk Virginia 23529}
\author{M.~Bukhari\orcidicon{0000-0003-3604-3152}}
\affiliation{Jazan University, Gizan 45142,
Saudi Arabia}
\author{R.~Capobianco}
\affiliation{University of Connecticut, Storrs Connecticut 06269}
\author{E.~Christy}
\affiliation{Thomas Jefferson National Accelerator Laboratory, Newport News VA 23606}
\author{S.~Diehl}
\affiliation{University of Connecticut, Storrs Connecticut 06269}
\affiliation{Justus Liebig Universitaet Giessen, Giessen, Germany}
\author{M.~Dugger}
\affiliation{Arizona State University, Tempe Arizona 85287}
\author{R.~Dupré}
\affiliation{Université Paris-Saclay, CNRS/IN2P3, IJCLab, 91405 Orsay, France}
\author{R.~Dzhygadlo}
\affiliation{GSI Helmholtz Centre for Heavy Ion Research, Darmstadt, Germany}
\author{K.~Flood}
\affiliation{Nalu Scientific, Honolulu Hawaii 96822}
\author{K.~Gnanvo}
\affiliation{Thomas Jefferson National Accelerator Laboratory, Newport News VA 23606}
\author{L.~Guo}
\affiliation{Florida International University, Miami Florida 33199}
\author{T.~Hayward}
\affiliation{University of Connecticut, Storrs Connecticut 06269}
\author{M.~Hattawy}
\affiliation{Old Dominion University, Norfolk Virginia 23529}
\author{M.~Hoballah}
\affiliation{Université Paris-Saclay, CNRS/IN2P3, IJCLab, 91405 Orsay, France}
\author{M.~Hohlmann}
\affiliation{Florida Institute of Technology, Melbourne, Florida 32901}
\author{C.E.~Hyde\footnote{chyde\MVAt odu.edu}
\orcidicon{0000-0001-7282-8120}}
\affiliation{Old Dominion University, Norfolk Virginia 23529}
\author{Y.~Ilieva}
\affiliation{University of South Carolina, Columbia South Carolina 29208}
\author{W.~W.~Jacobs}
\affiliation{CEEM, Indiana University, Bloomington Indiana 47405}
\author{K.~Joo}
\affiliation{University of Connecticut, Storrs Connecticut 06269}
\author{G.~Kalicy}
\affiliation{Catholic University of America, Washington D.C. 20064}
\author{A.~Kim}
\affiliation{University of Connecticut, Storrs Connecticut 06269}
\author{V.~Kubarovsky}
\affiliation{Thomas Jefferson National Accelerator Laboratory, Newport News VA 23606}
\author{A.~Lehmann}
\affiliation{Erlangen-Nuremberg University, 91058 Germany}
\author{W.~Li}
\affiliation{CFNS, Stony Brook University, Stony Brook, New York 11794}
\author{D.~Marchand}
\affiliation{Université Paris-Saclay, CNRS/IN2P3, IJCLab, 91405 Orsay, France}
\author{H.~Marukyan \orcidicon{0000-0002-4150-0533}}
\affiliation{A. Alikhanyan National Science Laboratory (Yerevan Physics Institute) Yerevan Armenia}
\author{M.~J.~Murray}
\affiliation{University of Kansas, Lawrence Kansas 66045}
\author{H.~E.~Montgomery}
\affiliation{Thomas Jefferson National Accelerator Laboratory, Newport News VA 23606}
\author{V.~Morozov}
\affiliation{Oak Ridge National Laboratory, Oak Ridge Tennessee 37830}
\author{I.~Mostafanezhad}
\affiliation{Nalu Scientific, Honolulu Hawaii 96822}
\author{A.~Movsisyan}
\affiliation{A. Alikhanyan National Science Laboratory (Yerevan Physics Institute) Yerevan Armenia}
\author{E.~Munevar}
\affiliation{Universidad Distrital Francisco Jos\'e de Caldas, Bogot\'a Colombia}
\author{C.~Mu\~noz Camacho}
\affiliation{Université Paris-Saclay, CNRS/IN2P3, IJCLab, 91405 Orsay, France}
\author{P.~Nadel-Turonski\footnote{turonski\MVAt jlab.org}}
\affiliation{CFNS, Stony Brook University, Stony Brook, New York 11794}
\author{S.~Niccolai}
\affiliation{Université Paris-Saclay, CNRS/IN2P3, IJCLab, 91405 Orsay, France}
\author{K.~Peters}
\affiliation{GSI Helmholtz Centre for Heavy Ion Research, Darmstadt, Germany}
\author{A.~Prokudin}
\affiliation{Thomas Jefferson National Accelerator Laboratory, Newport News VA 23606}
\affiliation{Penn State University Berks, Reading Pennsylvania 19610}
\author{J.~Richards}
\affiliation{University of Connecticut, Storrs Connecticut 06269}
\author{B.~G.~Ritchie}
\affiliation{Arizona State University, Tempe Arizona 85287}
\author{U.~Shrestha}
\affiliation{University of Connecticut, Storrs Connecticut 06269}
\author{B.~Schmookler}
\affiliation{CFNS, Stony Brook University, Stony Brook, New York 11794}
\author{G.~Schnell}
\affiliation{University of the Basque Country UPV/EHU \& Ikerbasque, Bilbao, Spain}
\author{C.~Schwarz}
\affiliation{GSI Helmholtz Centre for Heavy Ion Research, Darmstadt, Germany}
\author{J.~Schwiening}
\affiliation{GSI Helmholtz Centre for Heavy Ion Research, Darmstadt, Germany}
\author{P.~Schweitzer}
\affiliation{University of Connecticut, Storrs Connecticut 06269}
\author{P.~Simmerling}
\affiliation{University of Connecticut, Storrs Connecticut 06269}
\author{H.~Szumila-Vance}
\affiliation{Thomas Jefferson National Accelerator Laboratory, Newport News VA 23606}
\author{S.~Tripathi}
\affiliation{University of Hawaii, Honolulu Hawaii 96822}
\author{N.~Trotta}
\affiliation{University of Connecticut, Storrs Connecticut 06269}
\author{G.~Varner\orcidicon{0000-0002-0302-8151}}
\affiliation{University of Hawaii, Honolulu Hawaii 96822}
\author{A.~Vossen}
\affiliation{Duke University, Durham North Carolina 27708}
\author{E.~Voutier}
\affiliation{Université Paris-Saclay, CNRS/IN2P3, IJCLab, 91405 Orsay, France}
\author{N.~Wickramaarachchi}
\affiliation{Catholic University of America, Washington D.C. 20064}
\author{N.~Zachariou}
\affiliation{University of York, Heslington, York, YO10 5DD, UK}
\date{December 1, 2021}

\maketitle

\newpage

\tableofcontents

\section{Executive Summary}
\label{summary}
The COmpact detectoR for the Eic (CORE) Proposal is prepared in response to the ``Call for Collaboration Proposals for Detectors to be located at the Electron-Ion Collider (EIC)". The CORE detector is designed to satisfy the ``mission need" statement with a physics scope that completely and comprehensively covers the one outlined in the EIC Community White Paper \cite{Accardi:2012qut} and the National Academies of Science (NAS) 2018 report. The distinguishing theme of the CORE detector is that it exploits fully technological advances in detector precision and granularity to minimize the size of the overall detector.

The CORE central detector is constructed around a moderately high field, 3 Tesla, short (2.5 m) central solenoid. The tracking technology is essentially all silicon, and the electromagnetic calorimetry is based on the highest performance crystals available. Hadronic particle identification (PID) is achieved with a combination of compact gaseous and solid-radiator ring-imaging Cherenkov detectors. The central detector is complemented by an extended suite of forward detectors downstream on the hadronic side of the intersection region.

In general, the size and mass of collider detectors are significant cost drivers. The baseline CORE design therefore exploits this advantage both by reducing the overall cost, but also by choosing the highest performance technologies where they are beneficial for the physics performance. The approach adopted also provides design options, which could be made to further reduce costs or to further enhance performance in specific areas. These options are discussed in the body of the text.

The compact size of the central detector will ensure that the distance to the first collider magnets, the $\beta^*$, and hence the instantaneous luminosity can be maximized and chromaticity are minimized. This of the highest importance for all aspects of the EIC physics program. The forward detector is designed such that - if a downstream secondary focus were available - this property would be fully exploited.

The ability of the detector to achieve the EIC goals is supported by a body of simulations. We have considered a range of processes, including inclusive, semi-inclusive, and exclusive reactions. These cover and serve as a mnemonic for the gamut of electron-nucleon and light- and heavy-ion interactions. The demonstrated performance supports the delivery of results across the full spectrum of interest from the study of the internal properties of the nucleon to the approach to saturation in QCD interactions.

The $\sim$ 60 signatories of this proposal, form a proto-collaboration. They represent 25 institutions world-wide, which would form a core group of participants. This group has demonstrated its competence and understanding of the demands. Scrutiny of the total cost puts it in the range \$150-200M. There are no features of the design that would place an inordinate strain on a national laboratory procurement capability. All the technologies chosen are well established. With an appropriate level of funding and effort, the detector would be installed in time for initial operations of the collider. The collaboration anticipates that---with the approval of the proposal---it would expand both in terms of the total physicist count, but also in the level of commitment of the current and future collaborating institutions, including national laboratories.

Based on the Call for Collaboration Proposals for Detectors at the Electron Ion Collider, and the material presented herewith, the CORE detector design is a valid candidate for either ``Detector 1" or ``Detector 2". In particular, the proposal shows that the CORE design would match the requirements to deliver on the physics program laid out in the EIC White Paper~\cite{Accardi:2012qut} and NAS Report. In some interesting respects, for example forward detection, it would expand that physics scope. 

As a potential second detector, CORE satisfies all the criteria. Its innovative, compact designs and consequent technology choices are expected to be complementary in concept and approach to other detectors under consideration. The preferred intersection region design for CORE, with a secondary downstream focus, has been developed based on the constraints from the current machine design for IP6. Initial cost estimates and scale suggest that its construction schedule could be made compatible with completion by CD4.

\section{Introduction and Science Goals}

\label{introduction}
The CORE proposal is submitted in response to the ``Call for Collaboration Proposals for Detectors to be located at the Electron-Ion Collider (EIC).'' CORE is a hermetic, high-luminosity, general-purpose detector designed to support the full EIC physics program, and as such could be used as a basis for either Detector 1 or 2 in IP6 or IP8.

A science case for the EIC has been laid out in the EIC White Paper \cite{Accardi:2012qut}, the 2015 NSAC Long-Range Plan \cite{osti_1296778}, a National Academies of Sciences assessment \cite{NAP25171}, and the EIC Users Group Yellow Report \cite{AbdulKhalek:2021gbh}. The CORE detector is designed to support the full EIC program. Section 1.3 of the EIC White Paper \cite{Accardi:2012qut} summarizes the five key physics deliverables of the EIC.

\textbf{The proton spin} includes both quark ($\Delta \Sigma$) and gluon ($\Delta G$) contributions, where the latter is a crucial EIC deliverable.
However, a complete picture will require an understanding of partonic orbital angular momenta, for which measurements aiming at exploring the multi-dimensional structure of the proton are envisioned (and discussed below).

\textbf{The motion of quarks and gluons in the proton} and tomographic 3D-imaging in momentum space can be expressed within the framework of transverse-momentum distributions (TMDs). These are accessible through semi-inclusive measurements, which probe the correlation between the proton spin and the spin and transverse motion of the partons, creating a 3D image of the proton in momentum space. Both polarization and particle identification play a crucial role. Polarization of the colliding particles allows for measurements of spin asymmetries and therefore reconstruction of various contributions related to correlations of transverse motion and the spin of partons and/or the parent nucleon. The final particle identification of pions and kaons, allows one to clearly separate the contributions from the various quark flavors.

\textbf{Tomographic images of the proton}
in longitudinal momentum and transverse coordinate space can be obtained through generalized parton distributions (GPDs), which are obtained from measurements of deep virtual exclusive scattering (DVES). The luminosity, polarized beams, kinematic coverage, and forward detection combine to give the EIC a unique opportunity to create a map of the sea quarks and gluons, and explore how the wave function of the proton changes as a function of the proton momentum fraction carried by the parton. 
The critical channel to study GPDs is the exclusive production of a single real photon, called deeply virtual Compton scattering (DVCS). However, complementary channels such as exclusive meson production will also provide important input.
Even though it is experimentally more challenging, transverse imaging can also be done on nuclei. However, the high-resolution electromagnetic (EM) calorimetry proposed for CORE and the exceptional forward acceptance of IR8 will make it possible to study the DVCS process for a wide range of nuclei, creating opportunities for new discoveries and fully leveraging the capability of the EIC to deliver (almost) any nuclear beam from deuterium to uranium (spin-0 nuclei have a GPD structure that is particularly simple and easy to measure).

\textbf{QCD matter at an extreme gluon density} can be probed at the EIC, where nuclei can be used to enhance the effective gluon density. This opens up the possibility of approaching gluon saturation even though the e-p collision energy is lower than it was at HERA. There are several measurements with a sensitivity to gluon saturation. One is a comparison of diffractive and total DIS cross sections in e-p and e-A collisions. Another process is coherent diffraction (exclusive production of $\phi$ and $J/\psi$) on heavy nuclei, which also gives a way of probing the transverse gluon distributions and possibly its fluctuations. This measurement relies on effective vetoing of nuclear breakup, which is more challenging for heavy ions than for lighter ones. The ability of IR8 to detect nuclear breakup will be very important to ensure a clean measurement, which would be complemented by the high-resolution reconstruction of the produced mesons by the CORE central detector.

Studies of \textbf{quark hadronization} at the EIC benefit from the possibility to vary the nuclear radius, virtual photon energy, and produced meson. Adding detection of the nuclear breakup, which carries some information about the nuclear excitation energy, could provide one more unique opportunity to learn about hadronization.

Both the White Paper (WP)~\cite{Accardi:2012qut} and Yellow Report (YR)~\cite{AbdulKhalek:2021gbh} provide a comprehensive set of detailed studies made with some assumptions on detector capabilities. In the CORE proposal, we focus on the detector performance with the assumption that if this were similar or better than what was assumed in the WP and YR, then so would be the results of any particular analysis. We also wanted to use the available space to highlight important CORE capabilities, ({\it e.g.}, high luminosity for {\em all} energies, high-resolution photon detection, and high-purity lepton identification (including the scattered electron as well as muons), as well as synergies with an interaction region (IR) using a forward spectrometer with a $2^{nd}$ focus, which is currently envisioned for IP8 where Detector 2 would be located.

\section{Detector Overview}
This section provides a high-level description of the detector and discusses key design considerations. Physics simulations can be found in section \ref{simulations} and a detailed description of the individual subsystems is provided in section \ref{subsystems}. 

A compact detector layout offers a cost-effective solution that enables a number of synergies and unique capabilities. Most cost savings are organic and do not affect performance (the solenoid is smaller, calorimeters have fewer modules, RICH detectors have fewer photosensors, etc). But the smaller size of various subsystems also makes it affordable to invest in key physics capabilities such as excellent EM calorimetry. Smaller subsystems are also lighter, which in turn simplifies supports and engineering.

However, one of the most important aspects of the compact layout of CORE is the detector length. A shorter detector, made possible by a short solenoid, enhances the luminosity at {\it all} energies.

\begin{figure}[htbp]
\begin{center}
\includegraphics[width=\textwidth]{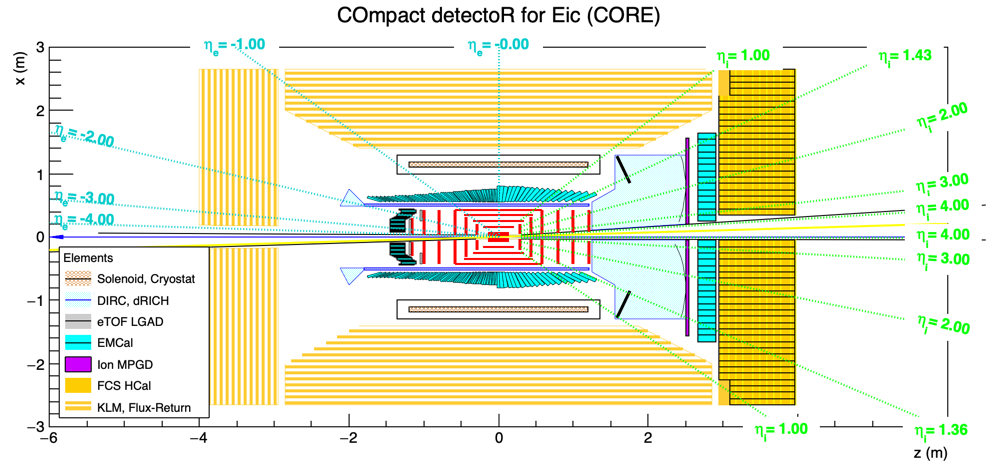}
\caption{Two-dimensional view of CORE in the horizontal plane. The electron beam travels in the $-\hat z$ direction, and ion beam direction is $\hat z \cos(0.035)+\hat x\sin(0.035)$. The negative pseudo-rapidity values $\eta_e$ are relative to the $z$-axis. The positive values $\eta_i$ are relative the ion beam direction.}
\label{fig:CORE-2D}
\end{center}
\end{figure}

\begin{figure}[htbp]
\begin{center}
\adjustbox{trim={0.0\width} {0.1\height} {0.0\width} {0.0\height},clip}
{\includegraphics[width=0.95\textwidth]{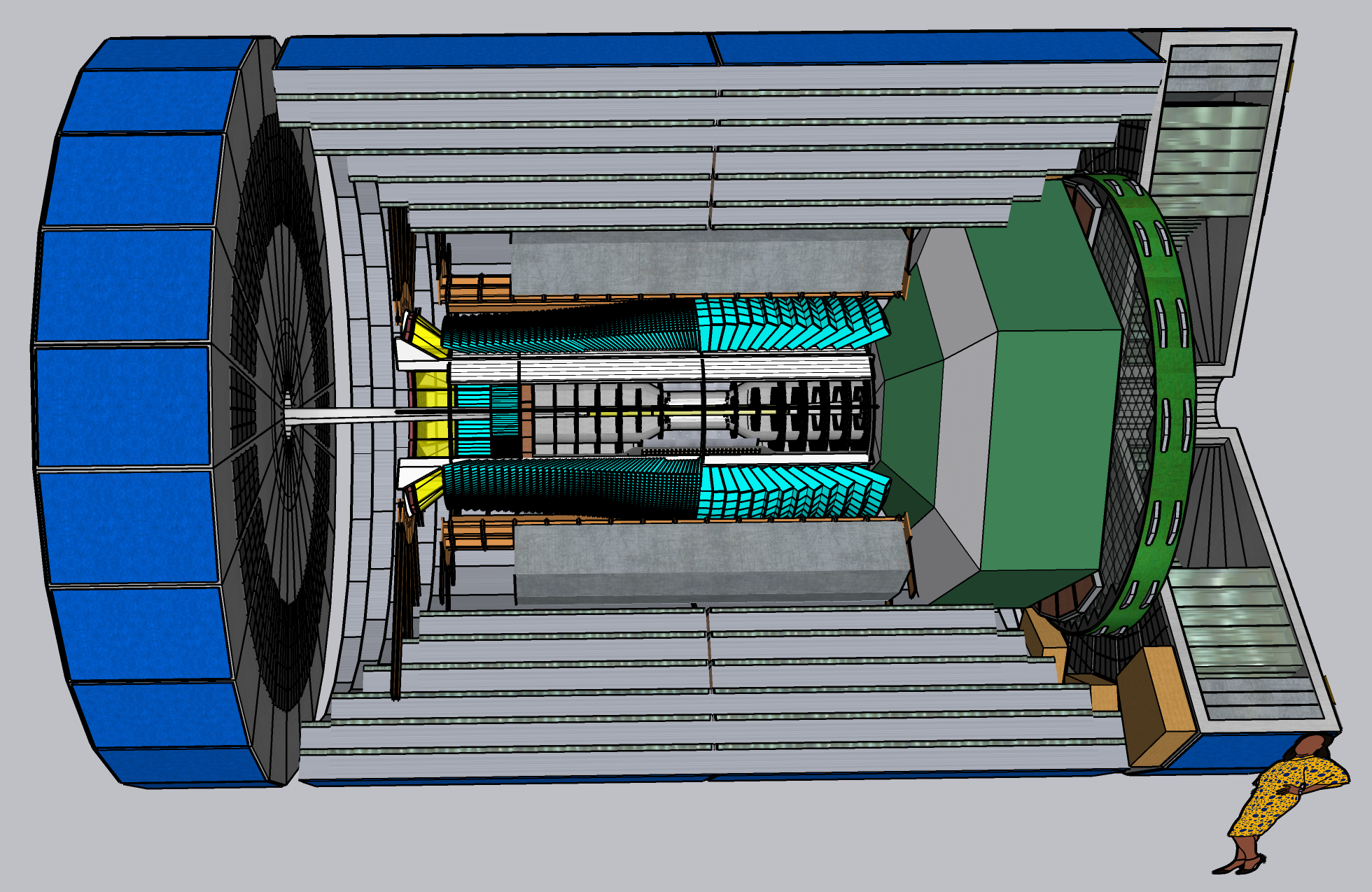}}
\caption{View of CORE created using ``SketchUp'' 3D modeling software.}
\label{fig:CORE-3D}
\end{center}
\end{figure}

\subsection{The CORE detector}
\label{detector}
The CORE detector uses a compact 3 T solenoid with a 2.5 m coil length and 1 m inner radius.
The field has been optimized for the requirements of the dual-radiator RICH (``projectivity'') and DIRC (low field on the photosensors). The fringe fields are in compliance with EIC machine design specifications.

The detector is built around a Si pixel tracker (2.4 m long and 44 cm in radius) with excellent vertex and momentum resolution (shown in Fig.~\ref{fig:tracker_resolution}).
The tracker makes efficient use of the magnetic field, taking up 19\% of the $7.8~\text{m}^3$ volume of the solenoid.
The small radius of the tracker and the proximity of the DIRC and barrel EMcal, which are located directly outside of the tracker, allow particles with lower $p_T$ to reach the PID systems even when operating at 3 T. This reduces the need to divide up the beam time between different solenoid field settings.
The silicon tracker is complemented by a MPGD tracker in-between the dual-radiator RICH and EMcal in the hadron endcap.

Particle identification is provided by three systems. For particles detected in the hadron endcap, the dual-radiator RICH (aerogel + gas) provides $3\sigma$ $\pi/K$ separation up to 50 GeV/c and $e/\pi$ up to 15 GeV/$c$ (with better separation at lower momenta). It has a gas depth of 1.2 m and ample space for the photosensors.
In the barrel outside the tracker there is a high-performance DIRC (0.5 m radius). The baseline option is to re-use the radiator bars from the BaBar DIRC \cite{BaBarDIRC:2005nhg}. The small size of the CORE DIRC would also make it affordable to build new, thinner bars---although this would require some additional R\&D for validation. The thin bar option is discussed in section \ref{hpDIRC}.
For the baseline bar option, the DIRC can provide $3\sigma$ $\pi/K$ separation up to 6-7 GeV/$c$, and better at lower momenta.
For momenta in the 0.2-0.5 GeV/$c$ range ($|p|$, not $p_T$), the DIRC can operate in a ``threshold mode'' (pions give a strong signal while kaons and protons are below Cherenkov threshold), extending the coverage down to the lowest momenta. A similar mode of operation is also available for the dual-radiator RICH, although the kaon threshold in the aerogel is 2 GeV/$c$ (due to the lower index of refraction). The DIRC also provides timing information with an $\eta$-dependent resolution, reaching about 20 ps closest to the electron endcap.
In the electron endcap, CORE will use an AC-LGAD TOF system with 25 ps resolution located in-between the Si-tracker and the endcap EMcal.
While a TOF system covers a smaller range in momentum than an aerogel RICH, it is light, compact, and easy to integrate with the endcap PbWO$_4$ EMcal. Hadron momenta in the electron endcap will be low when operating close to maximum luminosity with asymmetric beam energies ({\it e.g.}, 5-10 GeV electrons on 275 GeV protons).

The $4\pi$ electromagnetic calorimetry will comprise two technologies: PbWO$_4$ on the electron hemisphere $\eta<0$ and W-Shashlyk in the hadron hemisphere $\eta>0$.
The compact inner radius of the barrel EMcal (0.55 m) makes it affordable to make extended use of PbWO$_4$.
PbWO$_4$ is the best technology currently available in terms of resolution ($2\%/\sqrt{E}$, Moli\`ere radius, etc).
The advantages of W-Shashlyk are lower cost and better $\pi^0$ photon separation (due to the high-granularity readout).
The latter is most important in the hadron hemisphere where $\pi^0$ energies are higher.
The energy resolution ($6\%/\sqrt{E}$) is not as good as for PbWO$_4$, but twice as good as for W/SciFi.

There is 0.45 m of radial space reserved for the barrel EMcal inside the solenoid, which is sufficient to support either technology. The barrel EMcals will be projective, while the endcap ones will be non-projective. Projectivity in the barrel is essential for a proper integration of these two systems. Even if there are no angular gaps between them, the spatial separation means that the module of the barrel EMcal closest to the hadron endcap needs to be aligned with the line-of-sight to the IP to minimize corner hits that would lead to inefficiencies.
For the EMcal in the electron endcap we are proposing to re-use existing crystals.

The small area (0.6 m$^2$) and weight of the endcap PbWO$_4$ EMcal makes it possible to cantilever it from behind instead of supporting it from the barrel. This reduces the amount of supports in the EMcal acceptance and improves hermeticity. The large PbWO$_4$ coverage, in combination with supplementary eID from the DIRC, results in the best possible identification of the scattered electron and photon energy resolution.
The significance of the latter is illustrated below for the DVCS processes, where the final-state photon measurement enables the determination of the transverse momentum transfer to the target ion ($t \sim p_T^2$). Being able to do so with high resolution is essential for measurements where the $p_T$ of the ion is small. At mid- to low-$x$ and for higher ion mass $A$, this is the only way to measure the full $t$-distribution needed to obtain a complete 3D image of the nucleus (a small $t$ corresponds to a large impact parameter $b$).

The hadronic calorimetry consists of two systems. In the hadron endcap, the CORE baseline is to use Fe/Sci modules similar to those in the STAR FCS. The existing modules will be re-used for the periphery, while slightly longer new ones will be added in the center.
Such an Hcal is necessary for hadronic methods for kinematic reconstruction and high-$x$ jets. By contrast, at mid rapidity, jets at the EIC are best reconstructed from individual tracks. Thus, in the barrel CORE has adopted a neutral hadron detection and muon ID system based on the Belle II KLM ($K_L$ - $\mu$) detector, which is integrated with the flux return of the solenoid. As in the Belle KLM, this detector will focus on position rather than energy resolution for the neutral hadrons (making it possible to determine if there was a neutral hadron within the jet cone), and provide excellent muon ID (the Belle KLM can reach down to 0.6 GeV/$c$). The KLM is also used in the electron endcap as jets are still well reconstructed from the individual particles at large negative rapidities (small angle jets have low $p_T$ and a momentum comparable to the beam energy).

\subsection{Technical risk}
\label{risk}
CORE does not envision using any emerging technologies. Only mature technologies have been chosen that have either been used in previous experiments, or are developments of such technologies in an advanced stage of R\&D within the Generic R\&D for an EIC program, ongoing R\&D supported by the EIC project, or similar programs. 

Technologies used elsewhere that could easily be adapted for CORE include the HCals (STAR FCS and Belle KLM), the PbWO$_4$ EMcal ({\it cf.} PANDA, CMS, etc), and the MPGD.
Subsystems for which there is still ongoing R\&D are the high-performance DIRC, the dual-radiator RICH (dRICH), the AC-LGADs, the W-Shashlyk EMcal, and the silicon tracker.
In all cases the risk was deemed to be reasonable, as key aspects have already been demonstrated in R\&D and in other applications.

For the DIRC, ongoing R\&D is focused on evaluating the re-use of the BaBar bars and validating the performance improvements expected from the advanced optics, small pixels, and high-resolution timing.
The dRICH is a new design, but it is based on well-understood principles and components. The main risk is associated with the use of SiPMs as the preferred photosensor, which is being addressed in ongoing R\&D.
The AC-LGAD, on which R\&D is ongoing, offers a higher fill factor and better timing resolution than the resistive LGADs used for ATLAS and CMS, making it a slightly better option for the central detector and a more flexible one for the Roman pots (providing tracking and timing in a single device).
Shashlyk EMcals have seen extensive use. Replacing Pb with W to improve spatial resolution and reduce length could make machining a little more expensive, but should not pose a significant risk.
And finally, the Si tracker is based on ALICE ITS3 technology. While this is still in development and details may change, it seems reasonable to assume that with continued support at CERN, this project will be brought to a successful conclusion.
Although not part of the CORE baseline, the option of using new, thin DIRC bars would require some additional R\&D for validation. But since thin plates have been manufactured for TORCH \cite{Gys:2017dpe,TORCH}, the expected risk would be moderate.

For PbWO$_4$, the technical risk is very small since it is a widely used and well understood technology. Even though the CORE timelines are reasonable, with only two suppliers worldwide, there is always a risk of production delays. If this occurred, the production rates at Crytur could be increased through some investment in their capacity. However, since Uniplast can build relatively inexpensive, projective W-shashlyk modules for the barrel at a quick rate, W-shashlyk could be used in lieu of PbWO$_4$ at larger (less negative) $\eta$. It could then be replaced by PbWO$_4$ as part of a future upgrade.

\subsection{Complementarity}
\label{complementarity}
The call for proposals clearly suggests that the goal is to have two complementary EIC detectors, which can cross-check each other's results, but also provide optimizations towards different physics topics. Since we at this point don't know the details of the other detector, we will here briefly discuss some relevant aspects of CORE and its complementarity with the YR ``3T reference detector,'' the performance of which is summarized in Table 11.50 (p.588) of the YR.

As mentioned in section \ref{luminosity}, a shorter detector such as CORE will have a smaller chromatic contribution at any given luminosity, enabling the accelerator to operate with two detectors at a higher combined luminosity, simultaneously benefiting both (not just CORE).

Compared to the YR detector, CORE offers a dramatically improved photon and lepton detection. Both concepts use PbWO$_4$ (listed as $2.5\%/\sqrt{E}$ in the YR table) in the electron endcap, but the YR coverage is only $-3.5<\eta<-2$, while the CORE PbWO$_4$ extends throughout the full $\eta<0$ hemisphere. In CORE, the $\eta>0$ hemisphere is covered by W-Shashlyk with a resolution of $~6\%/\sqrt{E}$. In contrast, the YR detector offers 4-$8\%/\sqrt{E}$ for $-2<\eta<-1$, and 12-$14\%/\sqrt{E}$ for $\eta>-1$. In addition, CORE has a neutral hadron and muon ID system covering the electron endcap and barrel. This is based on the Belle II KLM, which was able to provide good muon ID even below 1 GeV/$c$.
The photon detection and electron ID capabilities of CORE are key to opening up new physics opportunities. An example, discussed in more detail in section \ref{2nd_focus}, is the ability to determine the $p_T$ ($\Delta_\perp$) of the target in DVCS using the photon rather than proton/ion, greatly extending the low-$p_T$ acceptance and mass range for ions (and avoiding many complications associated with beam effects).

The hadronic calorimetry in the hadron endcap is based on the STAR FCS and is generally similar to the YR detector. For a single EIC detector, we believe that the combination of this technology and a KLM-like system covering the barrel and electron endcap is the best solution. But the CORE layout offers a lot of flexibility. In the barrel, the radial compactness of CORE makes it possible to re-use the sPHENIX Hcal as the outer part of the flux return while adding a 0.5 m Hcal section in-between the sPHENIX Hcal and the cryostat. And in the electron endcap, the KLM could be replaced by an Hcal similar to the one used on the hadron side (as in the YR detector).

The hadron PID is similar to the reference detector, but there are important nuances making CORE perfectly fit the idea of cross-checking results while offering a different emphasis. In particular, while CORE can deliver excellent performance for all energies, it is optimized for high luminosity. This in turn means running with more asymmetric beam energies. In such configurations, the hadrons hitting the hadron endcap are highly energetic, while the ones on the electron side have low momenta.
In addition, many of the high-energy measurements ({\it e.g.}, related to gluon saturation using 18 GeV e on 110 GeV/A heavy ions, or $\Delta G$ with 18 GeV e on 275 GeV p) generally do not require high-momentum $\pi/K$ ID, and specifically not in the electron endcap.
In higher-luminosity configurations, with electron beam energies in the 5-10 GeV range, there are fewer hadrons in the electron endcap and the hadron momenta there are lower, which limits the benefits of high-momentum hadron ID. Thus, in the electron endcap CORE has a high-resolution TOF system instead of the aerogel RICH used in the YR detector. This makes the endcap shorter and lighter (which benefits photon and electron detection), and removes the technical risk of having to operate photosensors for an aerogel RICH in an environment that has both a high magnetic field (inside the solenoid) and high levels of radiation (near the beam).
Conversely, since high luminosities favor high hadron beam energies that give rise to many high-momentum hadrons in the hadron endcap, it is important to have a high photon yield in the gas of the dual-radiator RICH to maximize the reconstruction efficiency in the presence of backgrounds, which is why CORE allocates 1.2 m of longitudinal space for gas.

Finally, we also wanted to briefly discuss the question of general-purpose \textit{vs.} specialized detectors. While the latter could take many forms, a specialized detector would only be able to cross-check results for a narrow range of physics topics.
And while the lack of certain subsystem in a specialized detector might offer some cost savings, this has to be considered in the broader context of the total cost of setting up an additional IR. Even a simple specialized detector would thus constitute a major investment.
We also note that in the past, many specialized detectors were proposed to address various shortcomings of existing detectors. For example, the so-called Caldwell detector concept would have significantly improved the forward detection for diffractive physics at HERA.
But for the EIC, it is possible from the outset to build a $\sim$50 m long, dedicated spectrometer with a $2^{nd}$ focus, which would offer a forward-detection performance that is superior to that of any central detector, no matter how dedicated or specialized. The key issue then instead becomes how to best integrate the central and forward detectors and make sure to emphasize synergetic capabilities.

\subsection{Luminosity considerations}
\label{luminosity}
The EIC will have the unique capability to collide electrons with polarized protons and light ions, as well as unpolarized ions over the full mass range.
We anticipate that we will want to take full advantage of this opportunity to run in different configurations (polarization, ion species, c.m.~energy), and that beam time will be in high demand.
Supporting high-luminosity operations is thus a key requirement for an EIC detector. Since luminosity is linked to the individual beam energies, it is reasonable to assume that future PAC proposals will favor beam-energy configurations with the highest luminosity for the desired c.m.~energy, in particular as the 70\% beam polarization at the EIC will be energy independent (the 6 snakes will allow it to maintained the polarization from injection across all energies).

CORE supports high-luminosity operations in two ways. The first is by reducing the detector length. As a rule of thumb, the luminosity is inversely proportional to the focal length ($l^*$) of the quadrupole magnets of the accelerator, and thus increases if they can be moved closer to the interaction point (IP). By reducing the detector length to 4 m on the hadron side, CORE can provide a substantial luminosity increase for {\it all} energies. A shorter detector also has a smaller impact on chromaticity. This is important for two-detector running at the EIC, for which it is envisioned that different bunch pairs will collide at the two IRs. Thus, the luminosity of each detector will be limited \textit{individually} by the beam-beam tune shift, but there will also be a global chromaticity ``budget'' that will limit the combined luminosity when two detectors are operated in parallel. This scheme is popularly known as luminosity sharing. But two short detectors, or a combination of a long and a short one, will be able to operate at a higher combined luminosity than two long ones.
The key to reducing the detector length is the use of a short solenoid, which in CORE has a total coil length of 2.5 m and is centered on the interaction point.
The short CORE solenoid makes it possible to assign 1.2 metres to the gas in the RICH in order to enhance PID, while maintaining a 4 m length for the hadron side of the detector.

While a shorter solenoid improves luminosity for all energies, the choice of beam energies themselves is also an important luminosity consideration. For electron beams in the 5-10 GeV range the luminosity has some dependence on the energy due to beam-beam effects, but it drops rapidly between 10 and 18 GeV when the electron beam current becomes limited by the synchrotron radiation power. For protons and ions the luminosity is to good approximation proportional to the hadron beam momentum, except at the lowest energies where it drops rapidly due to space-charge effects.
Thus, in general, higher luminosities can be achieved with asymmetric beam-energy configurations. Some important EIC physics topics, such as gluon saturation and $\Delta G$, will require the highest ion and proton beam energies. Others may benefit from lower energies. But most experimental programs, and in particular ones that require high statistics ({\it e.g.}, exclusive reactions and 3D imaging), will likely favor a high luminosity. Thus, while an EIC detector needs to perform well across all energies, it is particularly important that it is optimized for collisions between 5-10 GeV electron beams and hadron beams close to the maximum energy.

\begin{center}
\begin{table}[htb!]
\begin{tabular}{ |c|c|c|c|c|c| } 
 \hline
 18x275 & 10x275 & 5x275 & 10x100 & 5x100 & 5x41 \\ 
 \hline
 1.65$\times 10^{33}$ & 10.05$\times 10^{33}$ & 5.29$\times 10^{33}$ & 4.35$\times 10^{33}$ & 3.16$\times 10^{33}$ & 0.44$\times 10^{33}$ \\ 
 \hline
\end{tabular}
\caption{Luminosities for electron and proton beam energy configurations (in GeV) listed in table 3.3 (p.102) of the CDR \cite{CDR}. The 5x275 setting was calculated by the BNL accelerator group using the same assumptions \cite{Vadim_private}. Supported proton and ion beam energies are in the continuous range of 100 to ($275 \times Z/A$) GeV/A, where $Z = A = 1$ for protons, and a discrete energy at 41 GeV/A. Electrons can have energies in the 5-18 GeV range. Electron and ion/proton energies are independent. Ion beams typically have luminosities per nucleon that are a little lower than, but comparable to protons. Due to its shorter length, for CORE all luminosities could be higher than the nominal CDR ones shown in the table.}
\label{luminosity_table}
\end{table}
\end{center}

\subsection{Impact of an IR with a second focus}
\label{2nd_focus}

The CORE detector design is compatible with that of the accelerator and interaction region layout of the CDR, and could be used in either IR6 or IR8. However, the performance (and complementarity with the other detector) would be enhanced by an interaction region design that has a secondary, downstream focus. This concept was developed by some of the CORE collaborators a decade ago. Currently such a design has been implemented for IR8. 
An IR with a secondary focus would be a desirable location for CORE regardless of whether it would be selected as Detector 1 or 2. Thus, although CORE is compatible with the IR6 layout, an IR8 location could be an attractive option for CORE also as Detector 1.

\begin{figure}[htb!]	
    \includegraphics[keepaspectratio=true,width=3.0in,page=1]{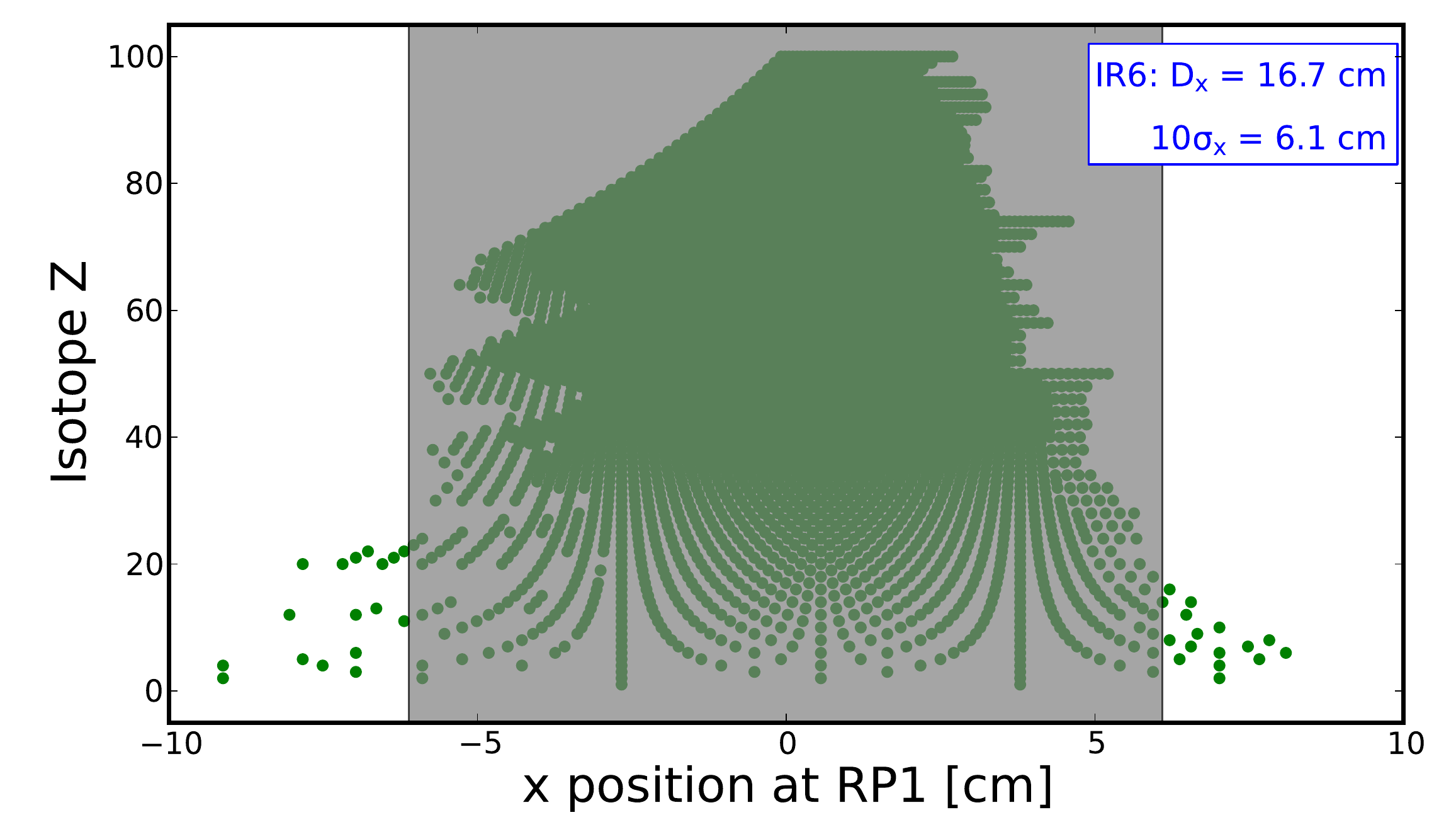}		
	\includegraphics[keepaspectratio=true,width=3.0in,page=1]{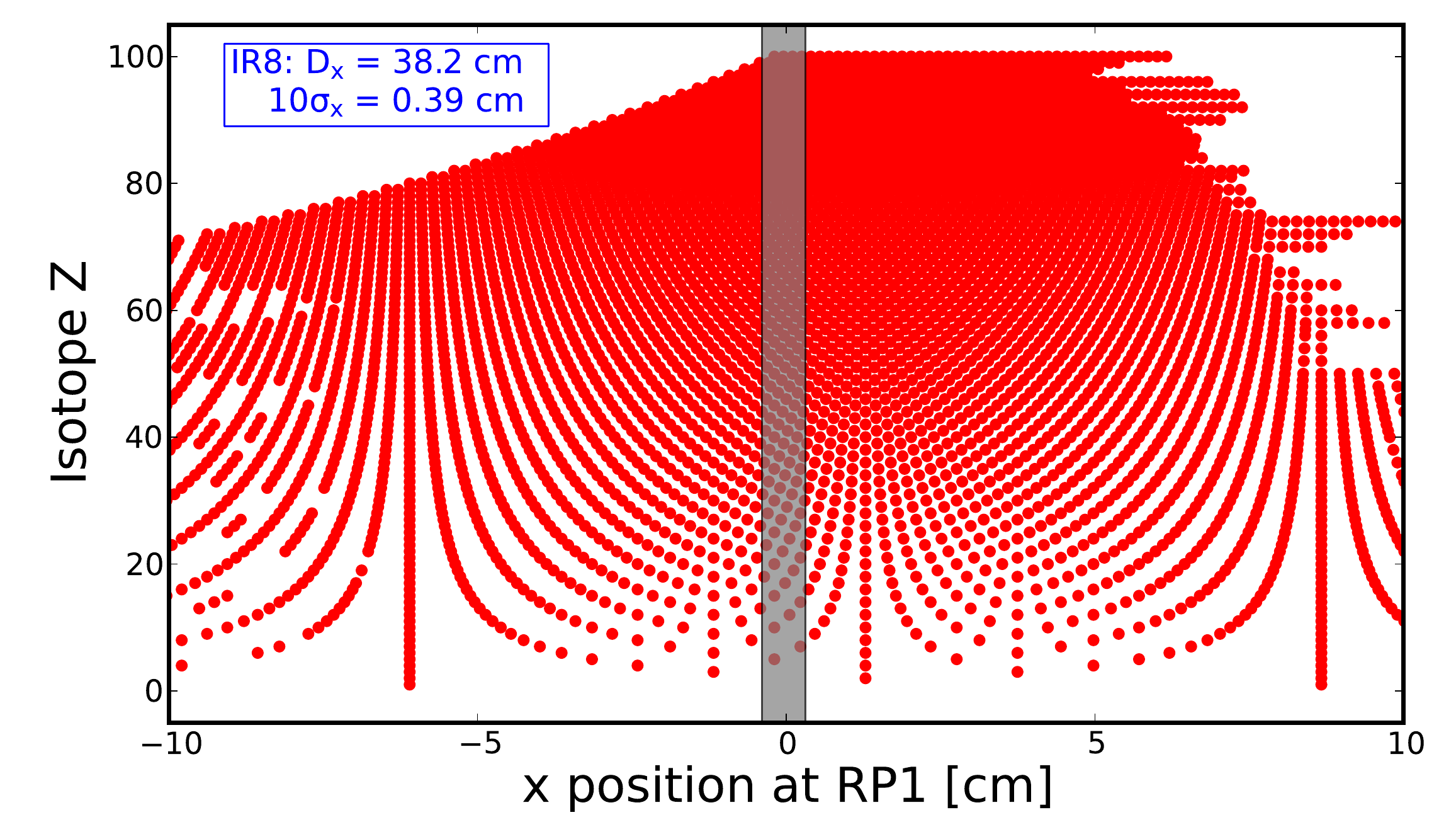}
	\caption{Isotope $Z$ vs. hit position in the best Roman pot in IR6 (left) and IR8 (right), respectively.
	The gray box on each plot shows the 10$\sigma$ beam size which prevents detection. This exclusion is much smaller in IR8 due to the $2^{nd}$ focus. The larger horizontal spacing in IR8 is due to a larger dispersion.
	The isotopes shown assume a $^{238}$U beam, but are representative for all heavy ions.
	The exceptional ability of IR8 to detect fragments with magnetic rigidities very close to that of the beam is also 
	indicative of the acceptance for recoil protons and nuclei that emerge from exclusive reactions with a low $p_T$ with respect to the beam.}
	\label{fig:rigidity}
\end{figure}

The $2^{nd}$ focus greatly enhances the low-$p_T$ and low-$x$ (large-$x_L$) acceptance for the recoiling target system in exclusive reactions on the proton and in coherent scattering by nuclei, and provides an unprecedented detection of nuclear fragments with magnetic rigidities close to that of the beam, including heavy ions that have lost only one nucleon ($A$-1 tagging). A comparison of fragments that can be detected in IR6 and IR8 is shown in Fig.~\ref{fig:rigidity}.

The excellent low-$p_T$ recoil proton acceptance even at the highest energies means that there is no need for a large-$\beta^*$, low-luminosity ``high-acceptance'' setting. Instead, the full $t$ distribution can be measured at maximum luminosity, without the need to match different data sets.

For light ions, two effects conspire to make low-$p_T$ acceptance for the recoiling target system crucial. First, the cross section peaks at lower $p_T$, and second, a given $p_T$ corresponds to a smaller scattering angle (which is determined by the $p_T$ {\it per nucleon}). At higher $x$, the second focus makes it possible to have essentially full acceptance down to $p_T$=0, and it significantly extends the low-$p_T$ coverage at lower $x$ (higher $x_L$). The ability to detect the scattered nucleus cleanly separates coherent and incoherent scattering. However, even when the target nucleus is detected, one still needs to reconstruct the four-momentum transfer to the target ($\sim p_T^2$), which is more challenging for a nucleus than a proton. This is why the ability of CORE to precisely determine the $p_T$ using the central detector is essential. But even more importantly, this method of reconstructing $p_T$ can also be applied when the nucleus is not detected and the forward spectrometer is instead used to veto nuclear breakup to suppress incoherent backgrounds and establish exclusivity.
For charged-particle production this is accomplished by the high-resolution tracker in combination with excellent muon ID (charmonium, time-like Compton scattering). And the PbWO$_4$ EMcal in the electron hemisphere provides the best possible photon detection (DVCS, $\pi^0$).

For coherent exclusive processes on light nuclei at low $x$ and medium-mass nuclei, it is also possible to use a hybrid method where the breakup is vetoed for the low-$p_T$ part while high-$p_T$ nuclei are detected. This method would greatly extend the kinematic coverage and mass range available for nuclei, while maintaining low levels of incoherent backgrounds. It is also a striking example of the synergy between the already exceptional low-$p_T$ acceptance for the recoiling target system provided by a forward detection utilizing a $2^{nd}$ focus, and a central detector providing the best possible resolution in the reconstruction of the $p_T$ when the ion is not detected (low $x$, high $A$). In particular, with the high-resolution EMcal proposed for CORE, the ability to measure DVCS for a wide range of nuclei could greatly expand our understanding of nuclear properties and pave the way for new discoveries.

Coherent diffraction on heavy nuclei (exclusive production of $\phi$ and $J/\psi$) is an important process for studying the onset of gluon saturation at the highest collision energies. However, vetoing of breakup becomes more challenging the heavier the nucleus is. If one has to rely only on evaporation neutrons detected in the ZDC and photons emitted from the final nucleus it is difficult to reach a suppression factor of at least 500, which is necessary to observe the coherent signal out to the $3^{rd}$ diffractive minimum. Thus, adding detection of nuclear fragments to the veto is essential to ensure a clean measurement. The most demanding case is single-nucleon knockout, which is not negligible in e-A collisions even at high energy. The ZDC efficiency for detecting a single neutron is far from sufficient. But the detection of the $A$-1 nucleus also requires a very capable forward spectrometer with large dispersion and a $2^{nd}$ focus, since the change in rigidity ($A/Z$) is small for heavy nuclei. The current IR8 layout is capable of tagging $^{89}$Zr in a $^{90}$Zr beam, which may already be sufficient for observing saturation, but with further refinements it should be possible for $A$-1 detection to still heavier masses.

Since heavy nuclei from the coherent events will never be detected, the $p_T$ will have to be reconstructed from the meson ($\phi$, $J/\psi$). This makes tracking resolution an important consideration for mapping out the diffractive minima.
And since the rate requirements are driven by the cross section in the high-$p_T$ tail, it is helpful to be able to use both the electron and muon final states for $J/\psi$ production.

While the introduction of a $2^{nd}$ focus greatly improves acceptance at small $p_T$ and low $x$ (large $x_L$), an important complementary capability is to have good acceptance at large $p_T$. In part, this can be accomplished by introducing superconducting magnet technologies that can support stronger fields in the final-focus quadrupoles of the accelerator and the large dipole magnets downstream of them (but before the $2^{nd}$ focus). This would in turn allow for larger apertures or higher gradients (making the magnet shorter) -- both of which would improve the transport of forward-scattered particles with larger $p_T$.
Such magnets are currently being pursued by the EIC project as an option for IR8. However, a shorter central detector also improves the high-$p_T$ part of the forward acceptance. Thus, a combination of a spectrometer with a large dispersion and a $2^{nd}$ focus, advanced magnets, and a short detector with the ability to reconstruct the $p_T$ for both charged particles and photons with high resolution, would create the best possible forward detection system for the EIC, enhancing its discovery potential.

And finally, fragments outside the shaded beam exclusion zone shown in Fig.~\ref{fig:rigidity} will include rare isotopes that can be studied using the forward detectors. In relativistic kinematics, the photons produced in the transition from excited states of the rare isotopes to the ground state will be emitted in the direction of motion of the ion, with an energy in the lab frame that is considerably higher than in the rest frame. Thus, a high-resolution EM calorimetry (\textit{e.g.}, LYSO) in front of the ZDC could allow for performing low-background gamma spectroscopy on rare isotopes, making the EIC complementary to low-energy facilities such as FRIB. If further studies confirm the feasibility of such measurements, they could become an important part of the EIC program in IR8 (and presumably also bring new funding opportunities). However, in this proposal we consider a LYSO EMcal in front of the ZDC to be an option and did not include it in the baseline cost. Rare isotopes are discussed in section 7.5.6 (p.241) of the Yellow Report \cite{AbdulKhalek:2021gbh}.

\section{Physics Simulations}
\label{simulations}

The user community has created an extensive body of physics simulations for an EIC detector described in the White Paper \cite{Accardi:2012qut} and Yellow Report \cite{AbdulKhalek:2021gbh}. The latter in particular used a recent detector model. While the CORE detector is in many ways complementary to this ``YR detector'' (as discussed in section \ref{complementarity}), both have also a lot in common in terms of general features and capabilities. For instance, they both are hermetic detectors with a 3 T solenoid and a high-resolution tracker, a high-resolution Hcal in the hadron endcap, as well as $4\pi$ EM calorimetry and PID. Thus, the focus of the proposal is not to demonstrate that this combination of capabilities is needed for the EIC physics program, as this has already been shown, but rather to focus on the {\em impact of the differences} between CORE and the YR detector benchmark. The differences fall into two main areas: EMcal resolution and electron endcap PID.

The CORE EMcal provides a much better resolution in the barrel and hadron endcap. In CORE, the resolution is $2\%/\sqrt{E}$ for $\eta<0$ and $6\%/\sqrt{E}$ for $\eta>0$, while the YR detector has $12\%/\sqrt{E}$ for the barrel and hadron side EMcals. This very significant difference has impact on, for instance, measurements of the DVCS process on nuclei and the purity of the reconstruction of the scattered electron.

Both, CORE and the YR detectors, have a DIRC in the barrel and a dual-radiator RICH in the hadron endcap. There is space for more gas in the CORE design.
The nominal momentum coverage in CORE is slightly better than that in the YR and importantly, the photon yield makes the measurement significantly more robust.
However, the main consequence of the CORE layout in terms of the hadron ID is that the $\eta<-1.65$ region, corresponding to a rather small electron endcap, is covered with high-resolution TOF rather than an aerogel RICH. While we strongly believe that the numerous advantages in replacing this RICH with TOF increase the overall capabilities of the detector, we will show the impact of this choice.

And finally, since the baseline solution for the barrel and electron endcap is a neutral hadron and muon (NHM/KLM) system, we include muon distributions for $J/\psi$ and $\Upsilon$ production. This system can also provide neutral-hadron (mostly $K_L$) detection with good angular resolution though limited energy resolution for jet reconstruction. Since EIC jets are best reconstructed from individual particles (except in the hadron endcap), the difference between the NHM/KLM and a traditional Hcal will be small in the barrel and electron endcap regions.

Most of the CORE performances presented here were obtained by means of processing generated data samples through DELPHES~\cite{deFavereau:2013fsa}, with the implementation of CORE adapted from \cite{Arratia:2021uqr}. DELPHES is a fast multipurpose detector response simulation, which takes into account detector features such as momentum, tracking, and calorimeter resolutions and calorimeter granularity. Single-particle four-vectors, as well as jets, are accessible for analyses. For each reconstructed particle, DELPHES preserves information about the true identity and kinematics of the generated particle, so the comparison between measured and true quantities is straight-forward to quantify.  

\begin{figure}[htb!]
\begin{center}
	\includegraphics[width=\textwidth]{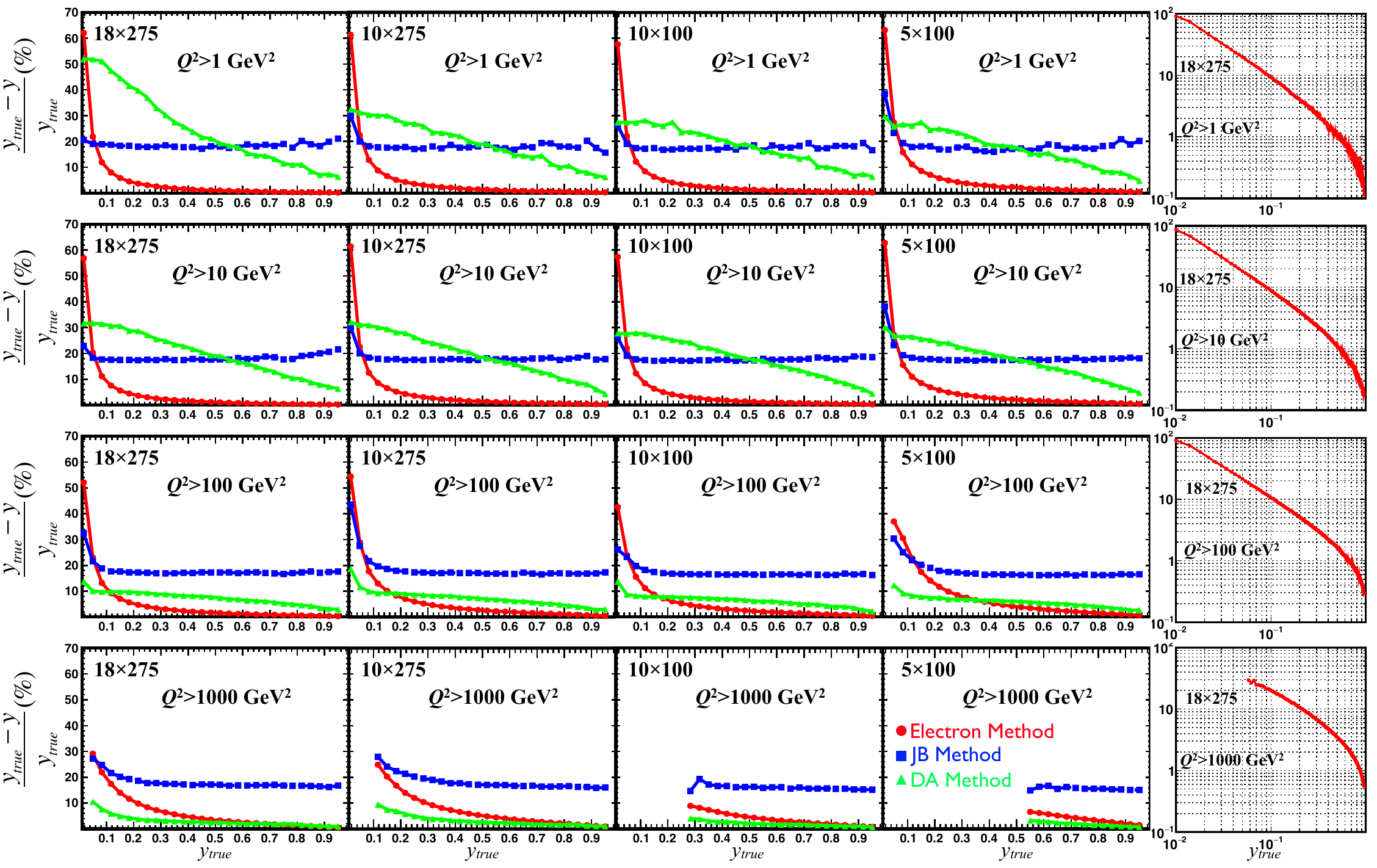}
	\caption{CORE relative $y$ resolutions (\textit{i.e.}, the standard deviation of event distributions over $(y_{true}-y)/y$, where $y$ is the reconstructed inelasticity) for DIS kinematics. The label $a\times b$ denotes the energy configuration of the colliding electron and proton beams, where $a$ and $b$ are in GeV. The resolutions are obtained by processing four different generated event samples through the CORE DELPHES simulation, each with a different minimum $Q^2$, in order to showcase detector performance in high- and low-cross-section kinematics independent of statistics. For each setting, resolutions obtained with the three standard methods, Electron, Jacquet-Blondel, and Double-Angle, are shown. The very right column shows only the resolutions obtained with the Electron Method on a log-log scale for a better visualization of the numerical values of the resolution of that particular method over the full range of $y$.}
	\label{fig:dis_sigmay_y}
\end{center}
\end{figure}

\subsection{Inclusive Reactions}
\label{inclusive}

One of the core objectives of the EIC physics program, and of inclusive DIS in particular, is the precise measurement of the virtual-photon--nucleon asymmetry $A_1$ aiming foremost at the gluon-spin contribution $\Delta G$. It also contributes strongly to constraining the quark-spin contribution $\Delta \Sigma$ to the proton spin. It does not require hadron PID as only the scattered lepton is required.
Thus, precise measurement of the scattered electron is fundamental, as for most of the EIC science program.
In particular, good reconstruction of the inelasticity $y$, which---in the target rest frame---is the ratio of the virtual photon’s energy to the beam-electron’s energy, is extremely important because the systematic uncertainties on $A_1$ from the measured longitudinal double-spin asymmetries depend on the reconstruction precision of the {\em depolarization factor} that is roughly proportional to $y$ at low to medium $y$.
The relative uncertainty on $y$, thus, translates to a contribution of similar size to the overall systematic uncertainty on $A_1$.

To evaluate the resolution effects of CORE on the reconstructed DIS kinematics $x$, $Q^2$, and $y$, neutral-current events generated with the Pythia8 generator, for several beam configurations, and $Q^2$ ranges were processed through DELPHES. Beam smearing and crossing angle of 25 mrad are accounted for in the generator. Radiative effects had been turned off. In DELPHES, the three-vector of a charged final-state particle is formed from angular measurements and momentum or energy measurement, depending on whether the resolution of the relevant calorimeter or the tracker is better for that specific particle. The reconstruction of the energy and angles of neutral particles is entirely based on calorimetric response.

\begin{figure}[htb!]
\begin{center}
	\includegraphics[width=0.8\textwidth]{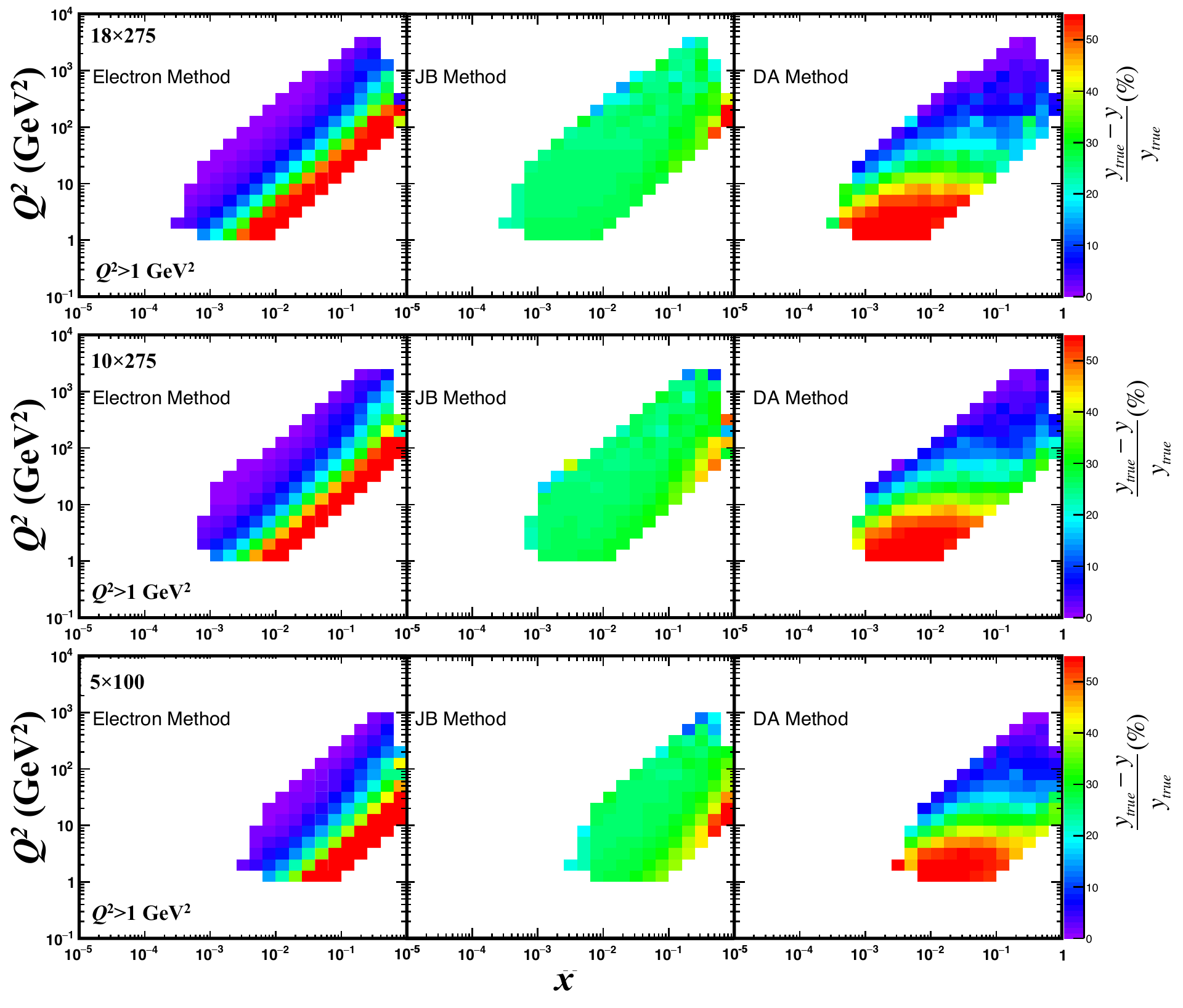}
	\caption{Relative $y$ resolutions for various $(Q^2,x)$ bins for several representative settings and reconstruction methods.}
	\label{fig:dis_sigmay_xQ2}
\end{center}
\end{figure}

While the reconstruction of the event kinematics from the measured energy/momentum and polar angle of the scattered electron (Electron Method) is the standard approach, it is well known that it yields quickly deteriorating resolutions at low $y$ and hadronic (Jacquet-Blondel~\cite{Jacquet:1979jb}) and mixed (Double-Angle~\cite{Bentvelsen:1992fu}) methods need to be used there. Here, we present the CORE performance in terms of the relative resolution of the inelasticity $y$ in bins of $y$ (see Fig.~\ref{fig:dis_sigmay_y}) as well as in bins of $(x,Q^2)$ (see Fig.~\ref{fig:dis_sigmay_xQ2}) obtained with all three methods. In the kinematics reconstruction with the latter two methods, we have used the energies and momenta of all detected final-state charged particles (excluding the scattered electron) and photons.

The excellent electron tracking resolution of CORE provides for a very good $y$ resolution (at percent level for most of the $y$ range, and even sub-percent level at large $y$) when using the Electron Method. The figures allow the reader to also assess the very-low-$y$ range, where the other method(s) take over.  

\begin{figure}[htb!]
\begin{center}
	\includegraphics[width=0.9\textwidth]{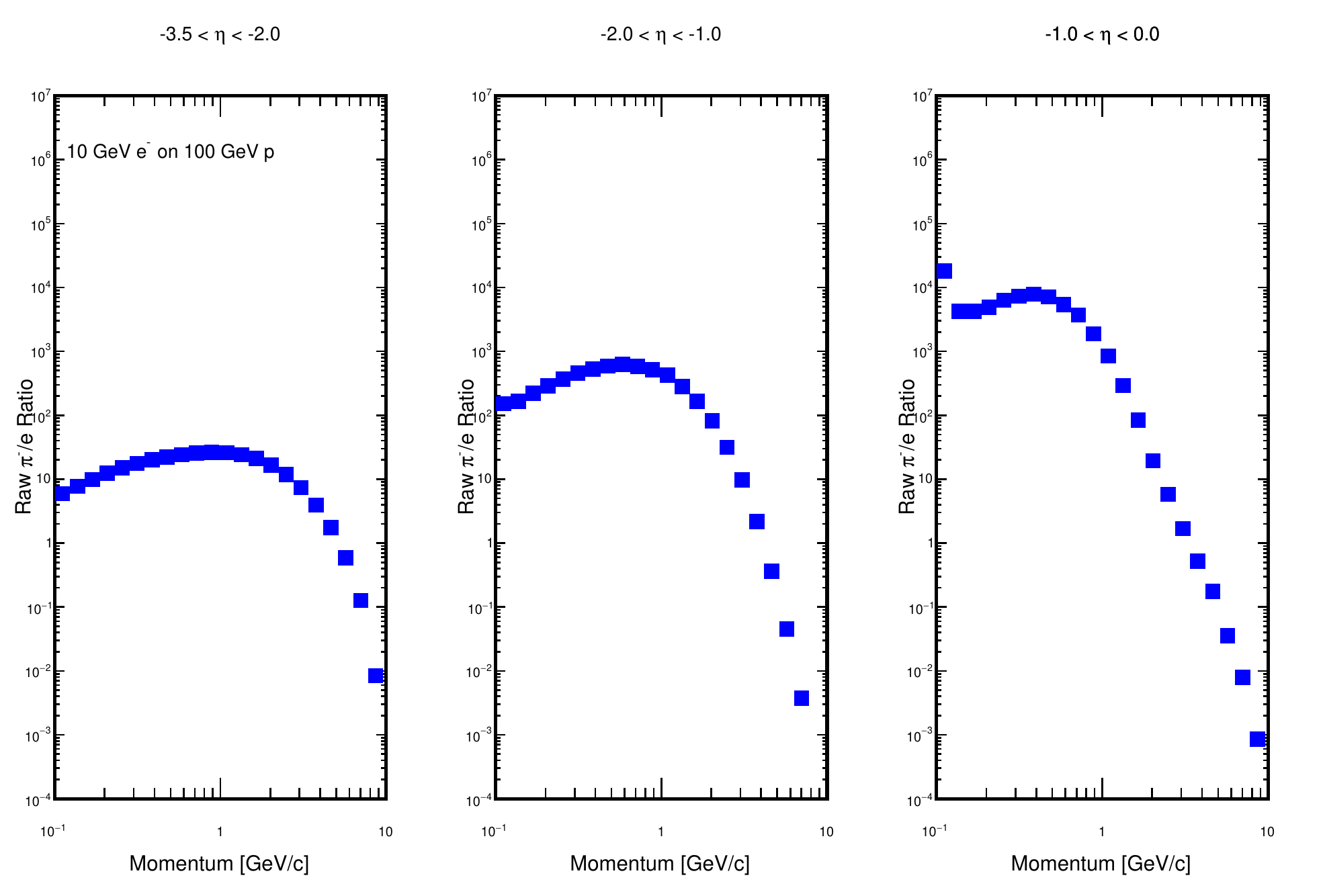}
	\caption{$e/\pi$ ratios in the electron hemisphere for 10x100 GeV. The left plot, covering $-3.5<\eta<-2$, corresponds to the CORE endcap (which extends to $\eta<-1.65$), while the other two correspond to the barrel. The relative $\pi^-$ background gets worse at increasing (less negative) values of $\eta$. There are very few scattered electrons going into the hadron hemisphere ($\eta>0$).}
	\label{fig:e_pi_ratio}
\end{center}
\end{figure}

Another important aspect of any EIC measurement is a reliable identification of the scattered electron. The main challenge is rejection of the $\pi^-$ background, which is large at low momentum. The $e/\pi$ ratio also has a strong angular dependence, and gets worse at larger (\textit{i.e.}, less negative) values of $\eta$, as shown in Fig.~\ref{fig:e_pi_ratio}.

\begin{figure}[tb!]
\begin{center}
	\includegraphics[keepaspectratio=true,width=2.1in,page=1]{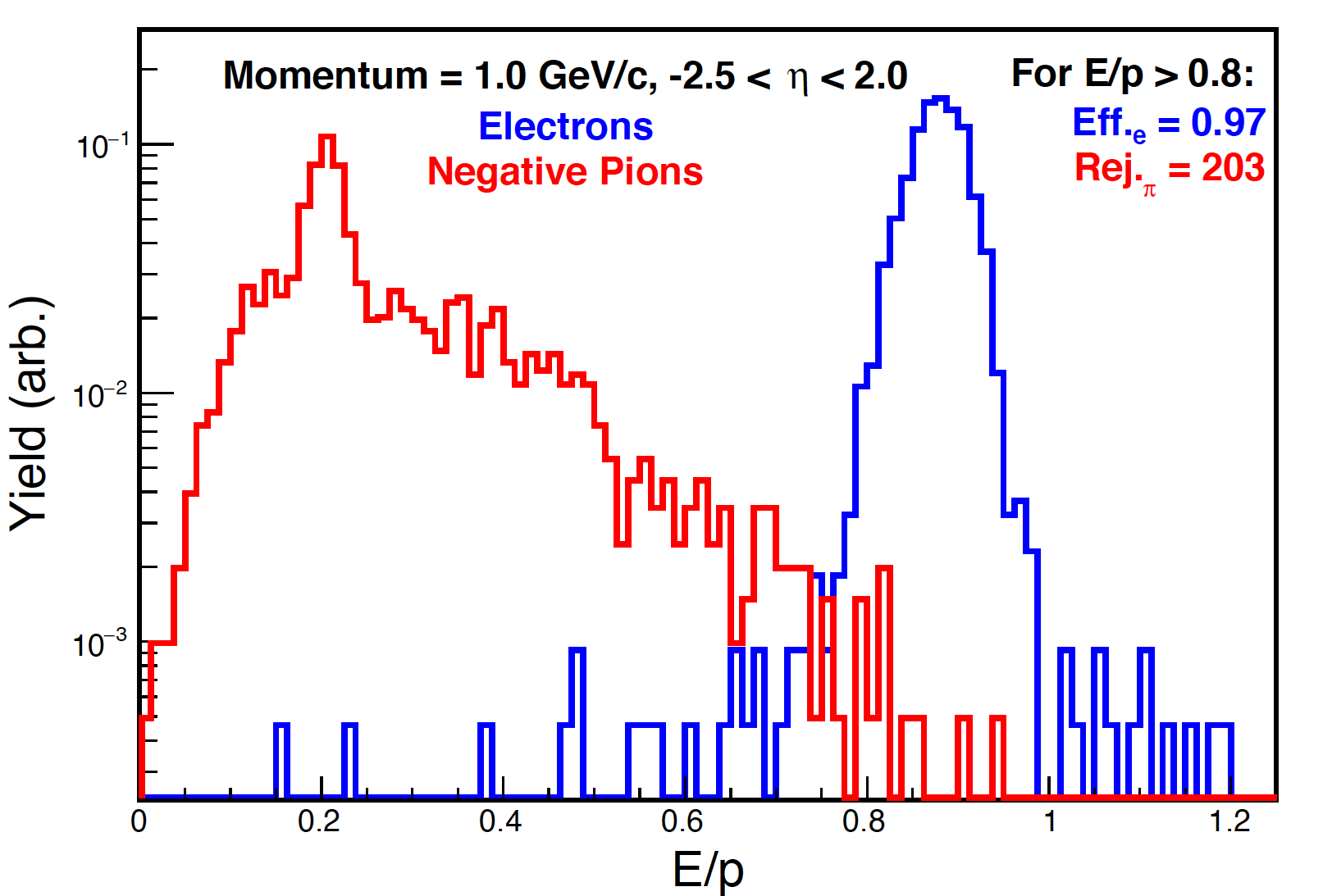}
	\includegraphics[keepaspectratio=true,width=2.1in,page=1]{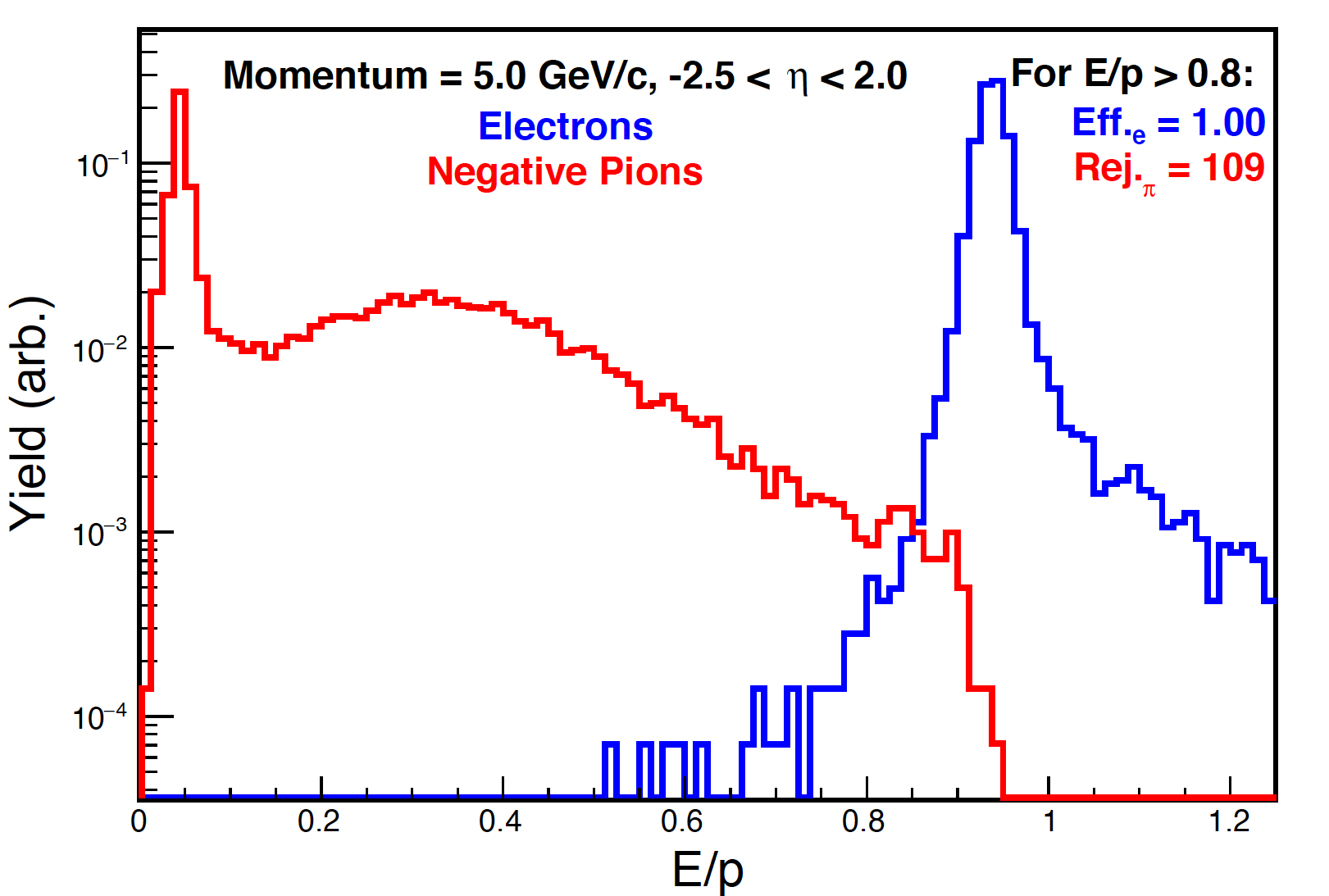}
	\includegraphics[keepaspectratio=true,width=2.1in,page=1]{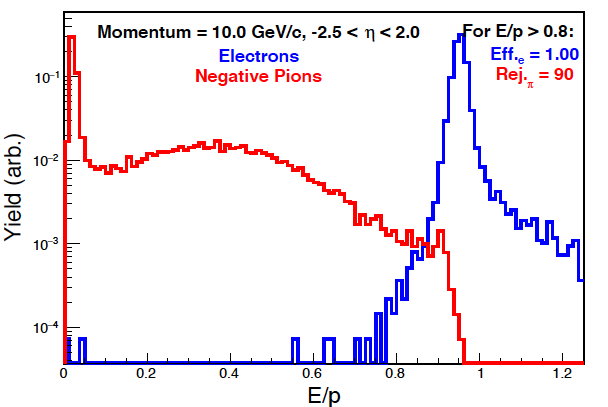}
	\caption{A Geant4 simulation of the $e/\pi$ identification in the PbWO$_4$ EMcal for three incident energies using the E/p ratio, where E is measured by the EMcal and p by the tracker. The simulation shows a segment ($-2.5<\eta<-2$) of the CORE endcap EMcal, but is also representative for the barrel. The E/p cut at 0.8 is arbitrary and can be adjusted as needed (and affects the three energies shown in a  different way since the electron peak shifts slightly). As shown in the last two plots (5 and 10 GeV), a rejection factor of 100 can be reached with almost no loss in efficiency. The first plot (1 GeV) shows a rejection factor of 200, but with an efficiency of 97\%, which could be a reasonable E/p cut for the lower energies.}
	\label{fig:E_over_p}
\end{center}
\end{figure}

Since electron identification is significantly more challenging for electrons going into the barrel than into the electron endcap, CORE extends the high-resolution PbWO$_4$ EMcal coverage up to $\eta$=0, and uses the DIRC as a supplementary detector for $e/\pi$ ID at low momenta.
Additional pion suppression will also come from kinematic constraints. A requirement on the total $E-p_z$ (summing all particles with $|\eta|<3.5$) can give a pion suppression factor of about 10-20.
Vetoing on electrons in the far-backward detectors, and selecting the best electron candidate when multiple candidates are present in the event may also help.

Figure~\ref{fig:E_over_p} shows the $e/\pi$ separation using the $E/p$ ratio, where the energy $E$ is reconstructed by the PbWO$_4$ EMcal and the momentum $p$ by the tracker. A pion rejection factor of 100 can be reached with a high efficiency.

The excellent electron ID provided by CORE would be particularly important for measurements of parity-violating DIS, which benefits from high luminosity and critically relies on a high-purity electron identification.
The $e/\pi$ capability of CORE could be further enhanced by using thinner DIRC bars, which is an option discussed in section \ref{hpDIRC}.

\subsection{Semi-Inclusive Reactions}
\label{SIDIS}
The semi-inclusive deep-inelastic scattering (SIDIS), where a final produced hadron is detected in coincidence with the scattered lepton, is one of the important physics channels to be explored by the EIC.
First of all, SIDIS allows one to employ the hadronization as the tool for parton flavor separation. This will be essential in order to separate the contribution of different flavors to various parton distribution and fragmentation functions that describe the structure of the nucleon and of the hadronization process.
SIDIS also provides the necessary additional degrees of freedom to tag partonic transverse momentum as well as to enable parton polarimetry in the final state. This can be exploited for disentangling various correlations~\cite{Accardi:2012qut,AbdulKhalek:2021gbh} of the parton motion and the spin encoded in the underlying TMD parton distribution and fragmentation functions.

Therefore, SIDIS contributes to three out of five major science goals the EIC is going to address: 
\begin{itemize}
    \item The origin of the proton spin: while the gluon helicity distribution ($\Delta G$) will be constrained best through inclusive DIS at the EIC (see above discussion of inclusive DIS), quark helicity distributions can be studied in more detail with SIDIS. Flavor separation and in this respect the contribution of sea quarks will profit strongly from the flavor-discriminating hadronization. Kaons, in particular, will be essential to decrease uncertainties on the strange-quark helicity, without having to rely on the SU(3) flavor symmetry in analyses of inclusive DIS. 
    \item The three-dimensional structure of the nucleon in the momentum space: SIDIS measurements will allow access to TMD distribution and fragmentation functions that encode the 3D nucleon structure. In particular, one of the driving physics cases, the Sivers functions, will be measured via specific azimuthal modulations of the transverse nucleon-spin Sivers asymmetries ($\propto\sin(\phi_h-\phi_S)$, for the definition of the angles see below), the nucleon transversity distributions and the nucleon tensor charge will be measured via another modulation, the Collins asymmetries ($\propto\sin(\phi_h+\phi_S)$). This requires, in addition to the requirements for the helicity distributions, measurement of azimuthal angles and the transverse momentum $P_{hT}$ of the hadron with respect to the virtual photon.
    \item The understanding of quark hadronization is arguably the latest QCD frontier. Not only is it a topic of interest by itself, it is also {\em the} tool for parton flavor separation discussed above. Hadron multiplicities or SIDIS cross sections from scattering by protons or very light nuclei will be crucial input in improving fragmentation function extractions. On the other hand, using various nuclear beams the EIC will be extremely important for unraveling details of the hadronization process and for the study of the in-medium hadronization effects. Furthermore, jets will play an important role as probes for hadronization, e.g., through hadron+jet observables (see, e.g., the YR~\cite{AbdulKhalek:2021gbh}).
\end{itemize}

\begin{figure}[b!]
\begin{center}
	\includegraphics[width=0.38\textwidth]{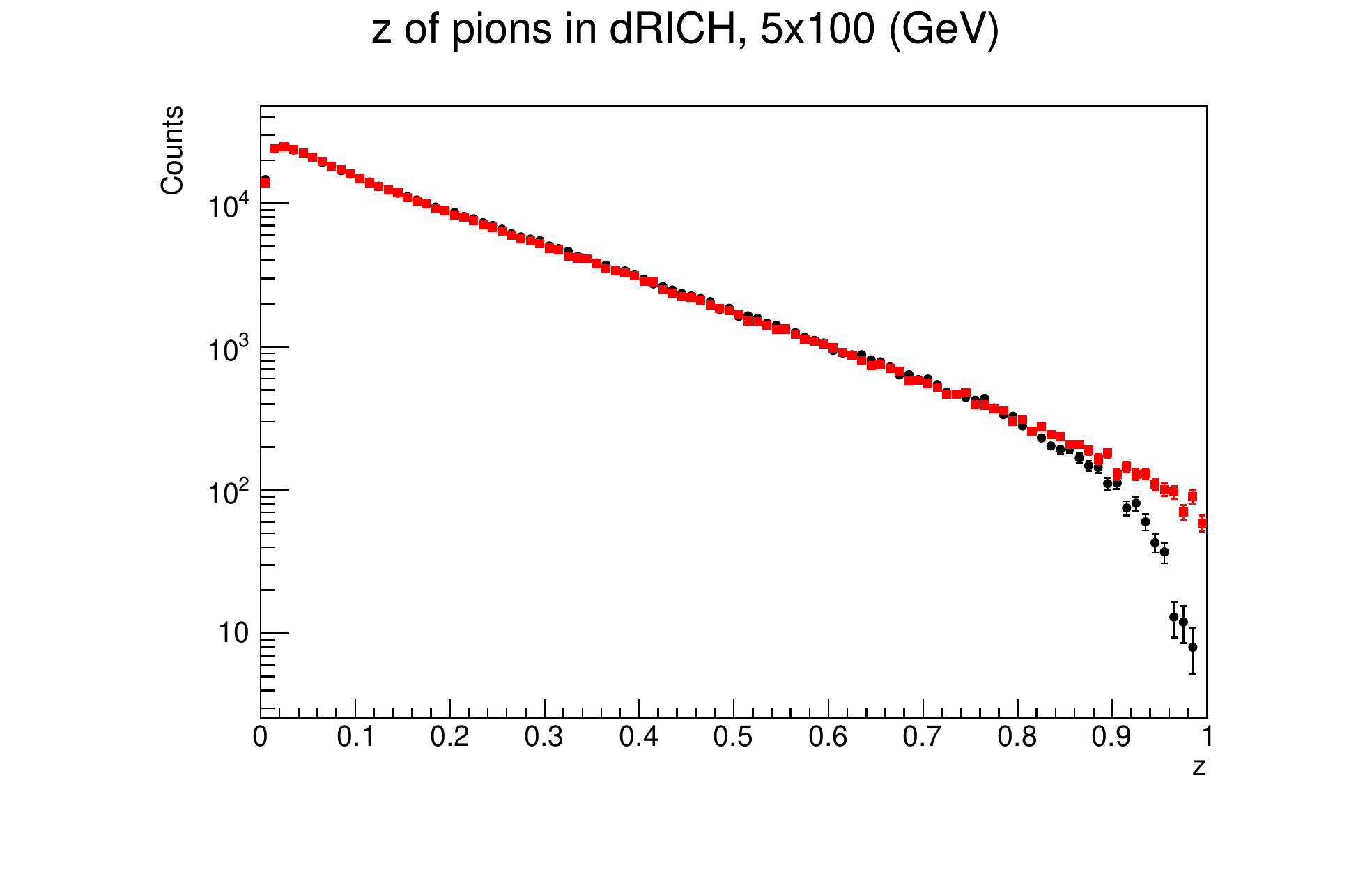}
	\includegraphics[width=0.38\textwidth]{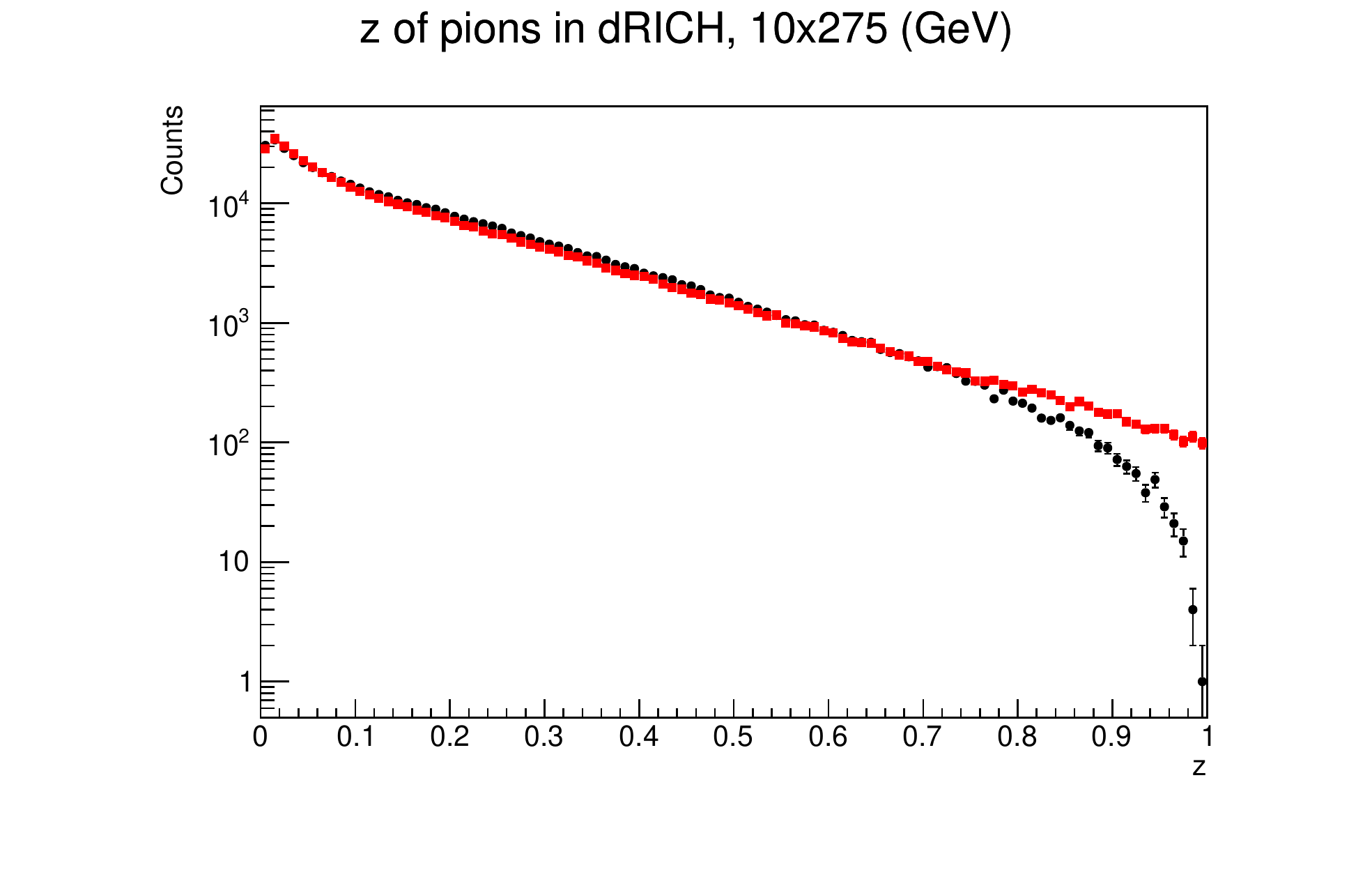} \\
	\includegraphics[width=0.38\textwidth]{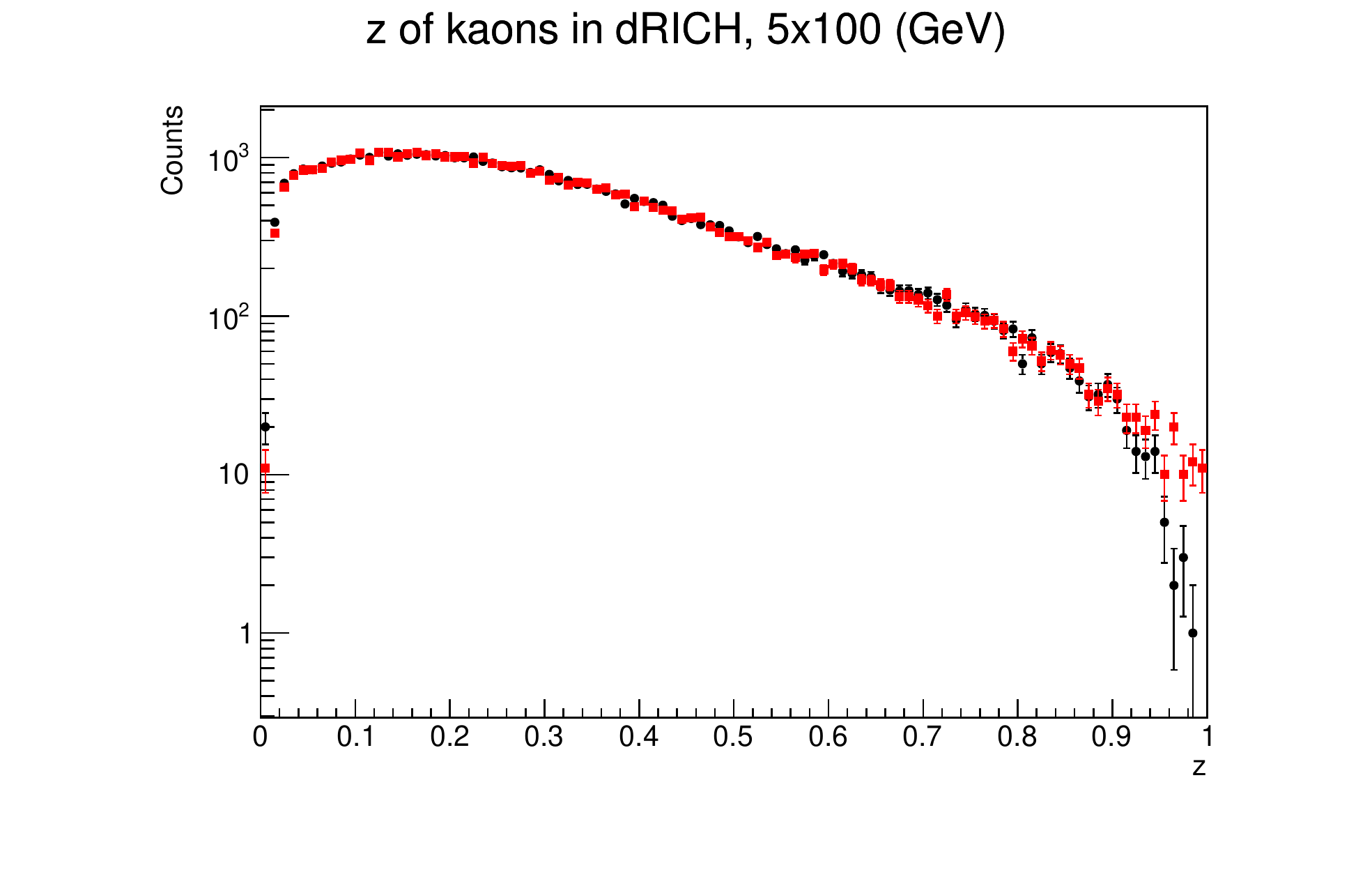}
	\includegraphics[width=0.38\textwidth]{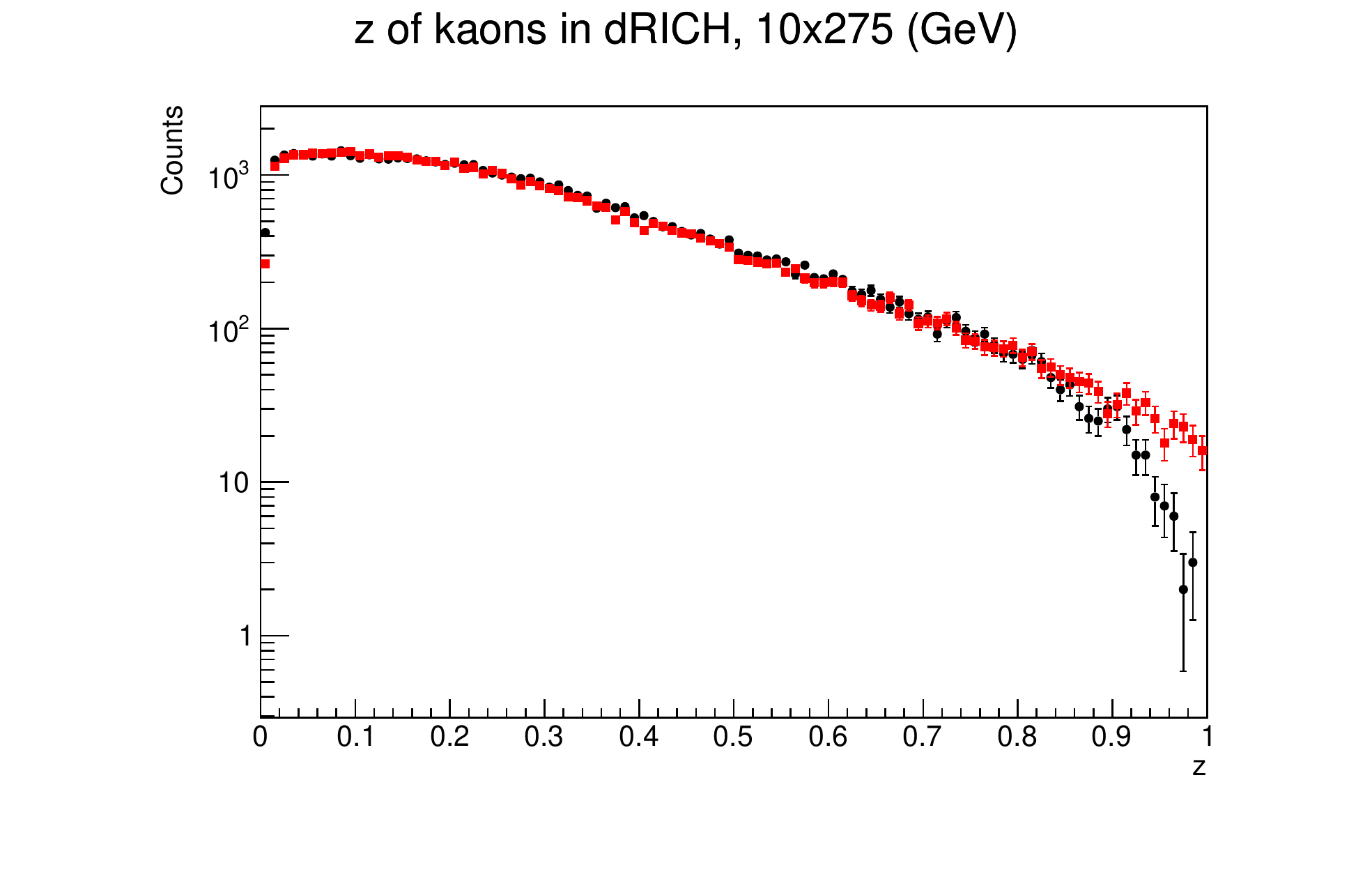}\\
	\includegraphics[width=0.38\textwidth]{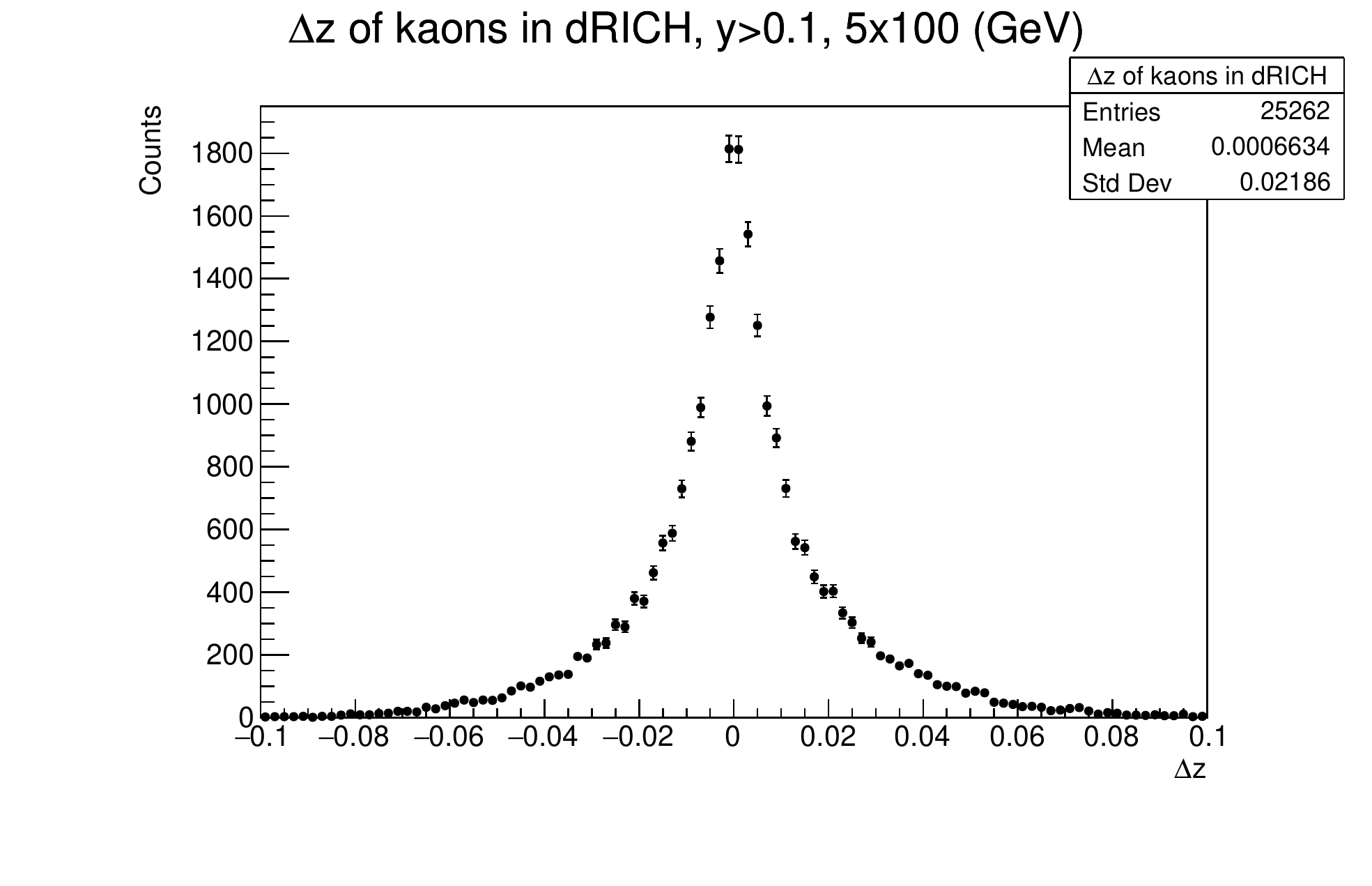}
	\includegraphics[width=0.38\textwidth]{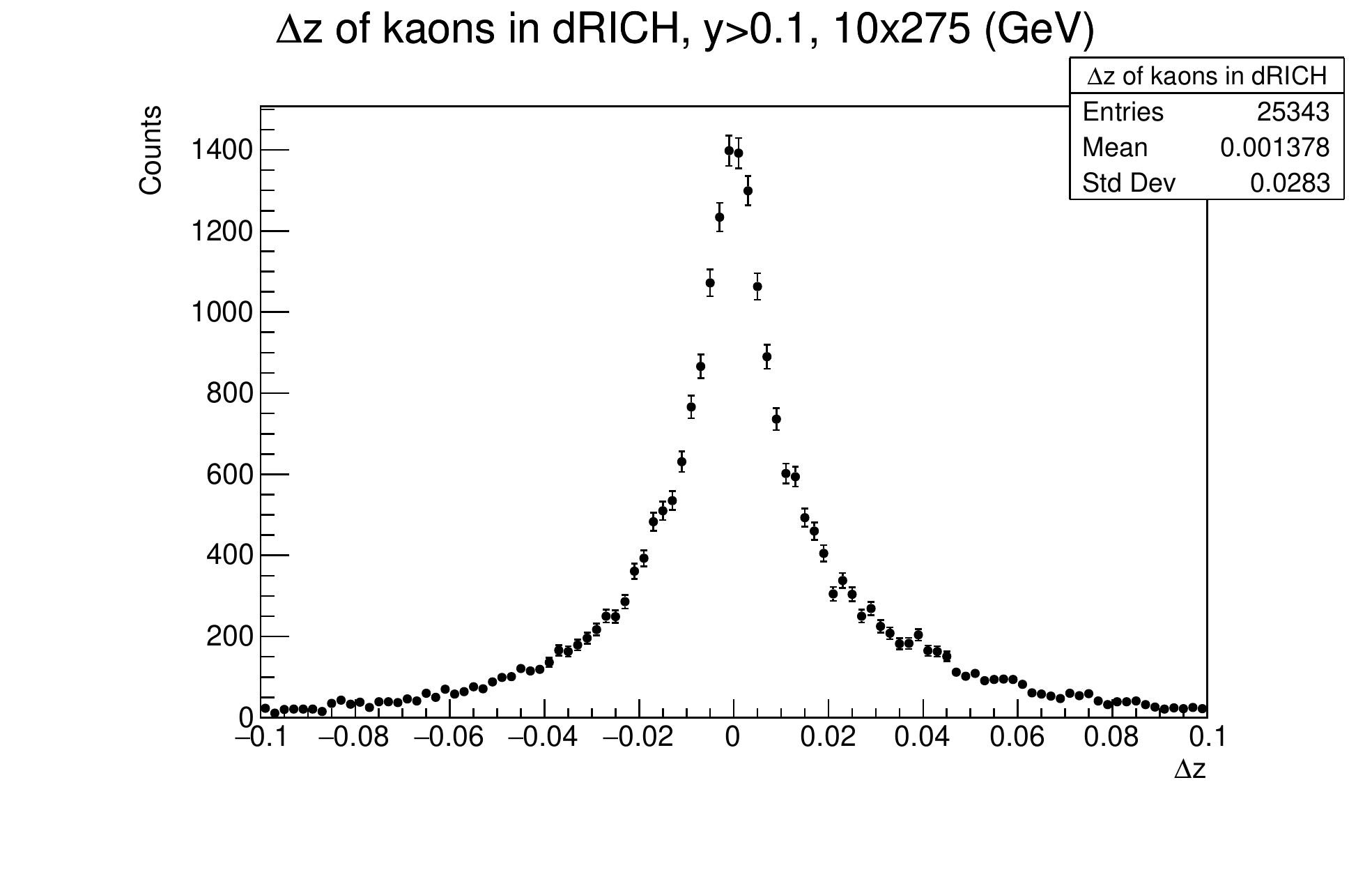} \\
	\includegraphics[width=0.38\textwidth]{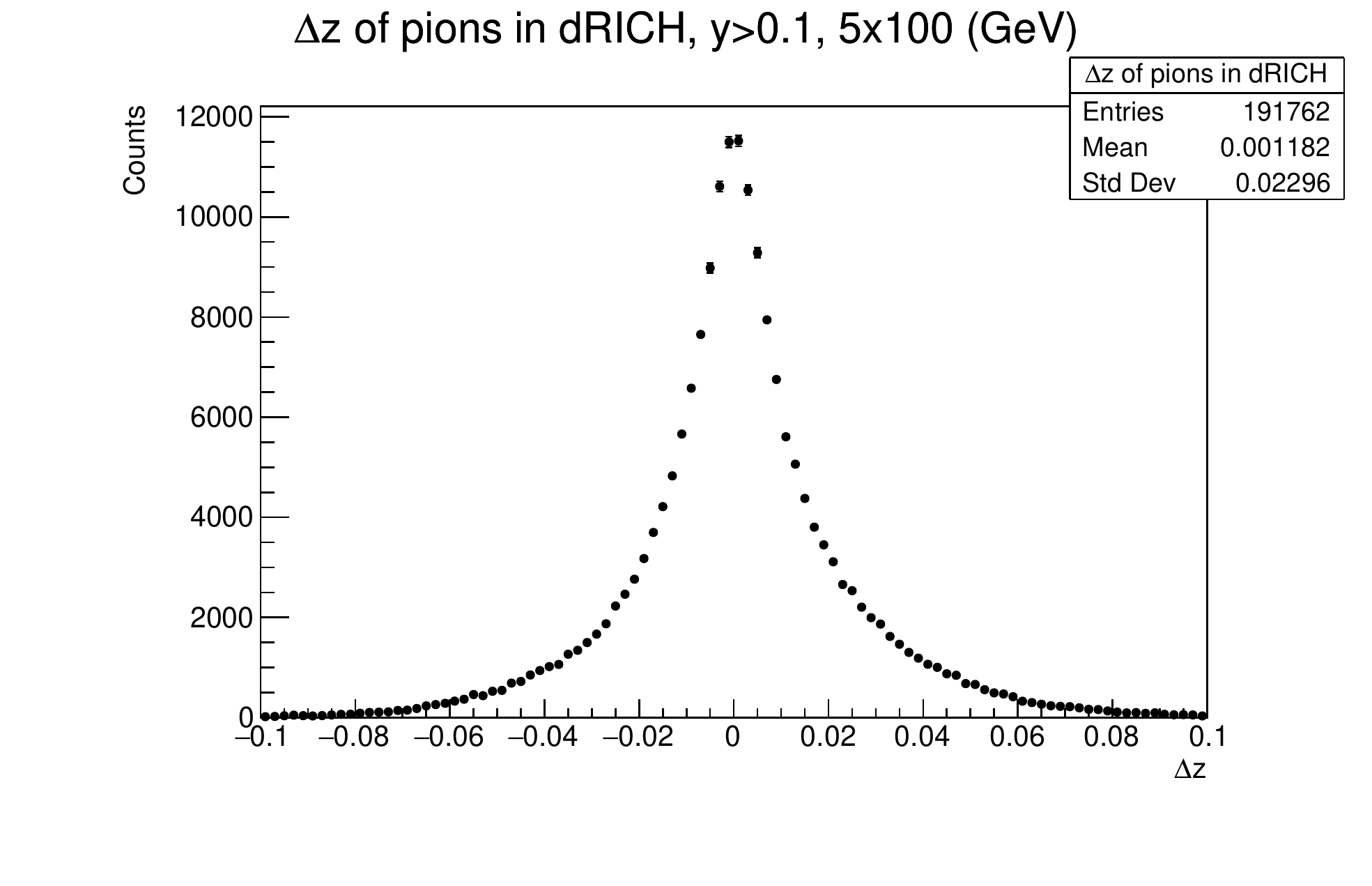} 
	\includegraphics[width=0.38\textwidth]{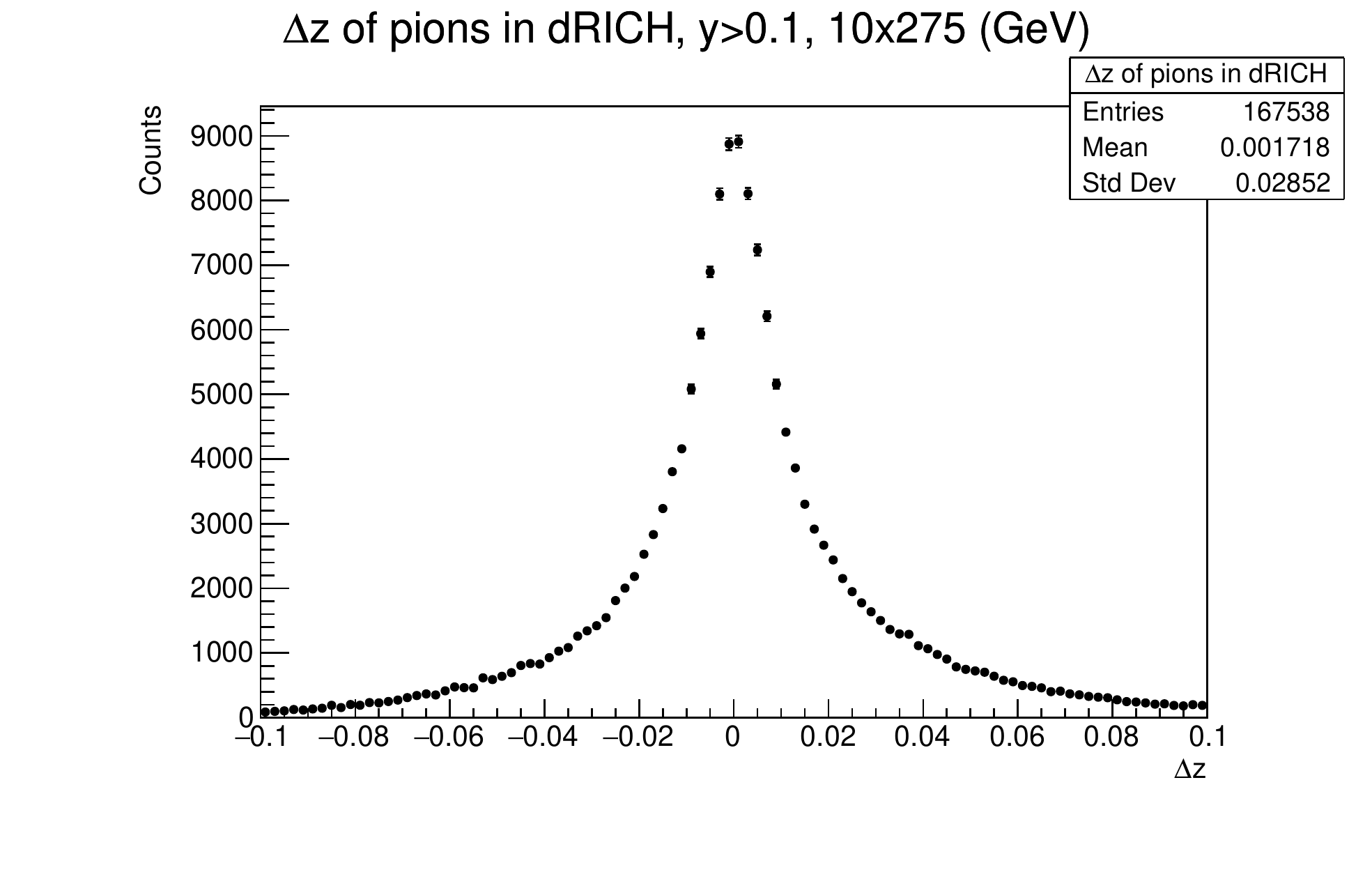}
	\caption{%
	\textbf{top 4 panels:} $z$ distributions of pions and kaons (as labelled) in the dRICH for the 5$\times$100 (left column) and 10$\times$275 (right column) beam-energy settings. Black points are generated kinematics; red squares correspond to kinematics as reconstructed by the CORE resolution simulated with DELPHES, and where the Electron Method is used for the reconstruction of DIS kinematics. At large values of $z$, migration into the unphysical region ($z>1$) is observed, arising, e.g., from $y$ smearing. 
	\textbf{bottom 4 panels:} Corresponding $z$ resolutions,
	avoiding the low-$y$ region where using the Electron Method is less appropriate and where other methods would be employed.}
	\label{fig:sidis_zCoverage}
\end{center}
\end{figure}
\begin{figure}[htb!]
\begin{center}
	\includegraphics[width=0.38\textwidth]{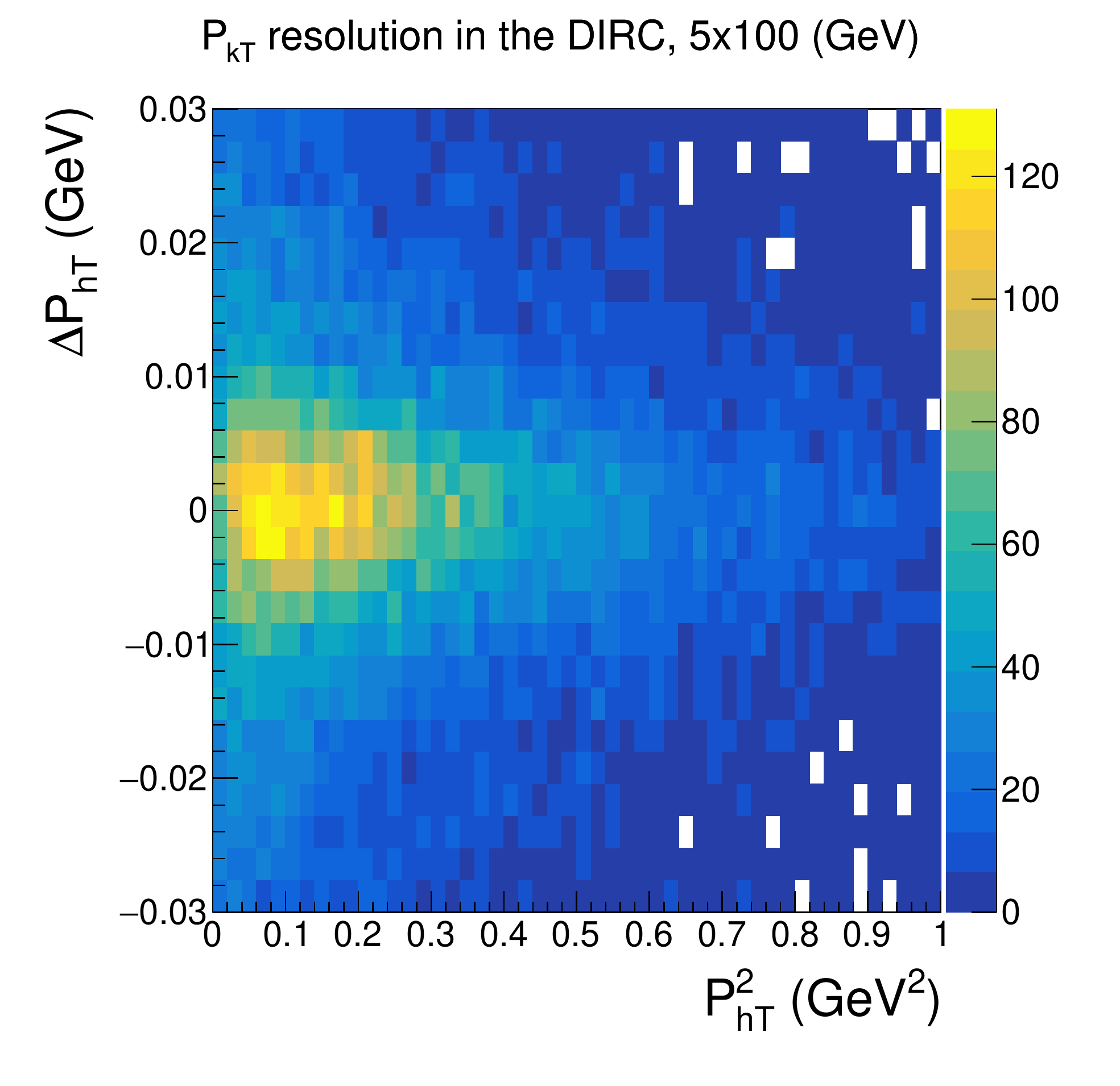}
	\includegraphics[width=0.38\textwidth]{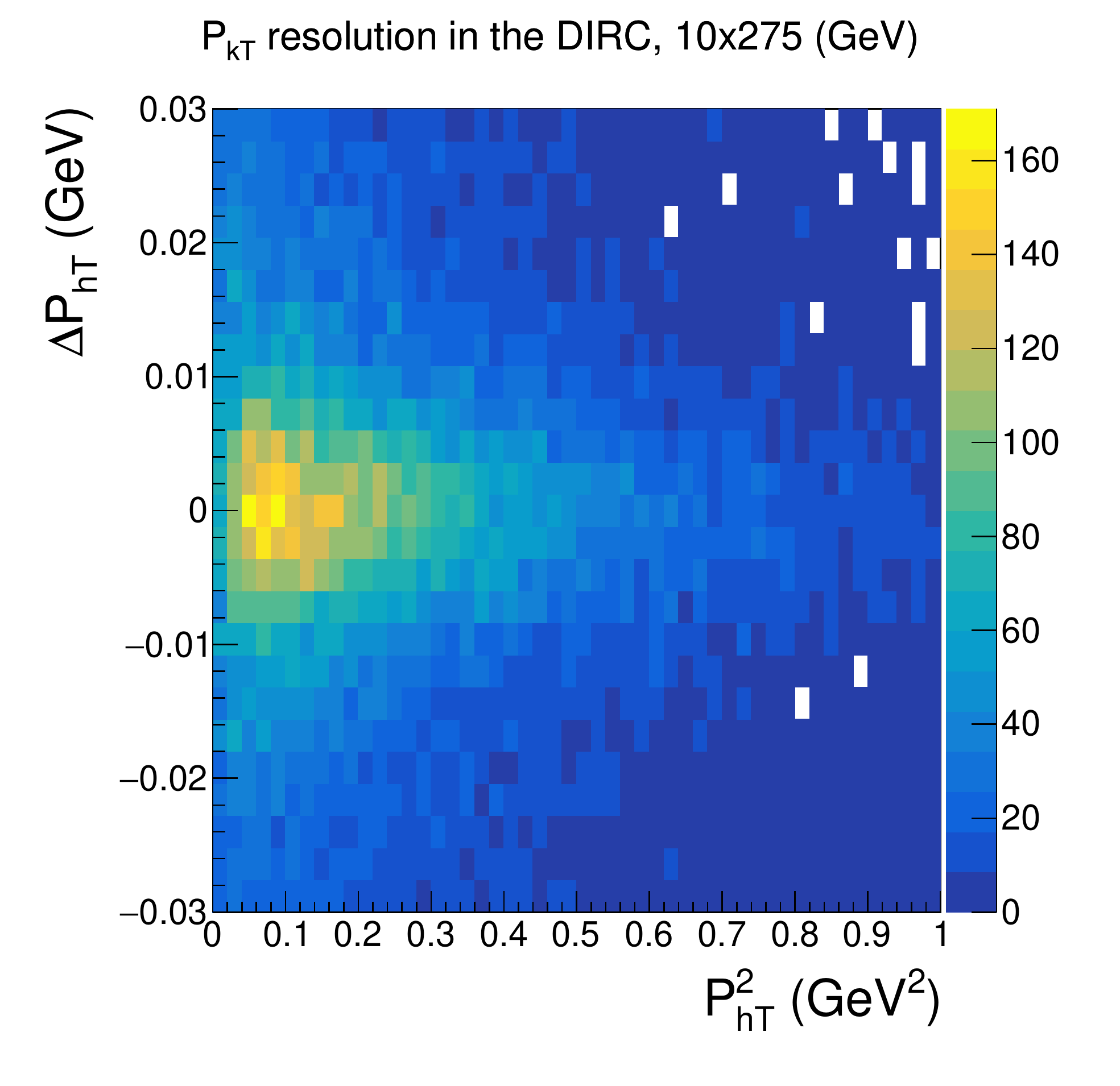}
	 \\
	 \includegraphics[width=0.38\textwidth]{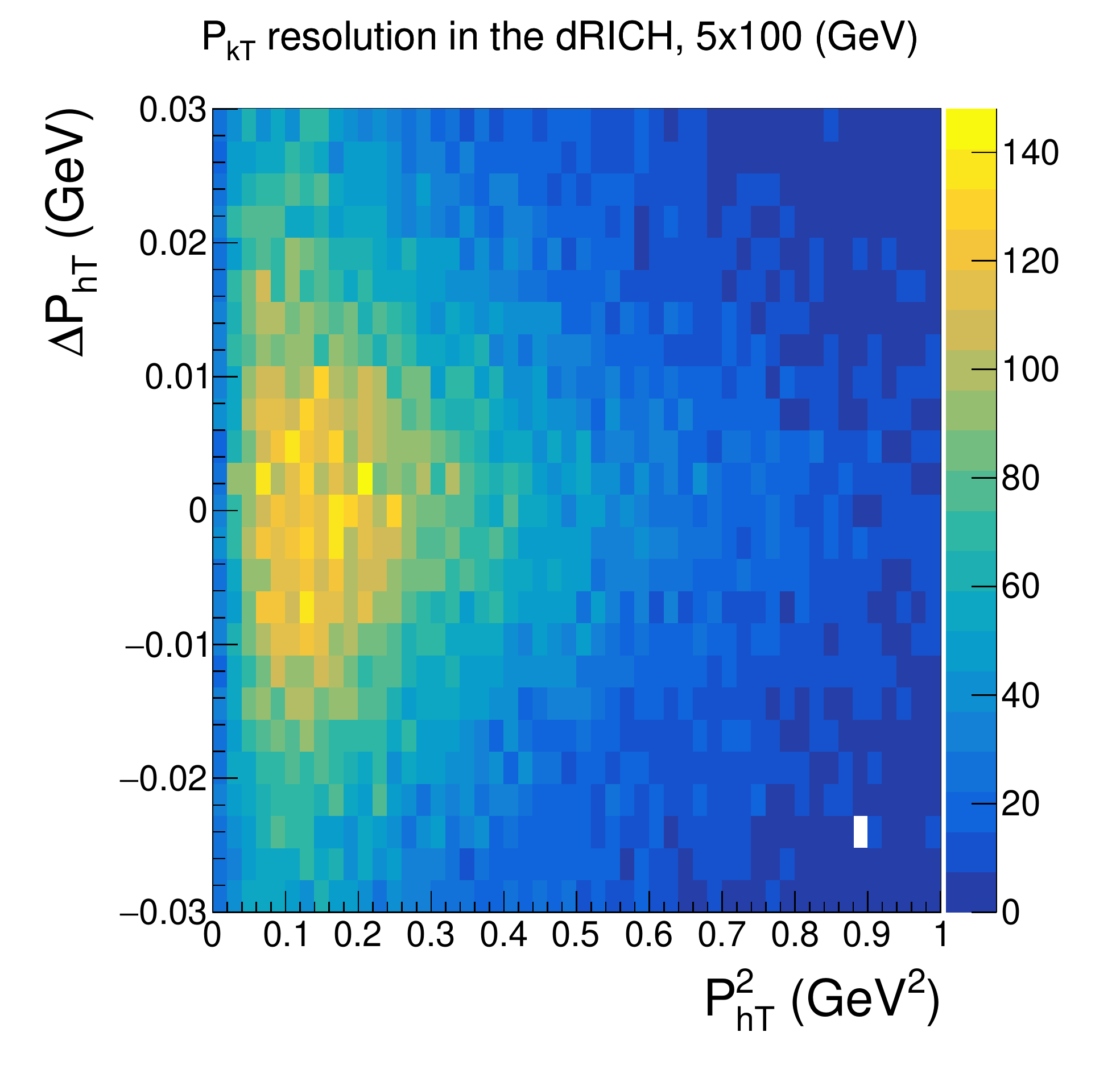}
	 \includegraphics[width=0.38\textwidth]{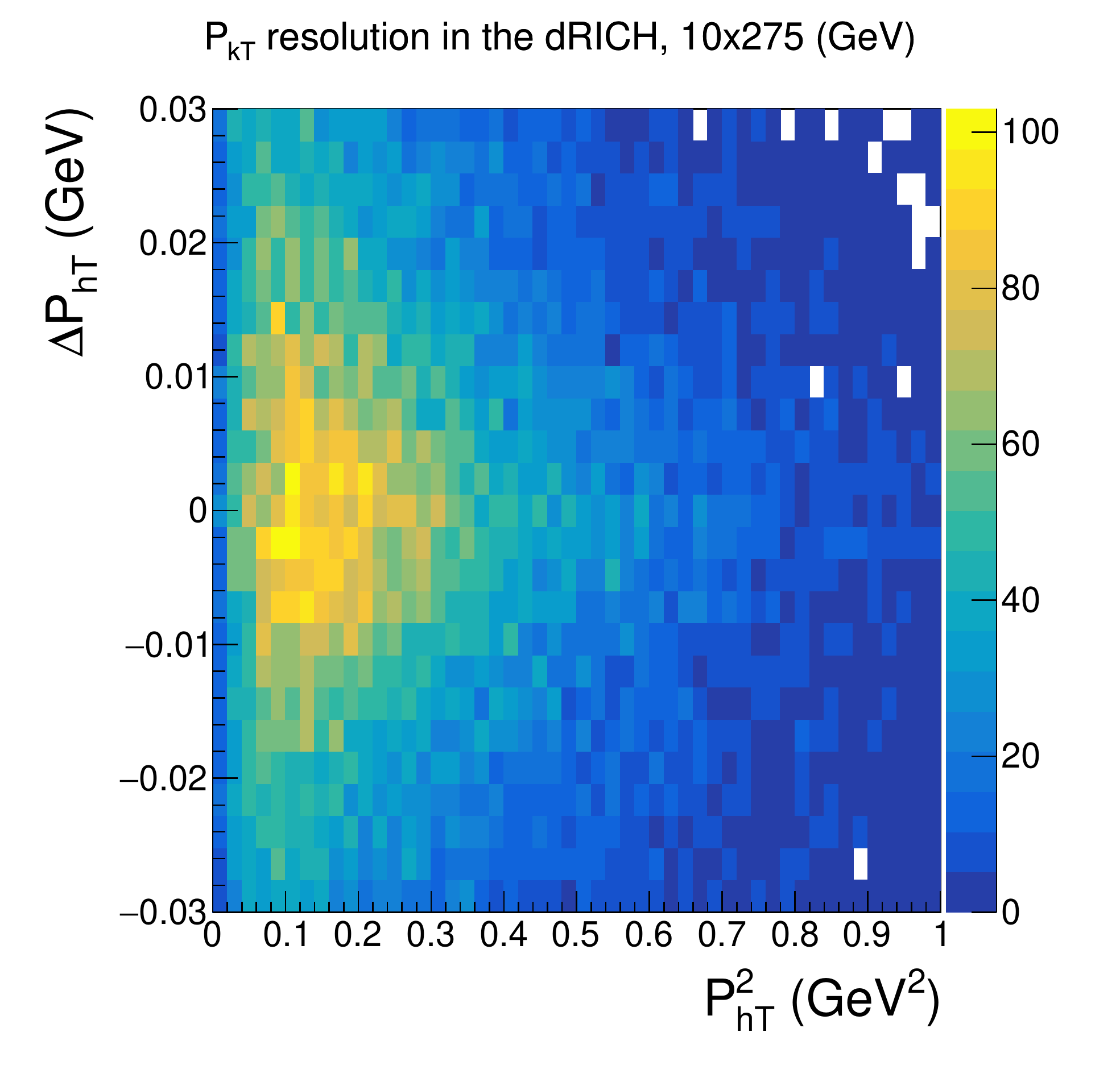}
	 \\
	\includegraphics[width=0.38\textwidth]{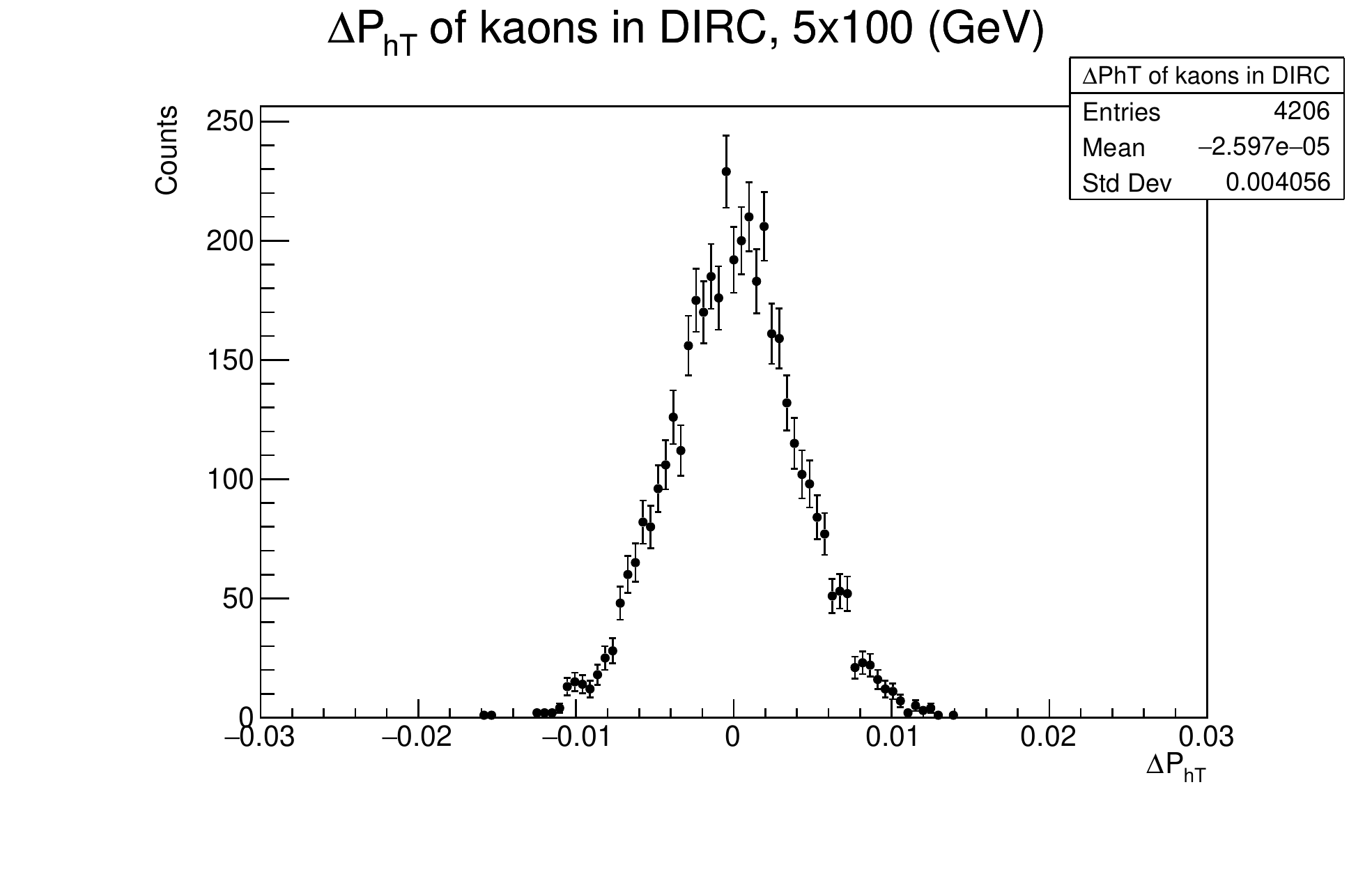}
	\includegraphics[width=0.38\textwidth]{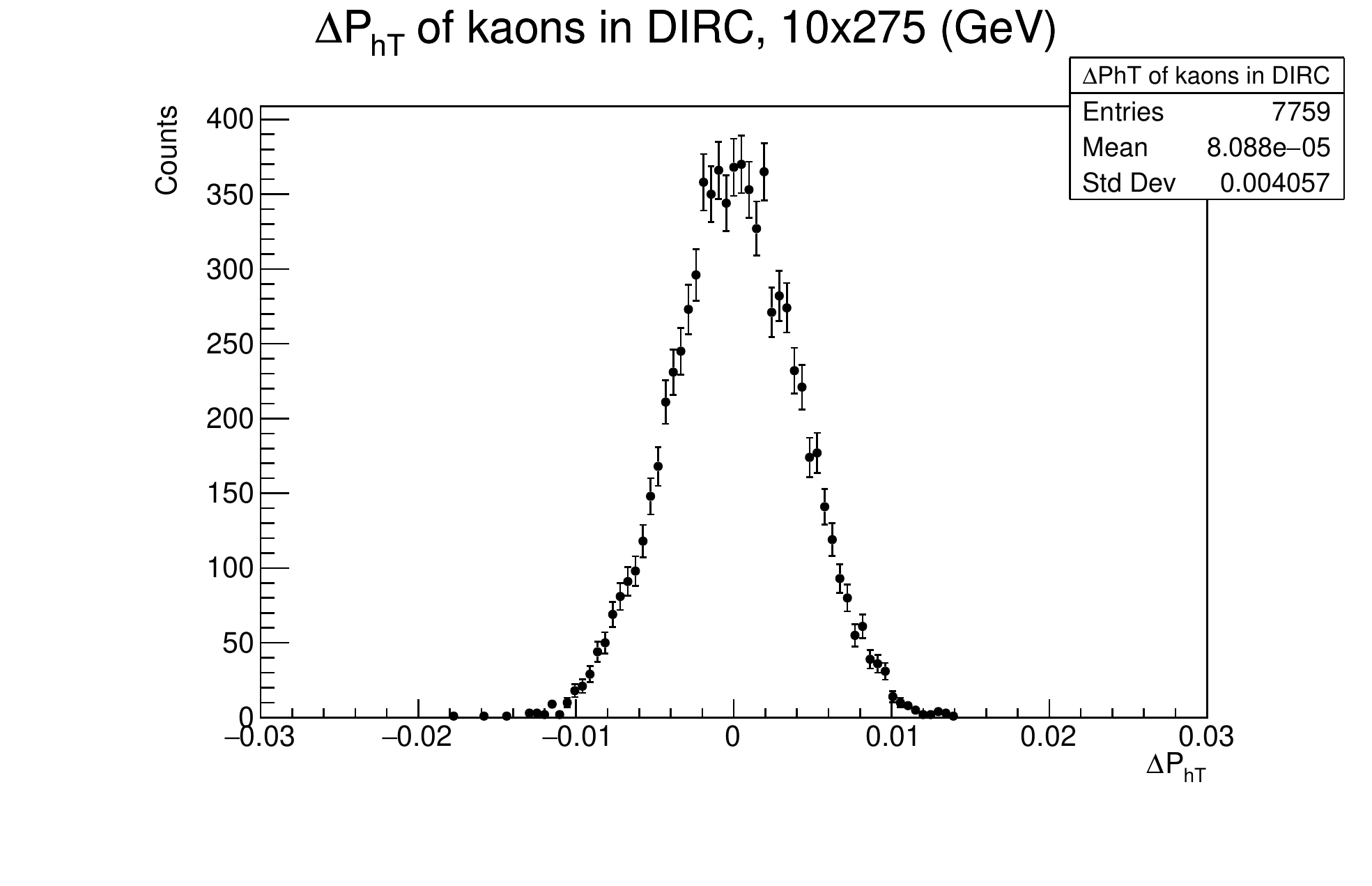} 
	\\
	\includegraphics[width=0.38\textwidth]{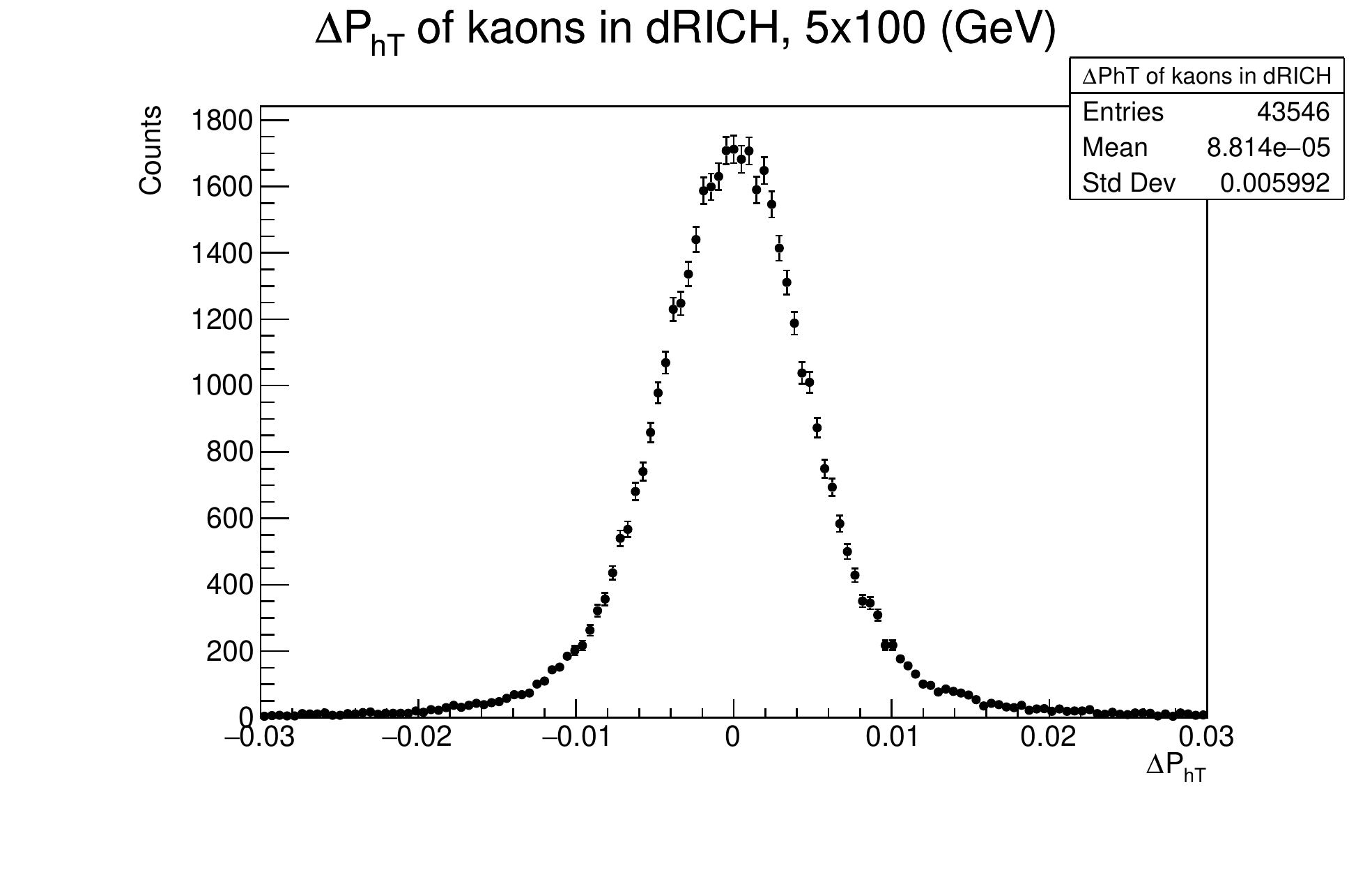}
	\includegraphics[width=0.38\textwidth]{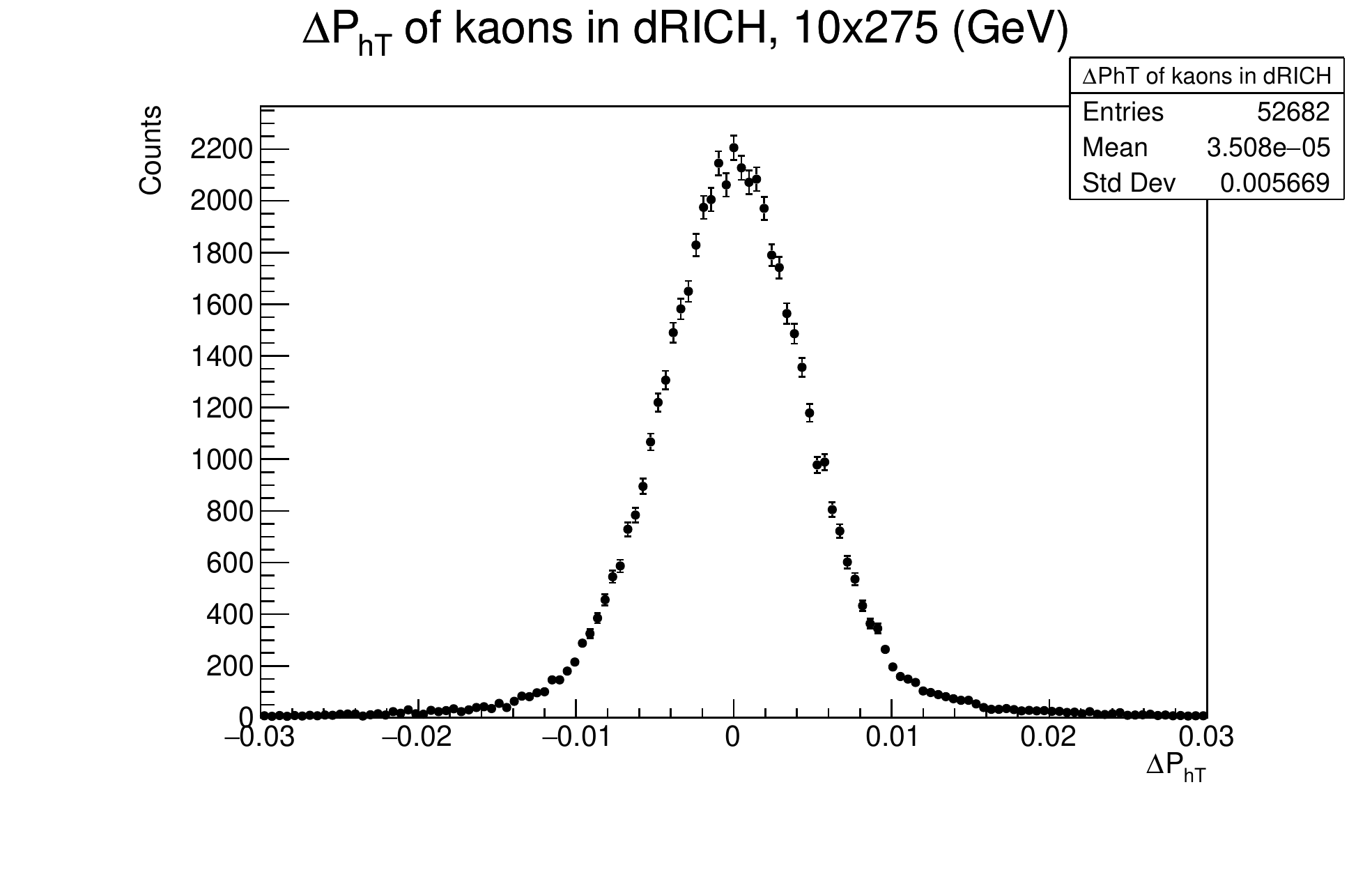}
	\caption{Resolution in the kaon's transverse momentum (w.r.t.~the virtual photon) for kaons in the DIRC or dRICH for the 5$\times$100 (left column) and 10$\times$275 (right column) beam-energy settings. Top two rows depict the resolution vs.~transverse kaon momentum; the bottom two rows show the canvas integrating over transverse momentum. The Electron Method is used for reconstruction of the DIS kinematics, and the CORE resolution is again simulated with DELPHES.}
	\label{fig:sidis_PhperpResolution}
\end{center}
\end{figure}

In order to fully realize the potential of SIDIS measurements, one needs to ensure a very good PID and momentum coverage and resolution, especially for the transverse momentum $P_{hT}$ of the produced hadron and for the azimuthal angles of both the target-spin orientation, $\phi_S$, and of the produced hadron, $\phi_h$, with respect to the lepton scattering plane. Furthermore, very good reconstruction of the DIS kinematics is vital. As discussed already in Section~\ref{inclusive}, for the inclusive-DIS double-spin asymmetry $A_1$, the resolution in $y$ directly enters the scale uncertainty on $A_1$ and hence on the helicity distributions extracted from it. The situation is the same for measurements of longitudinal SIDIS double-spin asymmetries, again to be used for the determination of helicity distribution and in particular of quark-flavor--separated helicity distributions.
This, however, is less relevant for most of the other leading spin asymmetries aimed at mapping out the three-dimensional (spin)structure of the nucleon in momentum space, except for the double-spin asymmetry related to the worm-gear TMD $g_{1T}$ and several sub-leading asymmetries~\cite{Bacchetta:2006tn}.
It is important to repeat the observation made in Section~\ref{inclusive} that the excellent electron tracking resolution of CORE will allow for a 
clearly sufficient $y$ resolution 
for basically the complete $y$ range.

The main requirements for SIDIS measurements are laid out by the WP~\cite{Accardi:2012qut} and YR~\cite{AbdulKhalek:2021gbh}. PID for SIDIS is one of the crucial aspects. CORE will be equipped with various hadron-PID detectors, which will be discussed in more detail in Section~\ref{sec:PID} where Fig.~\ref{fig:kaons_10x275_5x100} demonstrates very good coverage of the main kinematic space of hadrons by the PID detectors, allowing for efficient (at least 3$\sigma$) separation of kaons from pions. The DIRC in the central region and dRICH in the hadron endcap provide kaon ID up to very large kaon momenta. This is especially important as those momenta correspond to a large value of the fractional hadron energy $z$ and thus to kaons that have a large sensitivity to fragmenting strange quarks. This will be beneficial in probing the strange sea of the nucleon, which is hard to probe otherwise in neutral-current DIS. The $z$ distributions for pions and kaons in the acceptance of the dRICH are shown in Fig.~\ref{fig:sidis_zCoverage}, demonstrating for both hadron species coverage over the whole $z$ range and very good resolution. The reconstructed distributions are obtained from the DELPHES fast Monte Carlo \cite{deFavereau:2013fsa}, with the implementation of CORE adapted from \cite{Arratia:2021uqr}.

The resolution in $P_{hT}$ is presented in Fig.~\ref{fig:sidis_PhperpResolution} for kaons identified by the DIRC or dRICH, where two representative beam-energy configurations (5$\times$100 and 10$\times$275) are chosen. In all cases, the resolution is well suited to probe TMD physics, allowing also for a fine binning in $P_{hT}$ in the interesting low-$P_{hT}$ region.

\begin{figure}[htb!]
\includegraphics[width=0.99\textwidth]{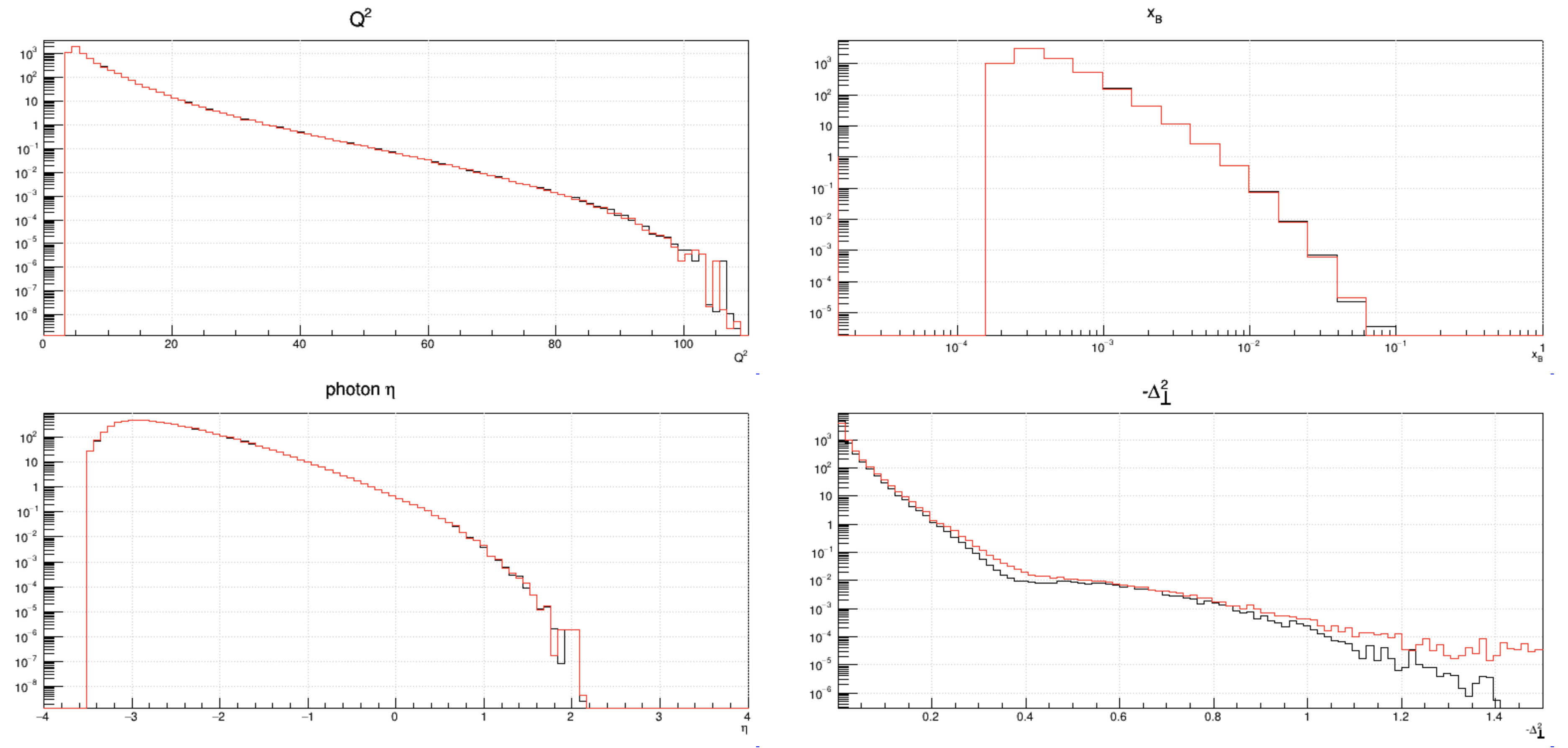}
\caption{\baselineskip 13 pt
Kinematic distributions of the DVCS $\alpha(e,e'\gamma)\alpha$ reaction for 10 x 137.5 GeV per nucleon.
The $\eta$-histogram is the distribution of the the final photon pseudo rapidity.
Vertical axes values are the cross sections integrated over the bin (and integrated over all other variables), in units of nano-barn. Statistics for a 10 fb$^{-1}$ run are ten million times the cross section values.  
The black histograms are the events within the central CORE acceptance, and the red histograms are the distributions, as reconstructed by the CORE resolution simulated with DELPHES.}
\label{fig:TOPEG-alpha-kin}
\end{figure}

\subsection{Exclusive Reactions}
\label{exclusive}

Deep virtual exclusive scattering (DVES) reactions probe the transverse distributions
of quarks and gluons, as functions of their light-cone momentum fractions, in the proton, neutron, and atomic nuclei.
Although the gluon distributions will dominate the DVES amplitudes at low-$x_B$, it is still of great interest to 
distinguish the transverse sizes of the quark and gluon distributions.
The vector meson $(e,e'V)$ and Compton $(e,e'\gamma)$ amplitudes have different sensitivities to the quark and gluon distributions, and are therefore essential for a fuller understanding of
nucleon and nuclear structure.  In nuclei, the evolution of the transverse density profiles with decreasing $x$ is an important signature of long distance QCD correlations, including shadowing and the approach to saturation.


\begin{figure}[htb!]
\includegraphics[width=0.95\textwidth]{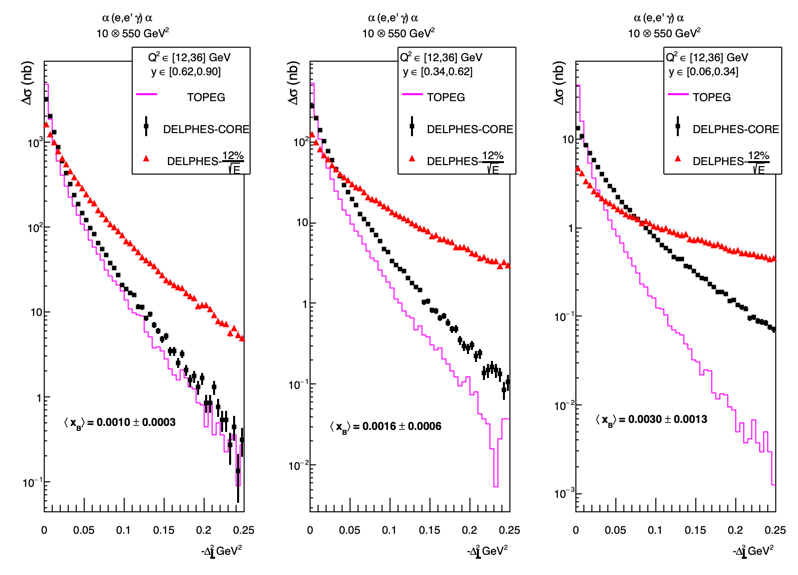}
\caption{\baselineskip 13pt 
DVCS $\alpha(e,e'\gamma)\alpha$ cross sections (integrated over the bin) as a function of 
	$\Delta_\perp^2$. The magenta solid lines are the generated TOPEG distributions\cite{Fucini:2021psq} for 10 x 137.5 GeV per nucleon
	in a band $12<Q^2<36\text{ GeV}^2$ and three bands in $y=q\cdot P/(k\cdot P)$. The black histogram is the result of the DELPHES Fast MC smearing with CORE tracking and EMCal resolution.
	For the red histogram, the CORE EM calorimeters for $\eta>-1.8$ are replaced with a 
``Yellow Report reference'' detector, \textit{i.e.}, a W-powder calorimeter with a resolution of $\sigma(E)/E = [2\%\oplus 12\%/\sqrt{E/\text{GeV}} \oplus 2\%/(E/\text{GeV} ]$. The mean and rms values of the $x$-Bjorken distributions for each sample are indicated on the plots. The integrated luminosities of the three MC samples, from left to right are 2395., 1339, and 22.3 nb$^{-1}$, respectively. An actual data run is expected to have an integrated luminosity of at least $10\text{ fb}^{-1}$.
}
	\label{fig:TOPEG-alpha}
\end{figure}

\subsubsection{Deeply Virtual Compton Scattering}

We used the TOPEG generator \cite{Fucini:2021psq} to demonstrate the power of CORE for studying deep virtual Compton scattering (DVCS) on both the proton and nuclei when the kinematics are reconstructed from the scattered electron and produced photon.
The DVCS kinematic distributions for a 10 GeV electron beam incident on a 137.5 GeV per nucleon $\alpha$ beam are displayed in Fig.~\ref{fig:TOPEG-alpha-kin}. The event sample spans the range $Q^2 \in [12,36]\text{ GeV}^2$ and three bins in $y$.
The reconstructed distributions are obtained from the DELPHES fast Monte Carlo \cite{deFavereau:2013fsa}, with the implementation of CORE adapted from \cite{Arratia:2021uqr}.

Fig.~\ref{fig:TOPEG-alpha} illustrates the excellent reconstruction of the $\alpha(e,e'\gamma)\alpha$ cross section. The histograms show the event distribution from TOPEG, and the reconstructed distribution from the DELPHES Monte Carlo.  Assuming break-up channels are vetoed by the far-forward detectors, the events are reconstructed from the $(e,e'\gamma)$ kinematics alone with $\Delta^\mu=(k-k'-q')^\mu$.  With the excellent EM calorimetry of CORE, the final state photon is reconstructed with resolution comparable to the tracking resolution. In particular, the resolution on $\Delta^\mu$ is independent of the ion beam momentum spread (both transverse and longitudinal). This is of great importance, since the rms transverse momentum spread of the $\alpha$-beam in this case is $\sim 1.1$ GeV.
The $\alpha(e,e'\gamma)\alpha$ cross section is plotted in 
Fig.~\ref{fig:TOPEG-alpha} as a function of  $|\Delta_\perp^2|$: the magnitude-squared of the transverse components of $\Delta$, in an event-by-event frame with $q$ and $P$ anti-colinear. The variable
$\boldsymbol \Delta_\perp$ is Fourier-conjugate to the transverse impact parameter of the active quark or gluon in the scattering amplitude.
The dramatic comparison of the CORE resolution
(black points, Fig.~\ref{fig:TOPEG-alpha}) \textit{vs.}
a more generic EMCal (red triangles)
illustrates the essential power of 
the high resolution CORE EM calorimeters for resolving the transverse quark and gluon structure of nuclei.

\begin{figure}[htb!]
\includegraphics[width=0.7\textwidth]{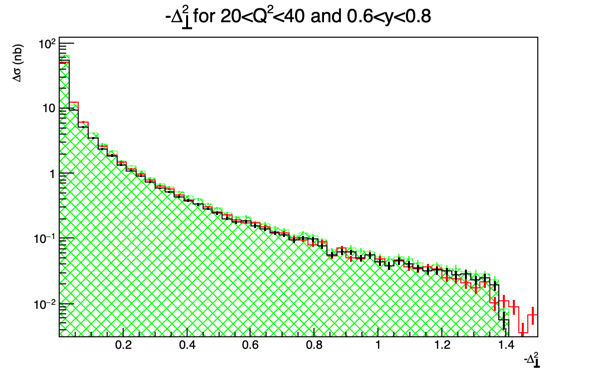}
\caption{\baselineskip 13pt
Proton DVCS cross section for 10 x 275
($k\otimes P = 10\otimes 275\text{ GeV}^2$)
as a function of $-\Delta_\perp^2$. Green hatched histogram is the generated distribution, based on the GK08 $H_g$ and $H_q$
generalized parton distribution model \cite{Goloskokov:2007nt}. The black histogram includes
all $p(e,e'\gamma)p$ events within the CORE acceptance.
The red histogram is the distribution, as reconstructed in the DELPHES monte carlo.}
\label{fig:pDVCS}
\end{figure}

The proton DVCS cross section is illustrated in Fig.~\ref{fig:pDVCS}.  The proton GPDs vary much more gently than nuclear GPDs.  Consequently the bin migration effects from the CORE resolution are almost negligible when reconstructing the event kinematics from the electron and photon. Thus, while tagging the forward proton is important for exclusivity, CORE will not rely on an accurate unfolding of hadron beam effects for  precise DVCS measurements.

\subsubsection{Exclusive Vector Meson Production}

\begin{figure}[hbt!]	
    \includegraphics[width=0.8\textwidth]{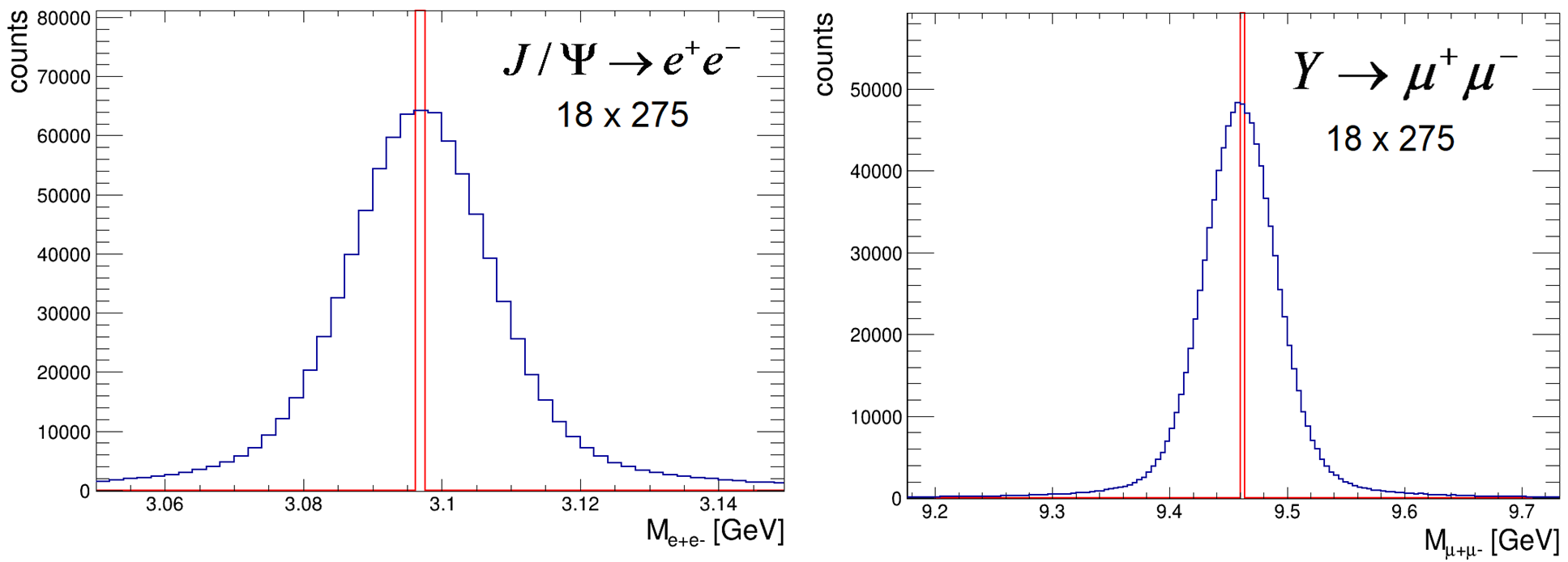}
	\caption{\baselineskip 13pt Generated (red) and reconstructed (blue) invariant masses in exclusive charmonium and bottomonium production on the proton at 18 x 275 GeV. The excellent invariant mass resolution of CORE is important since background from the Bethe-Heitler process which gives rise to the same di-lepton final state and thus cannot be suppressed through PID.
	\textbf{Left panel:} $J/\psi \rightarrow e^{+}e^{-}$. \textbf{Right panel:} $\Upsilon \rightarrow \mu^{+}\mu^{-}$.}
	\label{fig:invmass_jpsi_upsilon}
\end{figure}

\begin{figure}[htb!]
    \includegraphics[width=0.8\textwidth]{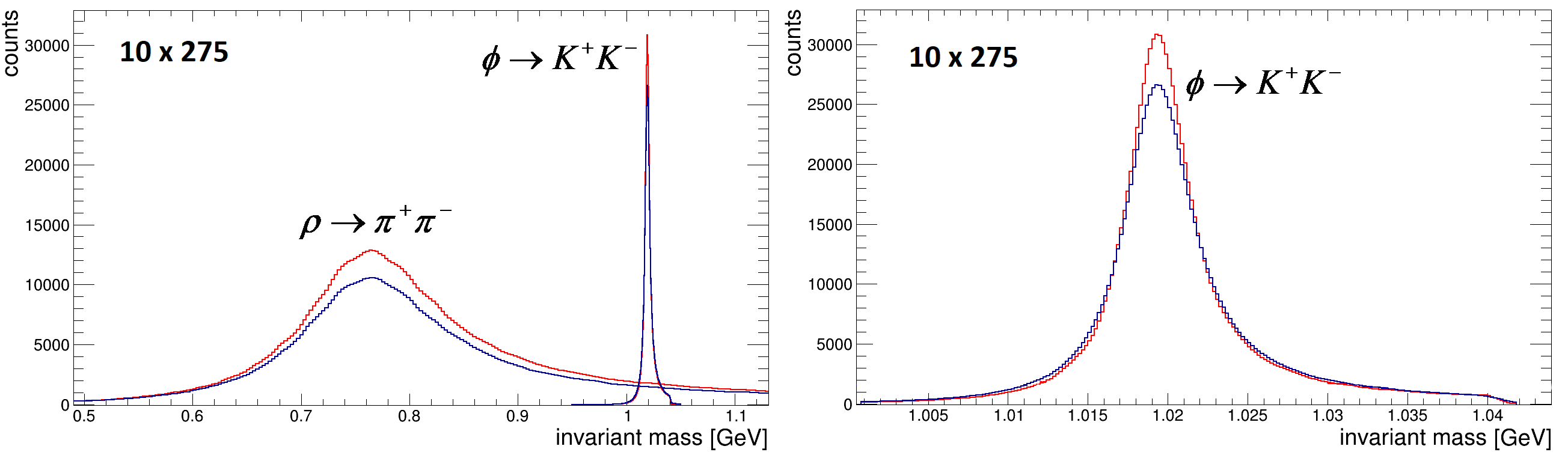}
	\caption{\baselineskip 13pt Generated (red) and reconstructed (blue) invariant masses from exclusive $\rho$ and $\phi$ production on the proton at 10 x 275 GeV. The left panel shows the $\rho$ and $\phi$ overlaid, while the right panel shows only the $\phi$. Two-pion production is the major background for the $\phi$, and this figure illustrates the impact of the excellent invariant mass resolution of CORE on the measurement of the $\phi$.}
	\label{fig:invmass_rho_phi}
\end{figure}

\begin{figure}[htb!]	
    \includegraphics[width=0.99\textwidth]{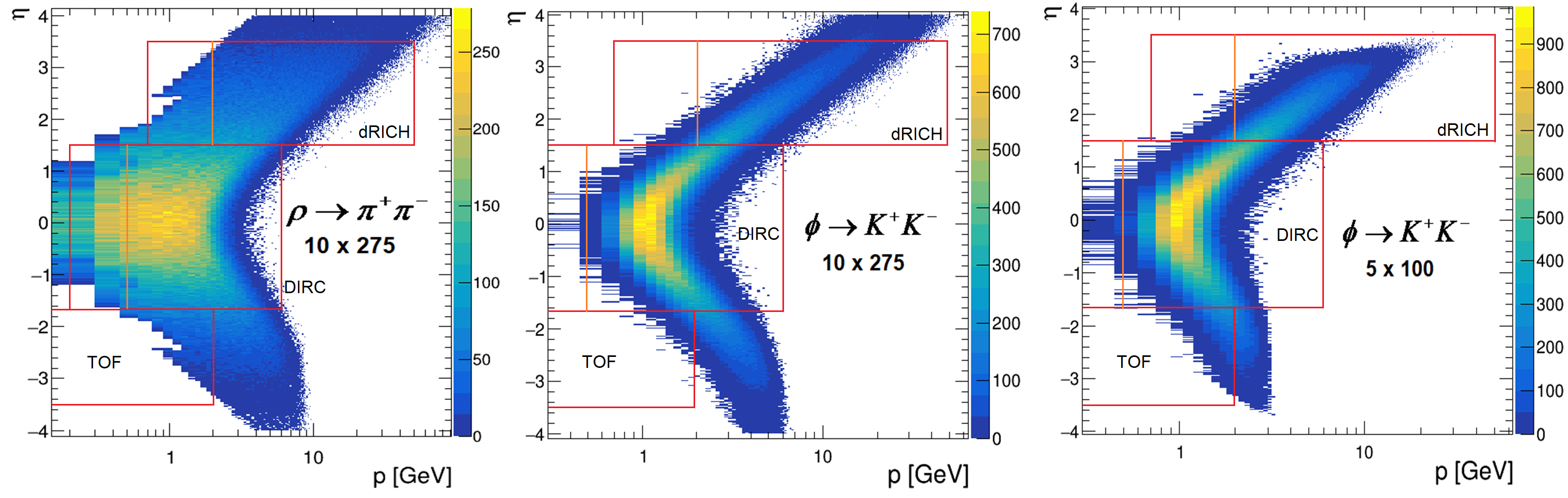}		
	\caption{\baselineskip 13pt $\eta$ vs p distributions for exclusive $\rho$ and $\phi$ production for beam energies of 10 x 275 GeV and 5 x 100 GeV. \textbf{Left panel:} $\pi^{-}$ from $\rho \rightarrow \pi^{+}\pi{-}$ at 10 x 275 GeV. \textbf{Central panel:} $K^{-}$ from $\phi \rightarrow K^{+}K^{-}$ at 10 x 275 GeV. \textbf{Right panel:} $K^{-}$ from $\phi \rightarrow K^{+}K^{-}$ at 5 x 100 GeV.}
	\label{fig:eta_p_rho_phi}
\end{figure}

\begin{figure}[htb!]	
    \includegraphics[width=0.8\textwidth]{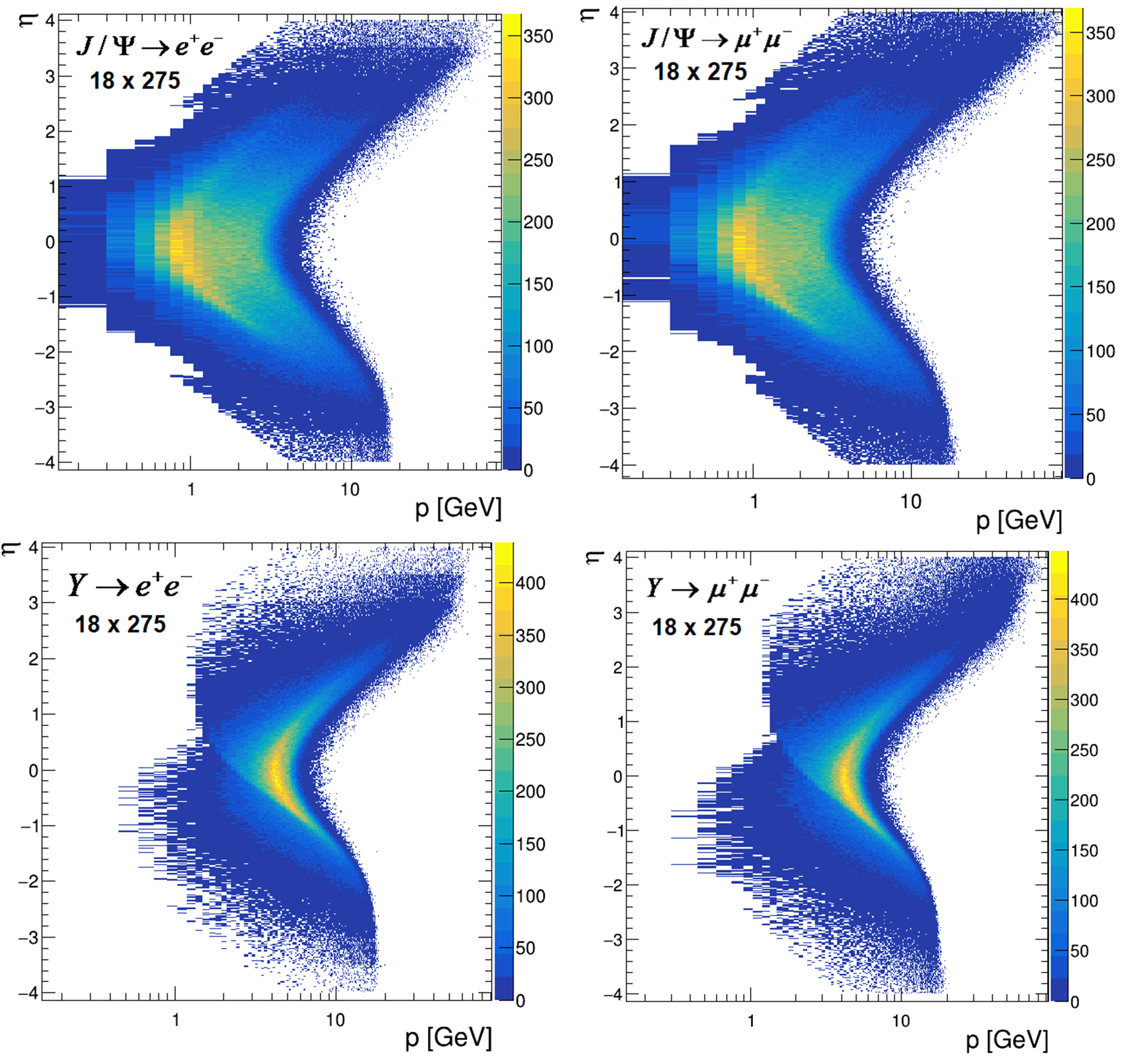}		
	\caption{\baselineskip 13pt $\eta$ vs p distributions for the exclusive $J/\psi$ and $\Upsilon$ production for a beam energy of 18 x 275 GeV. \textbf{Upper row:}  Electrons from $J/\psi \rightarrow e^{+}e^{-}$ (left) and muons from $J/\psi \rightarrow \mu^{+}\mu^{-}$ (right). \textbf{Lower row:} Electrons from $\Upsilon \rightarrow e^{+}e^{-}$ (left) and muons from $\Upsilon \rightarrow \mu^{+}\mu^{-}$ (right).}
	\label{fig:eta_p_jpsi_upsilon}
\end{figure}

The exclusive production of vector mesons has been studied with the {\tt estarlight} generator \cite{estarlight:2019}. The reconstruction is based on the DELPHES Monte Carlo.
Figs.~\ref{fig:invmass_jpsi_upsilon} and \ref{fig:invmass_rho_phi} illustrate the reconstruction of the invariant mass (here for vector mesons) based on their decays to charged mesons and leptons. For $\phi \rightarrow K^{+}K^{-}$ a mass resolution ($\sigma$) of 0.37~MeV can be achieved, while $J/\psi$ and $\Upsilon$ decaying to leptons can be reconstructed with 10.5~MeV and 31.1~MeV resolutions, respectively. The $\rho$ resonance has a large natural width. The main reason that the reconstructed peaks are lower is that particles were generated for the full $\eta$ range but the endcap acceptance only extends to $|\eta|>3.5$ (although it may be possible to reconstruct some of the very forward-going particles in the ``B0'' dipole behind the detector).

Since two-pion production is a major background for the $\phi$, Fig.~\ref{fig:invmass_rho_phi} shows the importance of the excellent CORE tracking resolution, which translates into an invariant mass resolution allowing for a clean extraction of the $\phi$.

Fig.~\ref{fig:eta_p_rho_phi} shows the $\eta$ vs momentum distributions for pions and kaons in the final state, with the PID capabilities of CORE superimposed. In exclusive reactions, kaon momenta are higher than in DIS (shown in Fig.~\ref{fig:kaons_10x275_5x100}). Thus, the so-called ``threshold modes'' of the dRICH in hadron endcap and the DIRC in the barrel are not used since all kaons from the $\phi$ decays fall within the ranges of the regular ``RICH'' modes.
The excellent invariant mass resolution for the $\phi$ shown in Fig.~\ref{fig:invmass_rho_phi} means that the background under the $\phi$ peak will be small without any PID. Identification of one of the charged kaons will reduce it even further. Identification of both will give a very clean signal.

Fig.~\ref{fig:eta_p_jpsi_upsilon} illustrates the $\eta$ vs momentum distributions for lepton final states. For $J/\psi$, excellent low-momentum muon ID is needed in the barrel (up to about 4 GeV/$c$), but a significantly wider range is needed for the electron-side endcap (1-15 GeV/$c$). Muon ID is particularly important for heavier charmonia, for which the cross section is small and a doubling of the statistics would be particularly impact-full. The muon distributions for the $\Upsilon$ show that a coverage of 1.5-10 GeV/$c$ is needed for the barrel and 3-20 GeV/$c$ for the electron endcap. In the hadron endcap ($\eta>1.2$, the momenta are very high for both the $J/\psi$ and the $\Upsilon$, which is more reminiscent of LHC experiments where muon detectors are placed behind the Hcal. In this region, it would be possible to add a supplementary muon ID system behind the Hcal if it was needed, but such a system is not part of the CORE baseline.

\section{Detector Subsystems}
\label{subsystems}
The CORE detector is built around a central idea of a compact set of core systems integrated with a short but strong solenoid inside of a spacious, instrumented flux return. The goal is to deliver the best possible performance in a cost-effective way while minimizing technical risk. To this end, the CORE subsystems were chosen to be mutually supportive and synergetic, both in term of geometry and technology choices.

Development of the key technologies that form the basis of the CORE detector (tracking, calorimetry, PID) was supported by the decade-long Generic Detector R\&D for an EIC program \cite{eRD}, in which CORE members were active from the beginning to the end of the program earlier this year. The progress was documented in bi-annual reports \cite{eRD}, which were reviewed by an advisory committee. A summary of this R\&D was included in the Yellow Report \cite{AbdulKhalek:2021gbh}. The consortia that were active within the Generic R\&D program have since supported the development of all EIC detector proposals, in line with the original intent of that program. Thus, the main focus of the detector proposals was to optimally choose and integrate these technologies into the detector concept, and to bring in some complementary ones that were not part of the Generic R\&D program. The Generic R\&D program has now been replaced by targeted R\&D (three-digit numbers) funded by the EIC project.

CORE subsystems that were supported by Generic R\&D program and are currently supported by targeted R\&D include the silicon tracker (eRD25, eRD104), the MPGD tracker (eRD6), the dual-radiator RICH (eRD14, eRD102), the DIRC (eRD14, eRD103), the PbWO$_4$ EMcal (eRD1), and the Fe/Sci HCal (eRD1, eRD107). The LGAD technology became part of the Generic R\&D at a relatively late stage (eRD24, eRD29) but AC-LGADs continue to receive support under the targeted program (eRD112), and are strongly supported by CERN.
For brevity, we provide the corresponding Yellow Report page numbers (referring to the 17 Mar 2021 version) for the respective technologies.

CORE also uses two systems that were not part of the Generic R\&D program: the the neutral hadron and muon system and the W-shashlyk EMcal. The former is directly based on the upgraded Belle II ``KLM,'' adapted to the flux return geometry of CORE. Since this system is basically an Fe/Sci sandwich, and has been tested at Belle, we do not foresee any significant technological challenges.

The W-shashlyk was not supported by the Generic R\&D program primarily because it is very similar to a Pb-shashlyk, even though it offers better performance in terms of both energy and spatial resolution. It also offers better energy resolution (6\% vs 12\%) than the  W/SciFi (tungsten powder) EMcal.
The main production challenge for the W-shashlyk is the need to drill holes in the W/Cu plates, which are more environmentally friendly but harder than Pb and thus requires a suitable vendor.

This section focuses on a general description of each subsystem, and how the technology was adapted for CORE. All changes and adaptations were made in collaboration with system experts from the R\&D consortia. The sub-systems, their provenance, and an assessment of their consequent technical risk, discussed in Section \ref{risk}, are given in Table \ref{table:detctor_technologies}. We also indicate the status of alternatives where appropriate.

\begin{table}
\resizebox{\textwidth}{!}{%
\begin{tabular}{ |c|c|c|c|c|c|c| } 
 \hline
 Component & Baseline Technology & Basis for Tech. & Risk & Alternative & Alternative seen as \\ 
 \hline
 \hline
 compact solenoid & NbTi & widespread use & medium & lower field & fallback \\
 \hline
 silicon tracker & MAPS (10 $\mu m$ pixels) & ITS3, eRD25 & medium & ITS2 (ALICE) & fallback \\
 \hline
 MPGD tracker & $\mu$RWELL & eRD6 & medium & GEM & fallback \\
 \hline
 LGAD TOF & AC-LGAD & eRD112 & medium & resistive LGAD & lower fill factor \\
 \hline
 dRICH PID & gas + aerogel & eRD14, HERMES & medium & non-CFC gas & eco friendly \\
 \hline
 DIRC PID & hpDIRC & eRD14, PANDA, BaBar & low & thin bars & improved $e/\pi$ ID \\
 \hline
 EMcal $\eta<0$ & PbWO$_4$ & PANDA, CMS, etc & low & W-shashlyk & fallback \\
 \hline
 EMcal $\eta>0$ & W-shashlyk & similar to Pb-shashlyk & low & W/SciFi & fallback \\
 \hline
 HCal $\eta>1.2$ & Fe/Sci towers & STAR FCS, eRD1, etc & low & eRD107 compensation & improved resolution \\
 \hline
 KLM $\eta<1.2$ & Fe/Sci 2D layers & Belle II & low & sPHENIX HCal & traditional HCal \\
 \hline
\end{tabular}}
\caption{Summary of detector technologies.}
\label{table:detctor_technologies}
\end{table}

\subsection{Solenoid}
\label{solenoid}
CORE uses a new solenoid with a maximum field of 3 T along the axis at its center (at 4700 A/cm$^2$). The solenoid supports variable field settings. The stored energy at 3 T is 38.8 MJ. The SC coil consists of three coils in series with a common main DC power supply. Separate power supplies can be ``floated'' across the end trim coils for small up/down adjustments relative to the main current. The inner radius of the coil is 110 cm and the outer radius is 116 cm. The coil is 250 cm long. The ``trim'' coils are 10 cm long and located at the edges of the coil (z of -125 to -115 cm and 115 cm to 125 cm, respectively). The inner radius of the cryostat is 100 cm. The outer radius of the barrel flux return iron is 265 cm. The design of the solenoid and flux return satisfies all BNL requirements.

The cryostat has a 5 cm inner vessel (Al), an 2 mm inner radiation shield (Cu), a 6 cm coil (10:1:1 mix of Cu, Nb, and Ti), a 7 cm coil support cylinder (Al), a 2 mm outer radiation shield (Cu), and a 5 cm outer vessel (stainless steel). The outer vessel acts as the first Fe-layer of the neutral hadron and muon detector discussed below and in section \ref{KLM}, which has its first scintillator layer in-between the cryostat vessel and the flux-return iron. The edges of the cryostat vessel will be a few cm thicker.

The short solenoid is a good geometric match for the silicon tracker, and makes it easier to combine a high central field with a low field on the DIRC photosensors and good ``projectivity'' in the gas volume of the dual-radiator RICH, both of which were also key requirements for the design.

\begin{figure}[htb!]	
    \includegraphics[keepaspectratio=true,width=2.9in,page=1]{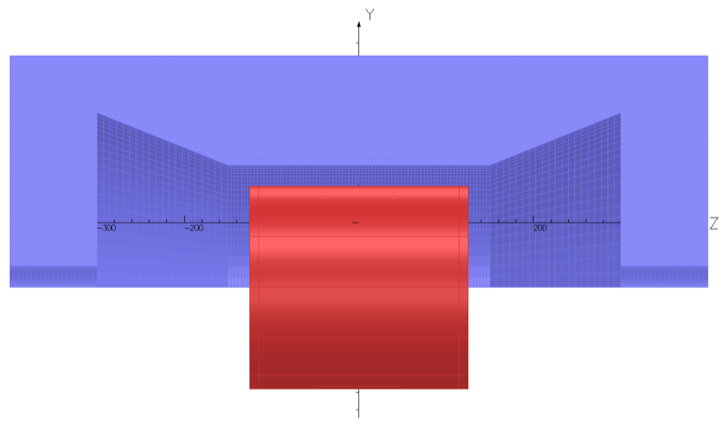}		
	\includegraphics[keepaspectratio=true,width=3.2in,page=1]{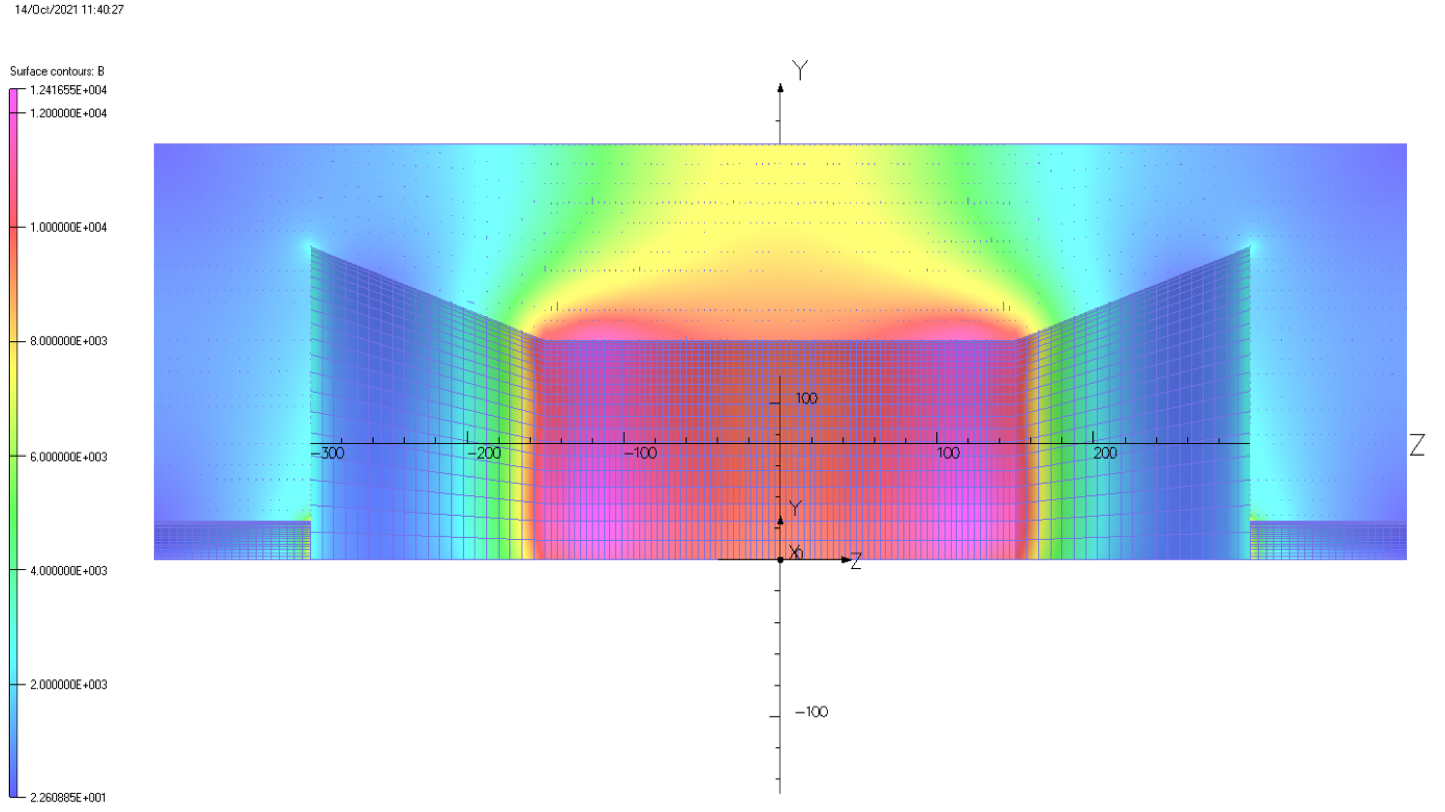}
	\caption{The CORE solenoid showing the distribution of the flux return iron (left) and the magnitude of the field in the iron when operating at 3 T (right). Note that the radial component of the field on the outer iron surface is lower than 0.014 T (140 Gauss) when the central field is at 3 T - and approaches such values only very locally along z. The saturation in the iron is very low.}
	\label{fig:solenoid}
\end{figure}

An advantage of a small solenoid is that it makes it easier to reduce the coil forces. The coil stress has a linear dependence on the radius, but is also affected by asymmetries in the iron distribution. While the latter can be mitigated by redistributing coil windings and flux return iron, a more robust solution is to make the flux return fully symmetric with respect to the solenoid since this leads to exact cancellations. Thus, while the CORE layout is compatible with an overall length of [-3.5 m, 4 m], we chose a symmetric [-4 m, 4 m] flux return. Making the electron side shorter than the hadron side offers only a limited benefit, and the extra space inside can be used for a more robust support structure for cantilevering the endcap EMcal on the electron side.

In CORE, all parts of the flux return are instrumented, and the endcaps are detachable. The small radius of the CORE solenoid makes it possible to build a flux return that contains a 3 T field within small outer radius. The model shown in Fig.~\ref{fig:solenoid} has an outer radius of 2.65 m assuming an average Fe content of 75\% in the barrel part of the flux return. The baseline is to instrument it a with neutral hadron and muon detector based on the ``KLM'' in Belle II, which is discussed in section \ref{KLM}.

\begin{figure}[htb!]
\begin{center}
    \includegraphics[keepaspectratio=true,width=2.9in,page=1]{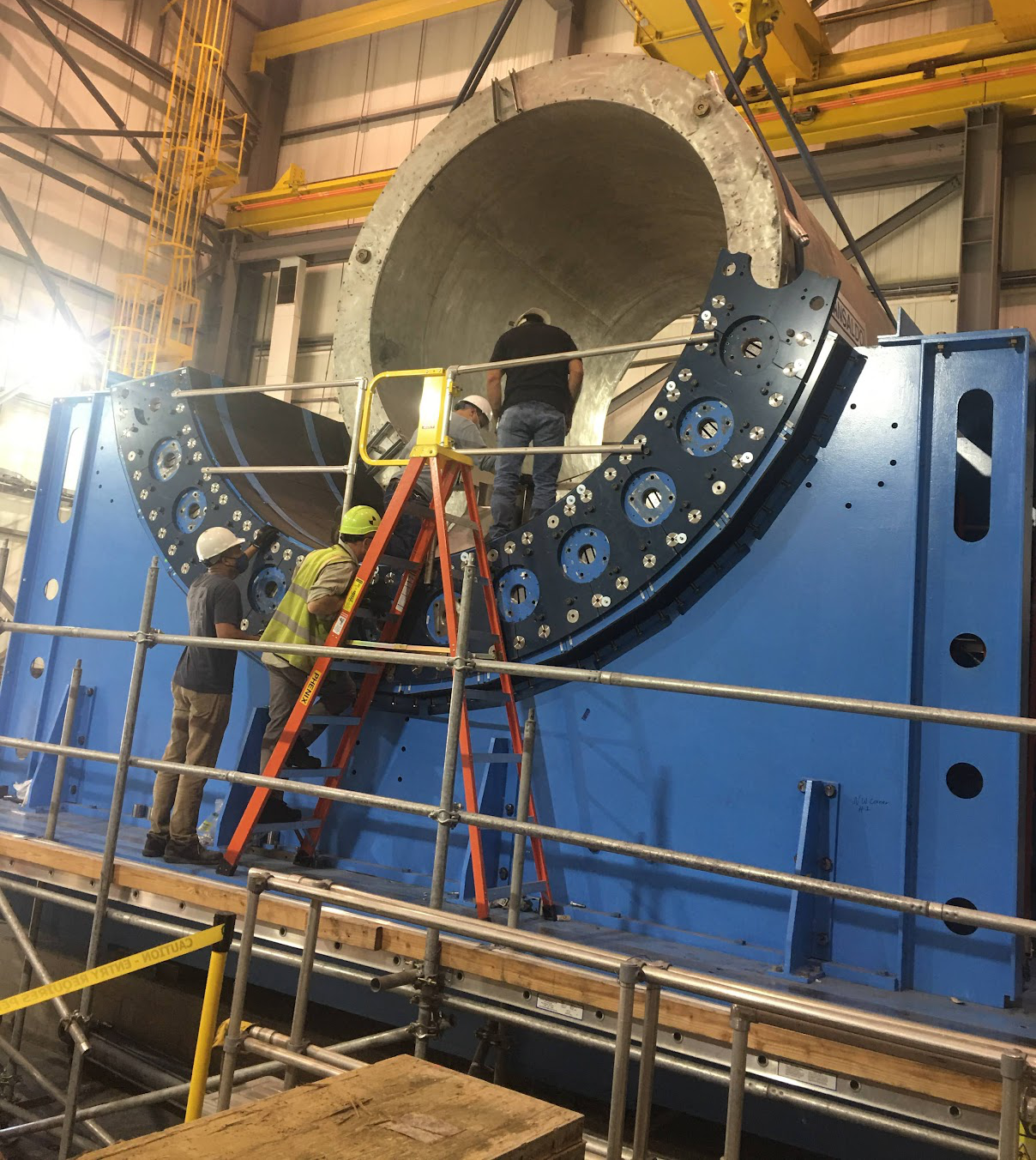}		
	\includegraphics[keepaspectratio=true,width=3.2in,page=1]{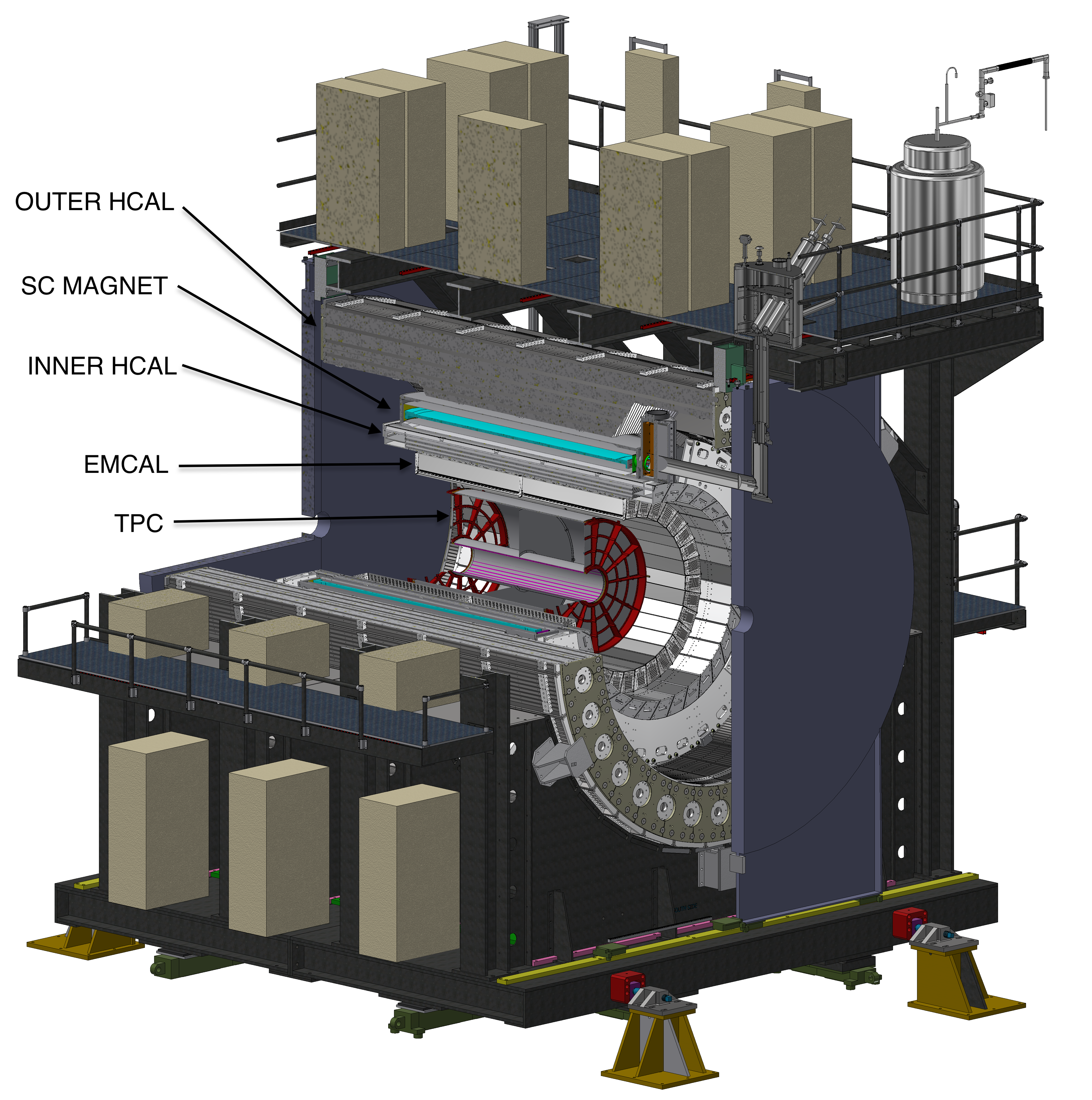}
	\caption{\textit{Left Panel}: Photo showing the BaBar solenoid being lowered onto the lower half of the sPHENIX HCal. The simple ``cradle'' consisting of four ``inverted'' steel arches is clearly visible, as is the platform to which the arches are attached. Both the platform and arches can be adapted for an EIC detector - although the platform would have to be shortened and adapted to the different angle of the EIC electron beam compared to the corresponding RHIC beam. \textit{Right Panel}: Drawing from the sPHENIX TDR showing additional structures planned for the platform. Such structures could be re-used for an EIC detector on this platform or a similar one.}
	\label{fig:sPHENIX_cradle}
\end{center}
\end{figure}

While this is not the proposed baseline, CORE could also provide an upgrade path for sPHENIX. The CORE solenoid is stronger but smaller than the old BaBar one, and would fit inside the sPHENIX barrel HCal, which is shown in Fig.~\ref{fig:sPHENIX_cradle}. This would leave 0.5 m of space in-between the re-used sHENIX HCal and the CORE cryostat. The space would be used for an inner HCal, which could be designed to incorporate an enhanced muon ID capability. Adapting the design for re-use of the sPHENIX HCal would be straightforward and reduce the cost to the project since the HCal and its iron would be an in-kind contribution.

If CORE is located at IR8, the rail-mounted platform and the ``inverted arch'' supports could be re-used. Both can be seen in the left panel of Fig.~\ref{fig:sPHENIX_cradle}.

However, the sPHENIX platform cannot be re-used without modification for any EIC detector. First, it is too long, and cannot accommodate the independently-supported detachable endcaps required by the EIC project. And second, an EIC detector has to be aligned with the electron beam, the angle of the arches with respect to the rail system the platform has to be adjusted by 12 mrad.

The best way to address these issues is to build a new platform with the correct length and angle with respect to the rails. It could re-use parts from the old one. The arches and additional structures from the old platform could be transferred to the new one.
Another option would be to shorten the existing platform and re-attach the arches at the correct angle. Simply cutting off part of the platform seems less structurally sound, but could work if the solution is well engineered.
For CORE, the conclusion is that the cradle at IP8 could be re-used with the same generic adaptations as for any EIC detector. And if the cradle were not available, building a completely new platform and set of four arches would not be very complicated or expensive.
If CORE were to be used in IR6, it could use a new cradle as described above, use the cradle from IR8, or be equipped with a sleeve to fit a modified STAR cradle.

\subsection{Tracking}

The tracking in CORE is provided by a silicon tracker integrated with an AC-LGAD layer on the electron side (discussed in section \ref{LGAD}). In addition, there is a MPGD tracker in-between the dual-radiator RICH and EMcal in the hadron endcap, which also seeds the ring finder for the RICH.

\subsubsection{Central Si-tracker}
\label{si_tracker}

The CORE silicon tracker was developed in collaboration with the Silicon Consortium. The starting point was the eRD25 design, which is described in detail in section 11.2.5 (p.454) of the Yellow Report \cite{AbdulKhalek:2021gbh}. The tracker is based on ALICE ITS3 technology using thin, air-cooled silicon with 10 $\mu m$ pixels, and provides both momentum and vertex reconstruction. The ITS3 vertex barrel layers are particularly innovative as they can be shaped into cylinders.

The changes made for CORE are primarily aimed at improving redundancy and track reconstruction efficiency, in particular in the 1.1$<\eta<$1.3 region. To this end, the layout was changed so that a track would pass through six layers of tracker at all values of $\eta$. This in turn meant that an additional disk was added on either side, and a $3^{rd}$ vertex layer was added.

\begin{figure}[htb!]
\begin{center}
	\includegraphics[width=1.\textwidth]{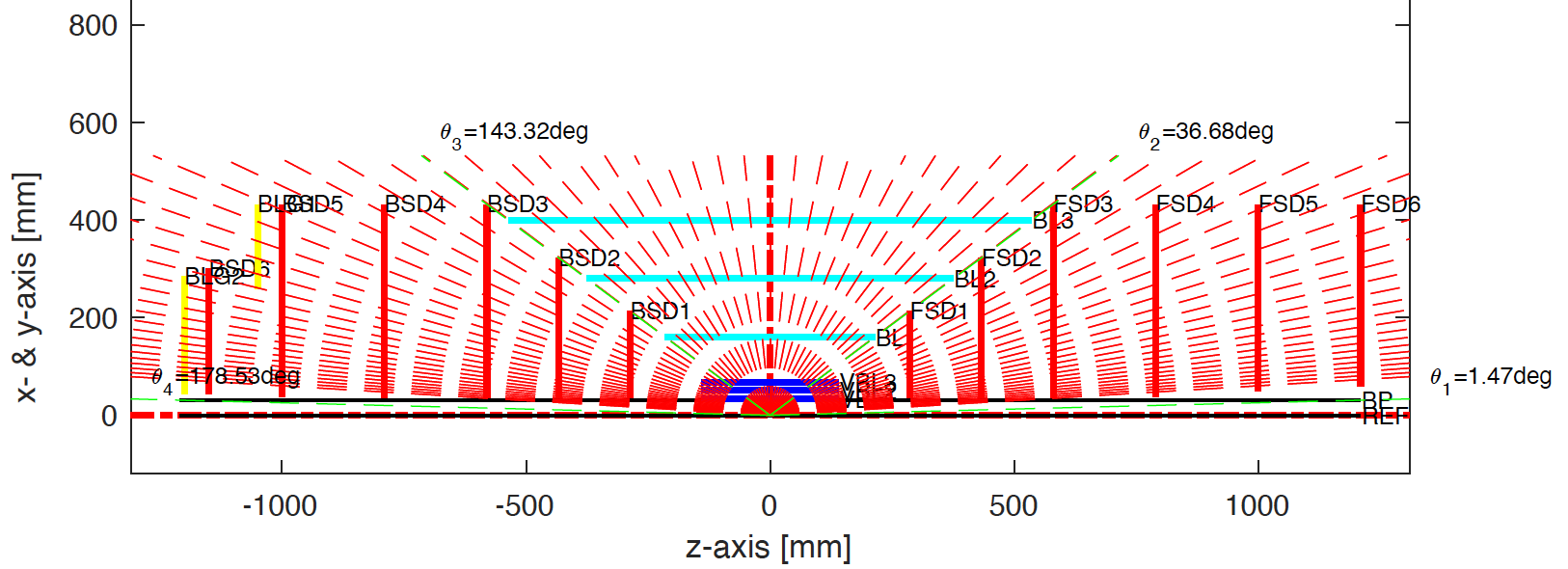}
	\caption{The layout of the MAPS disks (red) and barrel (blue) layers (all with 10 $\mu m$ pixels) used in the simulation. The vertex barrel layers are shown in dark blue. The AC-LGADs on the electron side (left) are shown in yellow. A single AC-LGAD disk could be added behind the last MAPS disk on the hadron side (right) as a future upgrade.}
	\label{fig:tracker_sim}
\end{center}
\end{figure}

The layout of the CORE tracker is shown in Fig.~\ref{fig:tracker_sim}. It has six MAPS disks on either side. The first two disks from the center on each side have smaller outer radii to match the barrel. The ``full-sized'' disks have outer radii of 43.2 cm. The $z$-locations of the first five disks on each side are: $\pm$28.6, $\pm$43.3, $\pm$58, $\pm$79, and $\pm$100 cm.
The last disk on the hadron side is located at a $z$ of 121 cm, while the one on the electron side is located at -115 cm. The latter has a smaller outer radius matching the AC-LGAD right behind it.

\begin{figure}[htb!]
\begin{center}
	\includegraphics[keepaspectratio=true,width=3.2in,page=1]{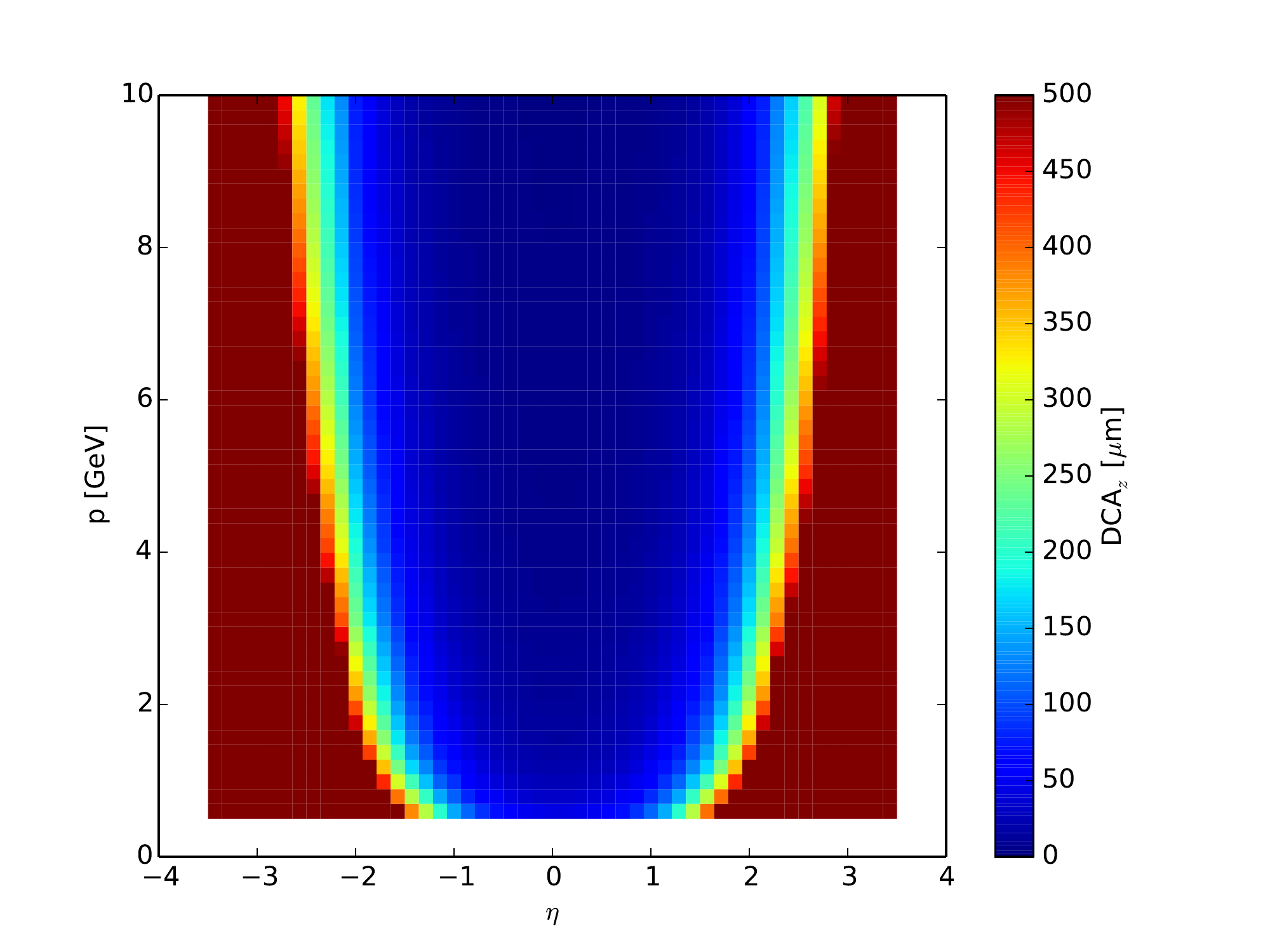}
	\includegraphics[keepaspectratio=true,width=3.2in,page=1]{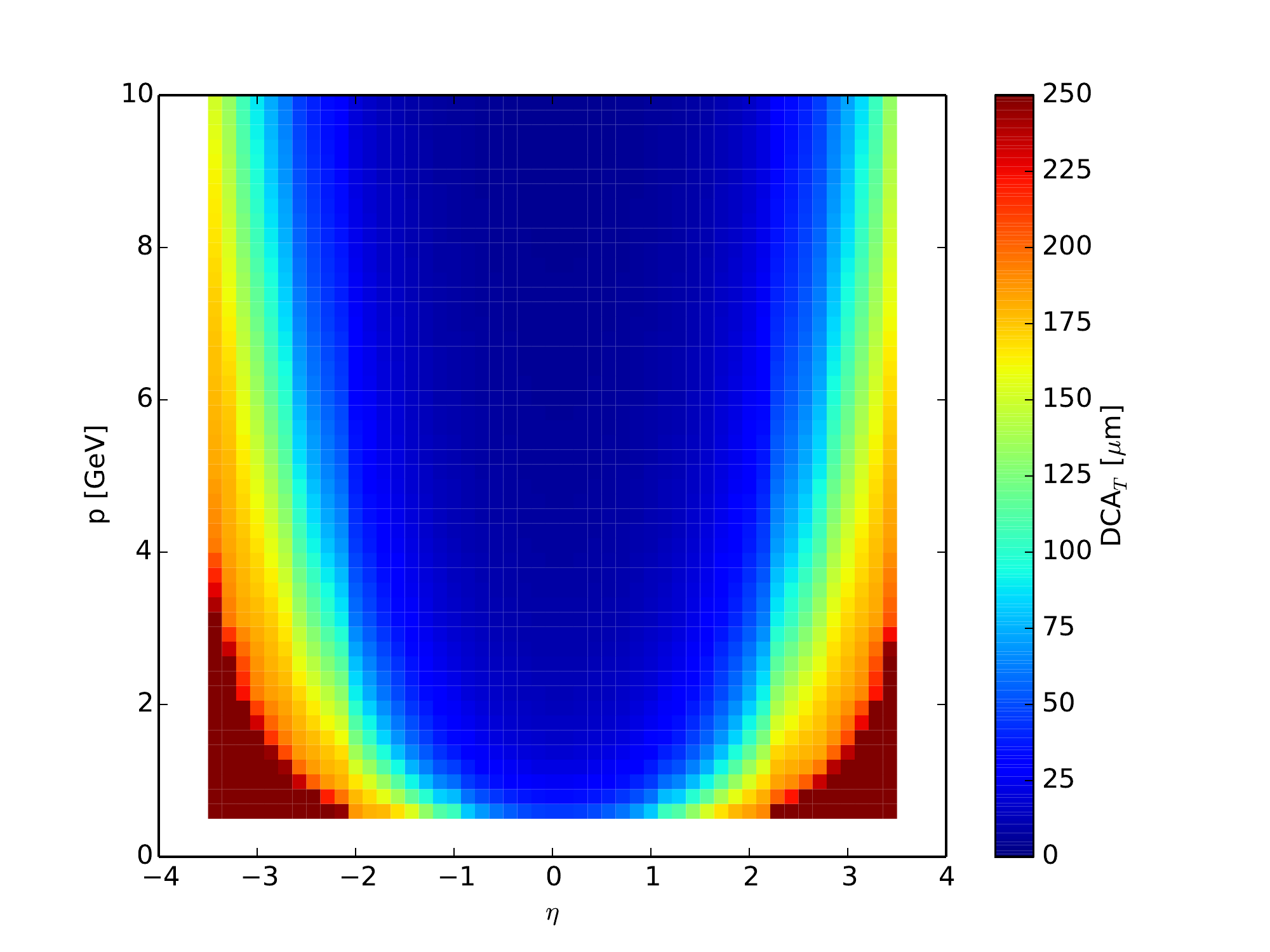}
	\includegraphics[keepaspectratio=true,width=3.2in,page=1]{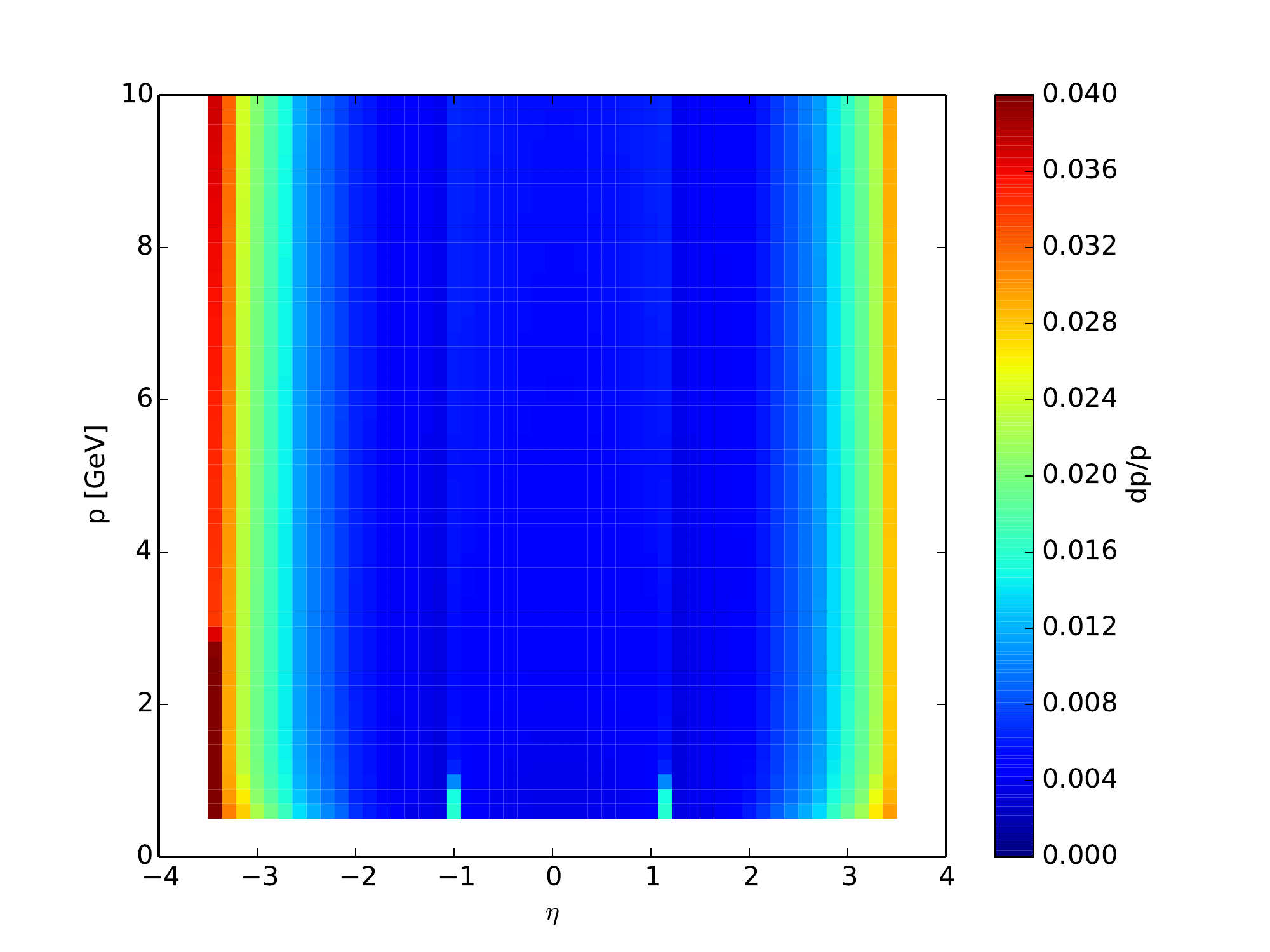}
	\includegraphics[keepaspectratio=true,width=3.2in,page=1]{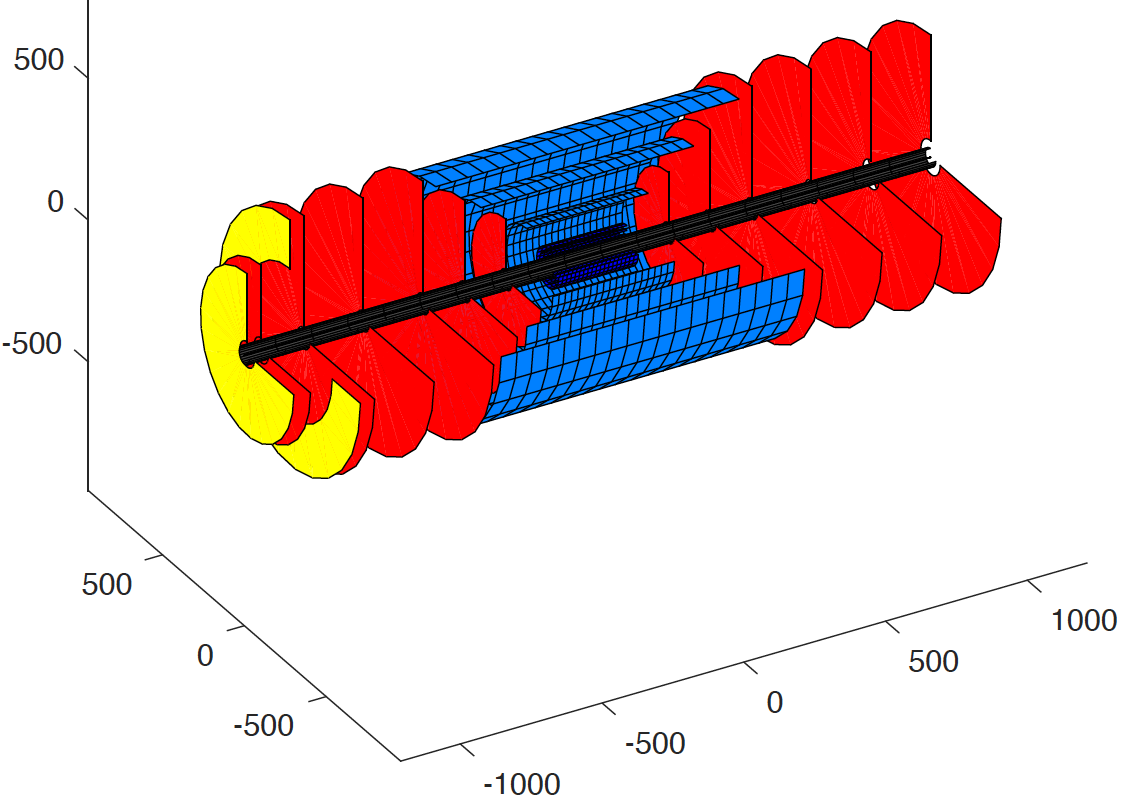} 
	\caption{Longitudinal (upper left) and transverse (upper right) vertex resolution, and (lower left) momentum resolution for the CORE silicon tracker at 3 T. When the solenoid is operated at lower fields, dp/p will deteriorate.
	The MPGD behind the dRICH was not included in this simulation. Including it will improve dp/p at forward (positive) rapidities. 
	A 3D view of layout used in the simulation is shown in the lower right panel. The MAPS layers are shown in red and blue, while the yellow layers are the LGADs. The MPGD is not shown here.}
	\label{fig:tracker_resolution}
\end{center}
\end{figure}

In the barrel, there are three single layers with radii of 16, 28, and 40 cm. The corresponding lengths are 42.9, 75, and 107 cm, respectively. The radius of the outer layer is slightly smaller than the radius of the large disks in order to facilitate integration and minimize edge gaps.
The tracker has three 28-cm-long vertex barrel layers located at radii of 3.4, 5.1, and 6.8 cm.
As can be seen in Fig.~\ref{fig:tracker_sim}, The acceptance of the vertex layers extends beyond the narrow support structure for the outer barrel layers, making it possible to reconstruct tracks that  hit the support structure. 
While such tracks will not have a good momentum resolution, being able to reconstruct them at all improves the hermeticity of the detector.
The coverage of the vertex layer is also matched to the location of the forward MAPS disks. In particular, the $4^{th}$ disk from the center is located at a position such that all particles that exit the tracker between the $3^{rd}$ and $4^{th}$ disks will have passed through all three layers of the vertex tracker. And all particles exiting before the $5^{th}$ disk will have passed through at least one layer of the vertex tracker.

\begin{figure}[htb!]
\begin{center}
	\includegraphics[width=0.5\textwidth]{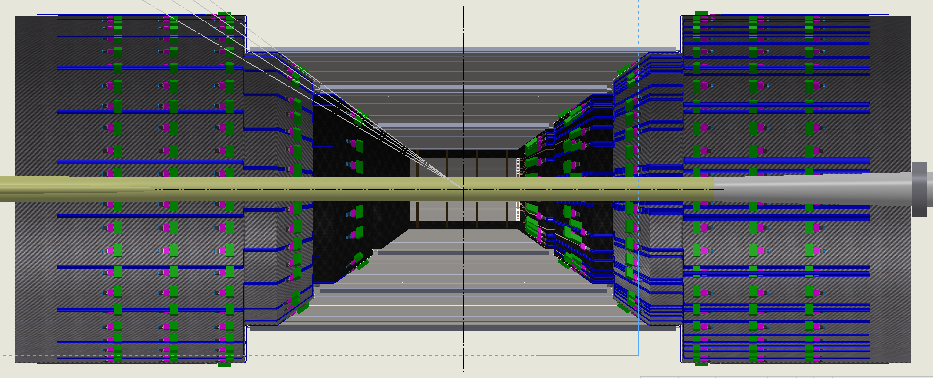}
	\caption{A realistic implementation of the CORE tracker with readout and services.}
	\label{fig:tracker_CAD}
\end{center}
\end{figure}

There are also additional tracking detectors on both sides of the MAPS disks. On the electron side, there is an AC-LGAD layer integrated with the last MAPS disks. This also serves as a Time-of-Flight PID detector, and is described in detail in section \ref{LGAD}. On the hadron side, there is a $\mu RWELL$ MPGD with 100 $\mu m$ position resolution behind the dual-radiator RICH ($X/X_0 \approx 2\%$). The MPGD is described in detail in section \ref{GEM}.
The resolutions of the CORE tracker are summarized in Fig.~\ref{fig:tracker_resolution}. A CAD layout of the CORE tracker with redaout and services is shown in Fig.~\ref{fig:tracker_CAD}. 

The 10 $\mu m$ pixel MAPS trackers are ideal to determine the local angle of incident particles at the DIRC bar. However, since an optimization of the angular resolution has not (yet) been made for CORE, we are using a conservative value of 0.5 mrad.

\subsubsection{Forward MPGD tracker}
\label{GEM}
\begin{figure}[htb]
\centering
\includegraphics[width=0.95\columnwidth,trim={0pt 5mm 0pt 25mm},clip]{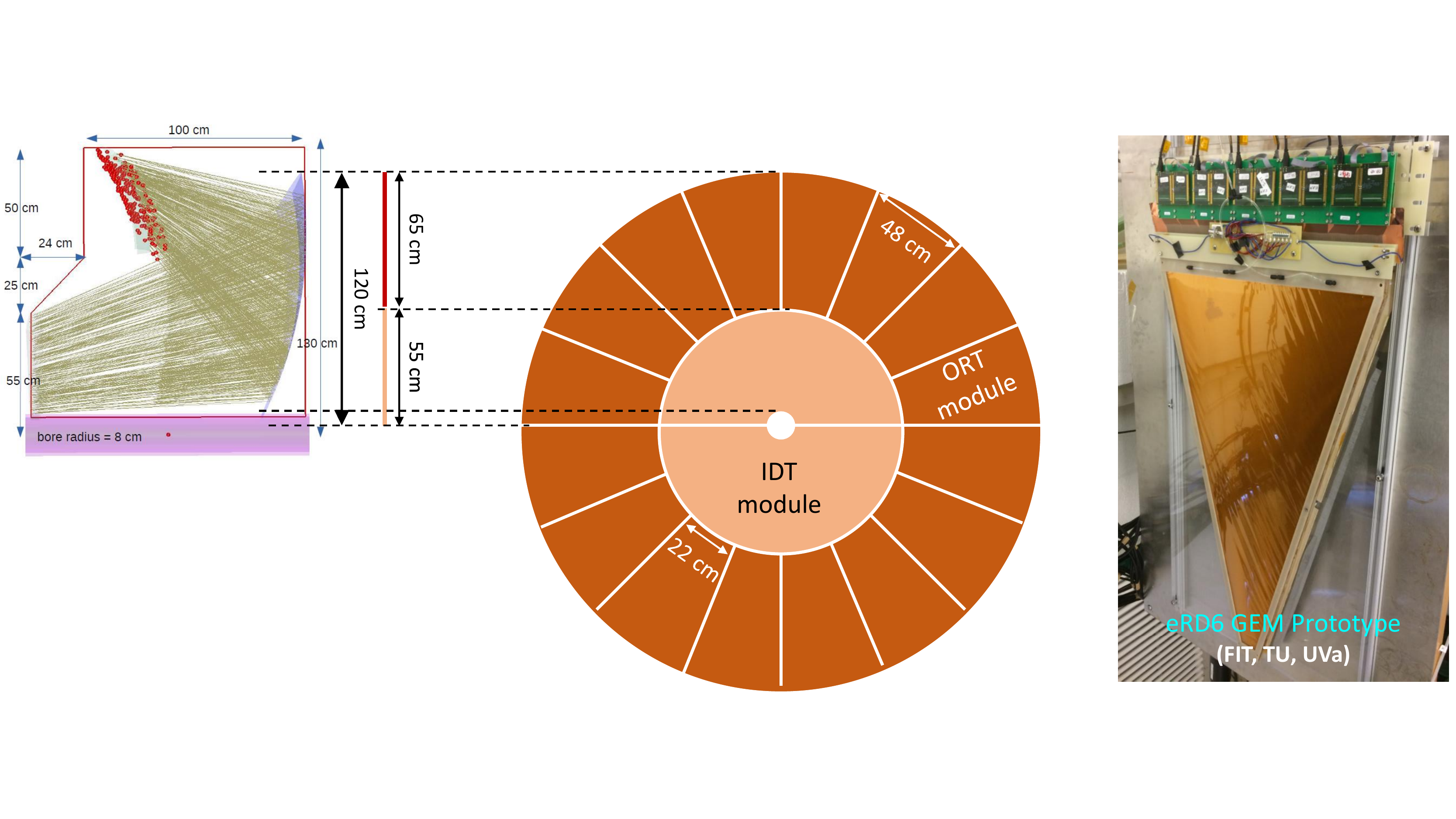}
\caption{\textit{Left:} Forward $\mu$RWELL (or GEM) tracking layer composed of the inner disk tracker (IDT) made of two half-disc modules  and the outer ring tracker (ORT) made of 16 trapezoidal modules matching to the dRICH geometry; \textit{Right:} A picture of the large EIC forward GEM tracker prototype with U-V strip readout layer developed and operated by the eRD6 consortium.}
\label{fig:mpgdDesign}
\end{figure}

The main purpose of the forward tracker is to provide directional information to the dual RICH detector to help seeding the Cerenkov ring reconstruction.
In addition, with the forward tracker located 1.25 m from the endcap Si-discs of the CORE central tracker, and with low-mass dRICH detector material in between the two sets of trackers, the forward tracker will provide an additional high-resolution space point for tracking thanks to the large lever arm that it offers.
The CORE forward tracker will be a Micro-Pattern Gaseous Detector (MPGD), which is being developed by the EIC tracking consortium (eRD6). MPGDs are described in more detail in section 11.2.4 (p.446) of the EIC Yellow Report~\cite{AbdulKhalek:2021gbh}.

A Resistive Micro Well ($\mu$RWELL) detector is the baseline choice for the CORE forward tracker because it is singularly suited for large-area and high-performance tracking in an environment with relatively modest rate as expected at the EIC.
A preliminary design of the CORE forward tracker configuration behind the dRICH is shown in the left panel of Figure \ref{fig:mpgdDesign}. The circular tracker has a radius of 1.2~m and is composed of an inner disc tracker (IDT) with two semi-circular $\mu$RWELL discs and an outer ring tracker (ORT) made of 16 trapezoidal $\mu$RWELL modules.

The IDT half-discs have a radii of 55 cm and the trapezoids have a maximum width of 48~cm. These dimensions are chosen to be compatible with the width of the kapton base material. The readout layer is based on 2D capacitive-sharing r-$\phi$ strips with a fine pitch to provide space point resolution ($\le 100 \, \mu$m) in $\eta$ and $\phi$, in the higher-rate environment of the high-eta region. The central discs have a small semi-circular opening at their centers to accommodate the beam pipe. The trapezoidal modules of the ORT have a height of 65~cm, and inner and outer bases of $\approx 48$~cm and $\approx$~22 cm, respectively. The readout layer for the ORT modules will be 2D capacitive-sharing U-V strips with all strips read out at the outer radius of the trapezoids along their wedge-shaped sides to avoid the presence of readout electronics or cabling within the active area. A U-V strip readout layer fully read out on the outer radius of the detector has been successfully demonstrated with a large GEM prototype developed by the eRD6 consortium for EIC as shown on the right of Figure \ref{fig:mpgdDesign}. The same design can be used as a readout with $\mu$RWELL technology. For the ORT modules, spatial resolution better than 100~$\mu$m and 250~$\mu$m are expected in azimuthal and radial directions, respectively. The exact final dimensions and configuration of the IDT and ORT modules will be optimized to allow overlap of the sensitive areas of the modules for full $\eta$ and $\phi$ coverage of the dRICH acceptance.

\begin{figure}[hbt!]
    \includegraphics[keepaspectratio=true,width=3.0in,page=1]{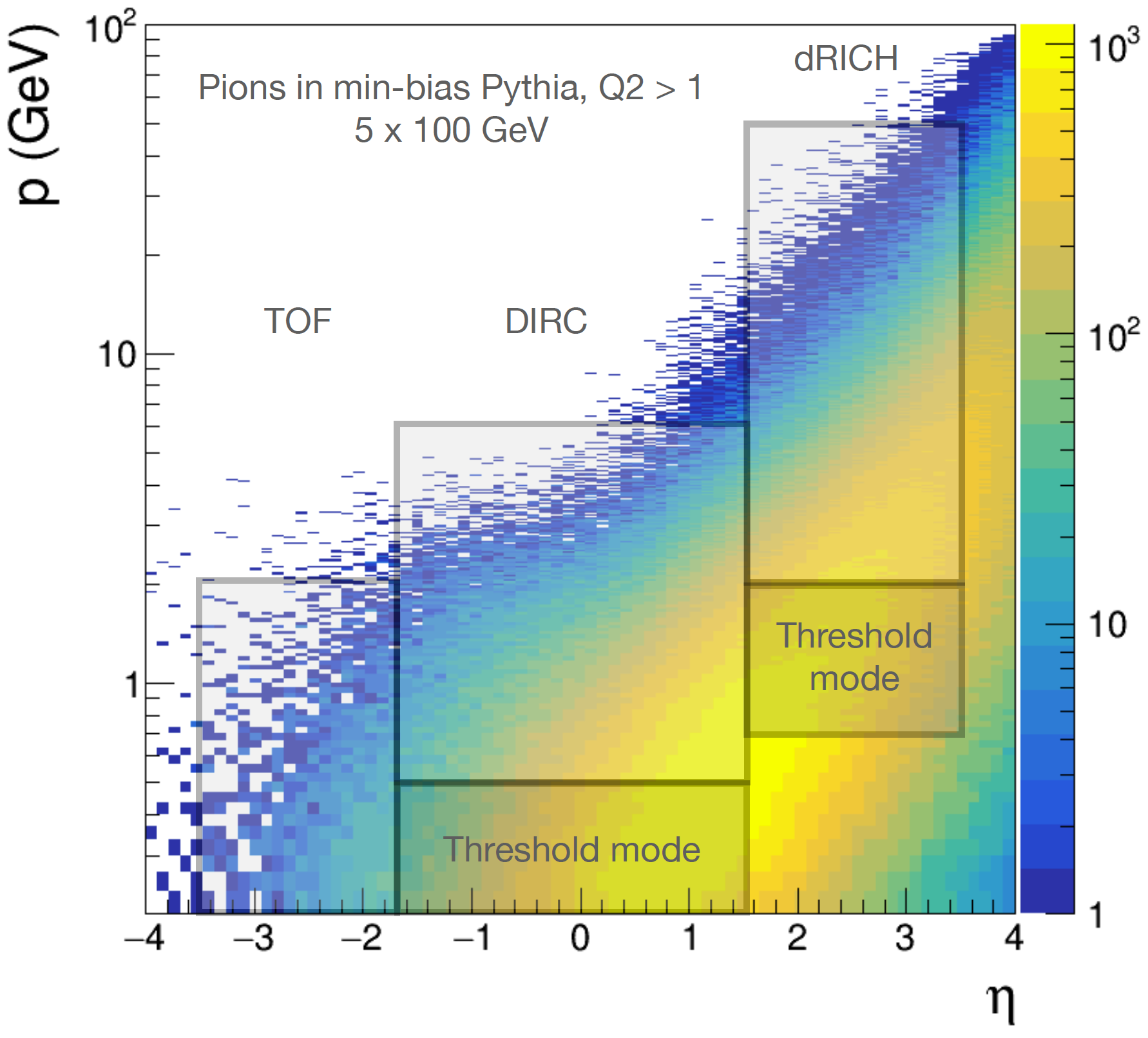}
	\includegraphics[keepaspectratio=true,width=3.0in,page=1]{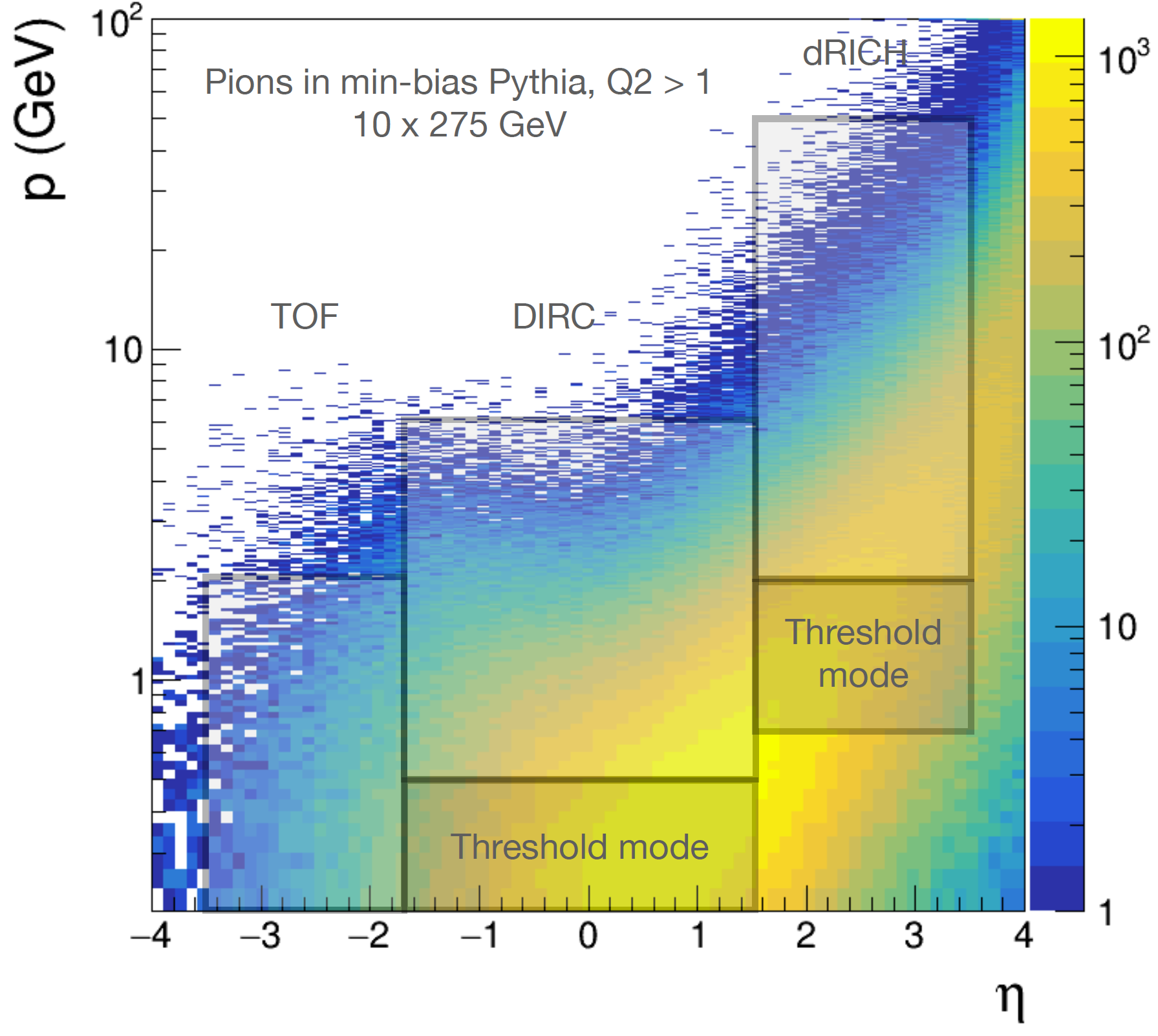}
    \includegraphics[keepaspectratio=true,width=3.0in,page=1]{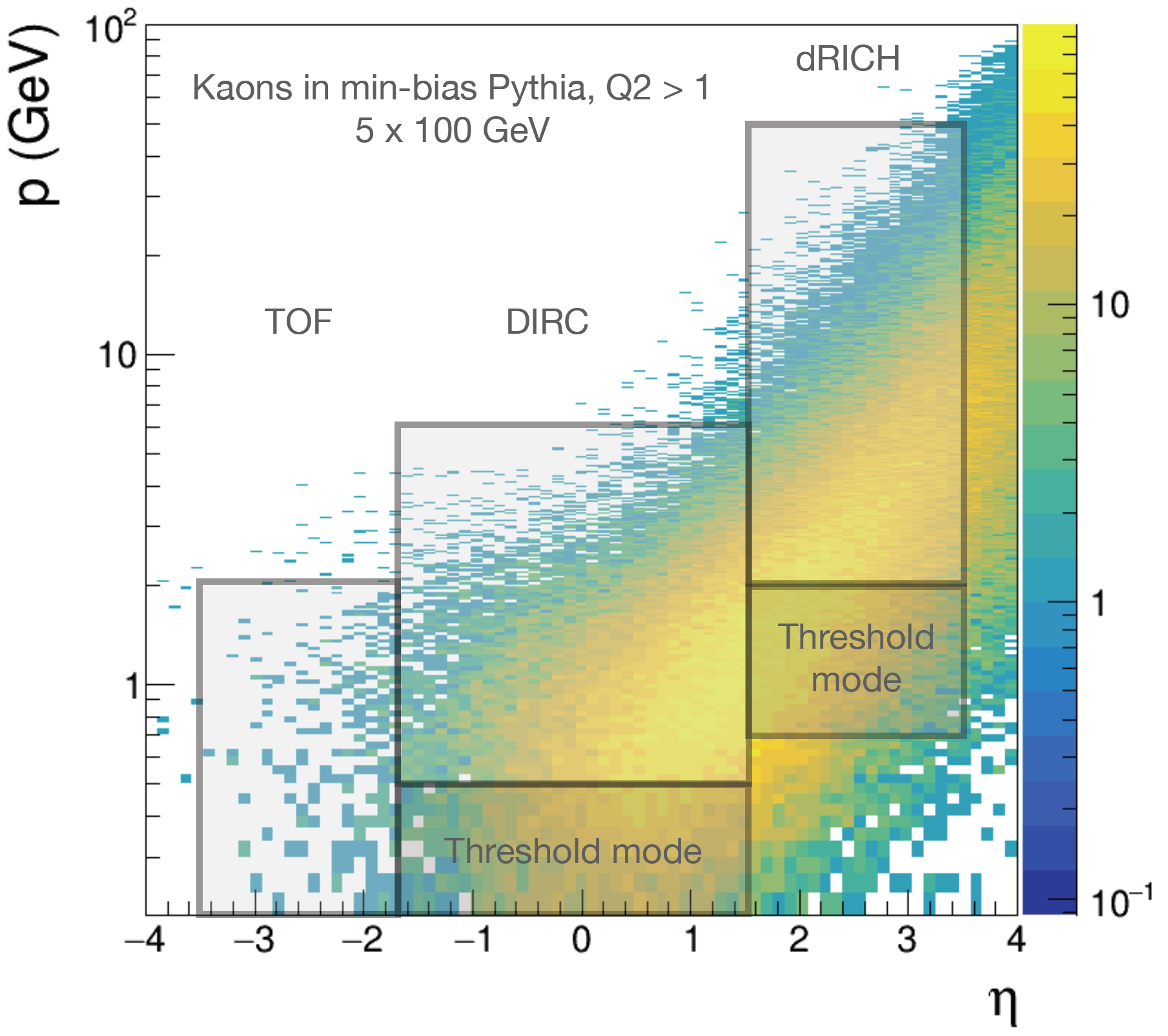}
	\includegraphics[keepaspectratio=true,width=3.0in,page=1]{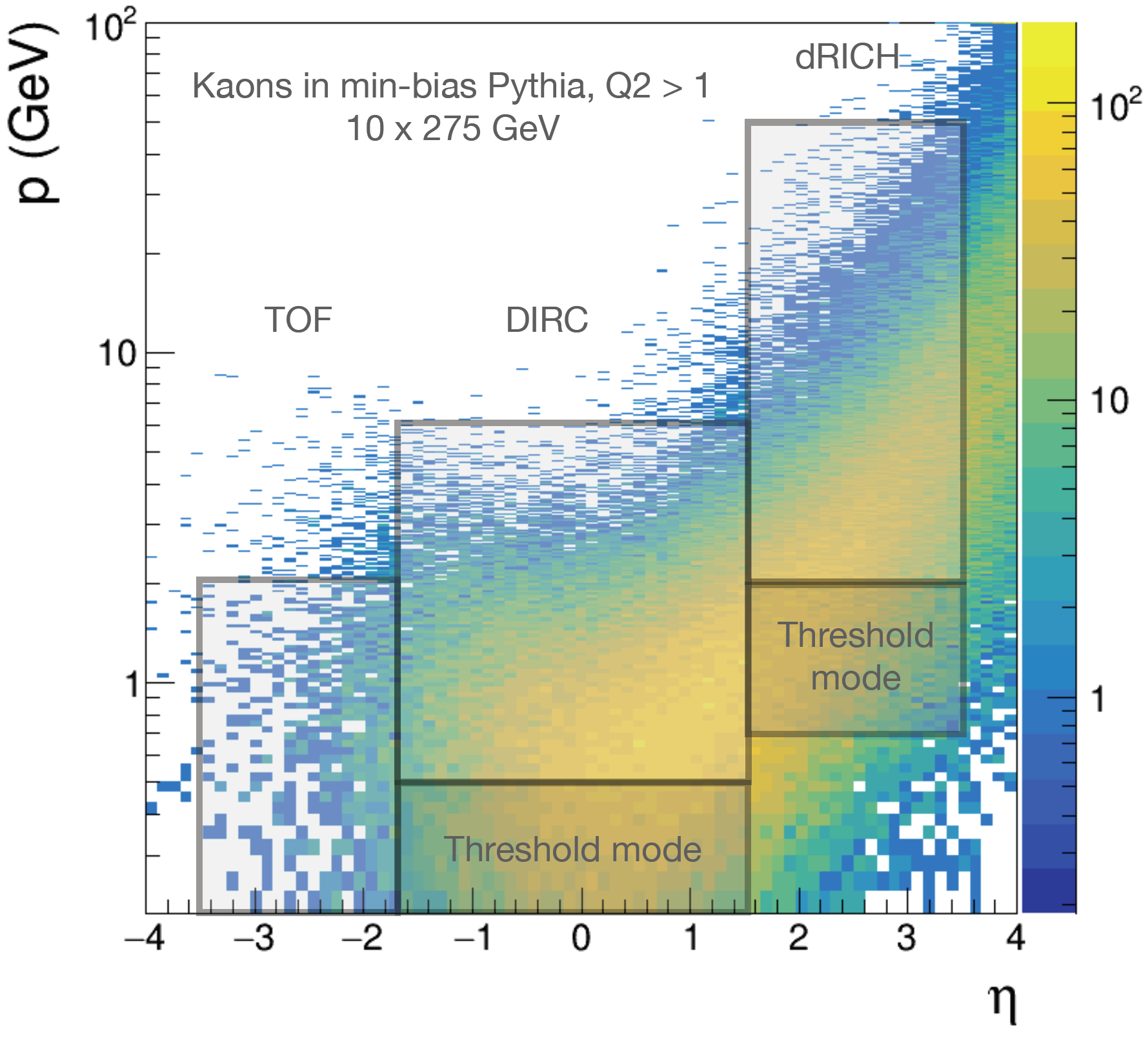}
	\caption{All DIS pions (upper panels) and kaons (lower panels) from the same data sets plotted separately as a function of $\eta$ for two beam energies, 5x100 GeV (left panels) and 10x275 GeV (right panels). The $3\sigma$ coverage of the CORE PID systems, as well as the region covered by the DIRC and dRICH ``threshold modes.'' The latter are discussed in the respective sections. A TOF similar to the one on the electron side (left) can be added to the hadron side (right) as a future upgrade.}
	\label{fig:kaons_10x275_5x100}
\end{figure}

\subsection{Particle Identification}
\label{sec:PID}

The hadron ID in CORE relies on three systems: a dual-radiator RICH in the hadron endcap, a DIRC in the barrel, and LGAD TOF in the electron endcap.
The distributions of pions and kaons with the PID coverage overlayed is shown for two beam energy settings in Fig.~\ref{fig:kaons_10x275_5x100}. Pion and kaon distributions in exclusive vector meson production are shown in Fig.~\ref{fig:eta_p_rho_phi} in section \ref{exclusive}.

\subsubsection{Dual-radiator RICH}
\label{dRICH}
The dual-radiator RICH (dRICH) is the primary hadron ID detector for CORE in the hadron endcap, covering $\eta>1.57$. It was designed and developed by the INFN group within the eRD14 EIC PID consortium, in which current members of CORE played a leading role.
The original size was matched to an EIC detector concept developed at JLab, which allocated more space for the dRICH than either the YR detector or CORE. However, while the CORE dRICH is smaller than the eRD14 design, sufficient space was allocated to preserve its excellent performance.

To achieve this, all dimensions were scaled uniformly except for the gas depth (along $z$), which was only reduced by 25\% from 1.6 m to 1.2 m. The latter also improved the aspect ratio compared with the original design, making more space for the photosensors, which are laid out so as to compensate for the non-flat focal plane created by the spherical mirrors. Thus, the CORE dRICH will be able to preserve the excellent characteristics of its slightly larger predecessor that was studied very thoroughly as part of the R\&D undertaken by eRD14.

\begin{figure}[htb!]
\begin{center}
	\includegraphics[width=0.45\textwidth]{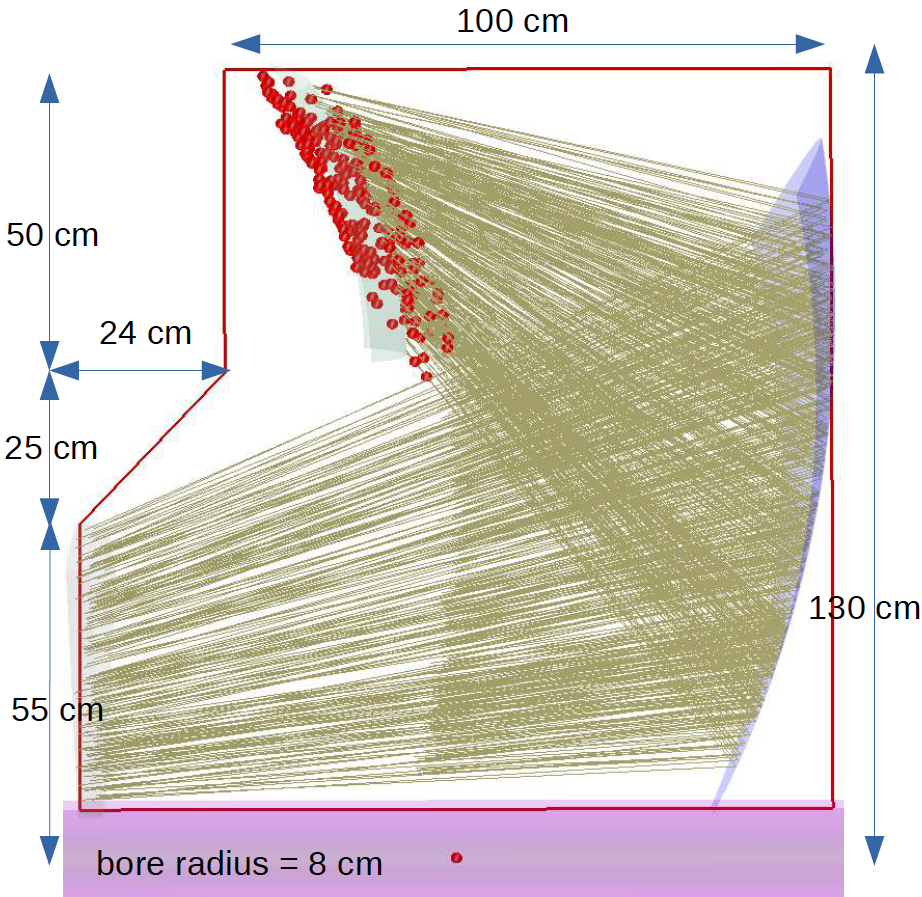}
	\caption{Dimensions of the CORE dRICH and ray-tracing of Cherenkov photons (yellow) emitted in the gas. Hits on the focal plane photosensors are shown in red and the mirror in blue.}
	\label{fig:dRICH_photons}
\end{center}
\end{figure}

The main requirement for the dRICH is to provide a continuous coverage (\textit{i.e.}, at least $3\sigma$ separation using either radiator) across the 2-50 GeV/$c$ momentum range, without leaving a gap in the middle as this would impact the physics. This meant matching an aerogel with a low index of refraction ($n = 1.02-1.03$) with a gas that would fall somewhere in-between $C_4F_{10}$ and $CF_4$. A good candidate for such a gas is $C_2F_6$, a gas mixture with a similar index of refraction, or even a pressurized gas such as Ar. If $C_2F_6$ is employed, which is the baseline assumption for the CORE proposal, an efficient recirculation system (which is included in the cost) will be required for environmental reasons.
Due to a higher photon yield, the mid-mass gas covers a momentum range that is close to what can be reached with $CF_4$.

Below kaon threshold in the aerogel, the dRICH can operate in a ``threshold mode,'' down to well below 1 GeV/$c$, where pions produce a ring (or ring segment) while kaons and protons do not.
The kaon / proton ambiguity has no impact on inclusive and exclusive measurements. For the latter, the target proton is tagged in the far-forward detectors.
Meson momenta in exclusive production are also higher than in DIS. As shown in Fig.~\ref{fig:eta_p_rho_phi}, in exclusive $\phi$ production there are no kaons with momenta low enough to be covered by the ``threshold mode.''
In the dRICH ``threshold'' region, for 10x275 the $\pi/K$ ratio is 9, and the $\pi/p$ ratio is 21. The dRICH will thus be able to provide excellent pion ID, but there will be a proton background for forward-going low-momentum kaons in SIDIS.
To address this, the dRICH could be supplemented by a TOF system. An option of placing an AC-LGAD layer in-between the last MAPS disk of the tracker and the aperture of the dRICH is discussed in the LGADs section \ref{LGAD}.

\begin{figure}[htb!]
\begin{center}
	\includegraphics[width=1.\textwidth]{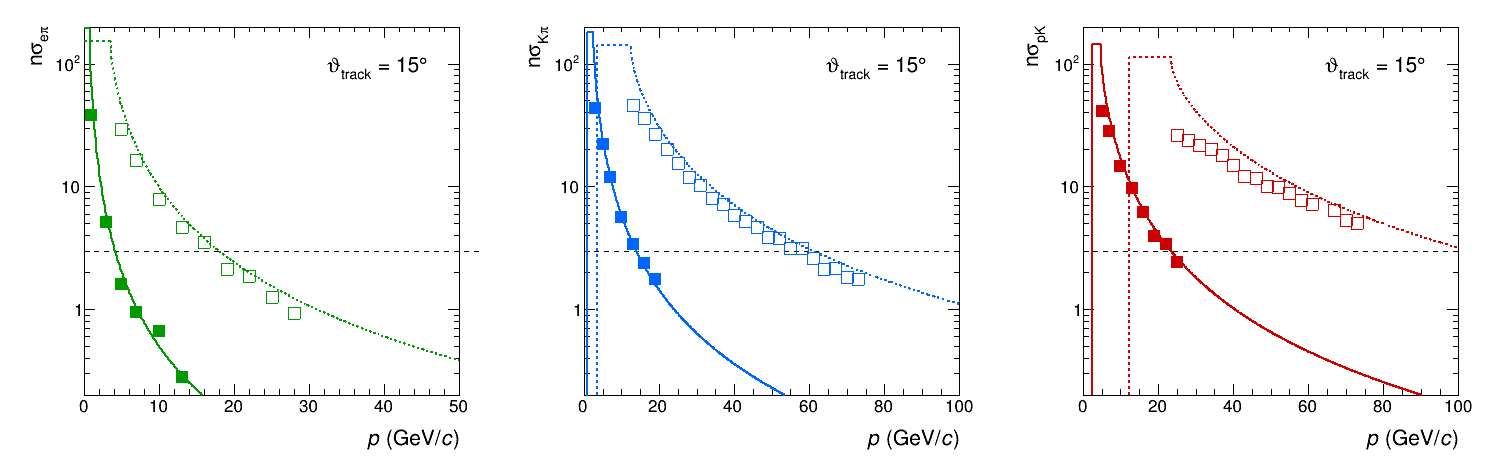}
	\caption{Separation power as a function of momentum for $e/\pi$ (green), $K/\pi$ (blue), and $p/K$ (red). The solid line is for the aerogel and the dashed line for the gas radiator. The flat section at low momentum indicates the threshold mode, not the $n\sigma$ separation (the discussion in the text refers to the solid blue curve). The plots show the performance for the original eRD14 dRICH with 1.6 m of gas (which reached $3\sigma$ $\pi/K$ separation at 60 GeV/$c$), but eRD14 studies suggested \cite{Alessio_private} that $3\sigma$ $\pi/K$ separation with 1.2 m of gas should still be well above 50 GeV/$c$. (The number of photons scales with the radiator length, and would thus be $\sim$25\% lower, but this would in part be compensated by smaller aberration effects).}
	\label{fig:dRICH_separation}
\end{center}
\end{figure}

One of the central design features of the dRICH are the outward-reflecting mirrors, which collect the light from both the aerogel and gas radiators onto the photosensors. This configuration makes it possible to move the latter out of the acceptance into an area where radiation levels are low, and where there is plenty of space for cooling infrastructure, making SiPMs a viable photosensor choice. It also means that Cherenkov light produced in the gas does not have to pass through the aerogel, where it can undergo Rayleigh scattering on its way to the photosensors. Instead, the design incorporates a filter in-between the aerogel and gas to minimize the impact of the former on the latter (as well as preventing exposure of the aerogel to the gas).

Aerogel offering the required performance is currently available from a Russian provider, and will soon likely also be sold by a Japanese company that inherited the Matsushita technology. Monitoring will be necessary to ensure the long-term stability of the aerogel.

The expected overall amount of material for a non-pressurized gas solution will be $X/X_0 \approx 2\%$, to which the vessel and mirror each contribute about 1\%.

Ongoing targeted R\&D supported by the project (eRD102) focuses on cost reduction for the mirror and vessel (which do no present a technological challenge), and on the definition of an optimal solution for the photosensors and readout electronics. 
For the CORE proposal, the baseline solution is to use cooled SiPMs in combination with time-over-threshold electronics based on the ALCOR chip.

A more detailed description of the dRICH can be found in section 11.5.3 (p.529) of the Yellow Report \cite{AbdulKhalek:2021gbh}.

\subsubsection{High-performance DIRC}
\label{hpDIRC}
The DIRC is the main system for hadron ID in the barrel, covering $-1.65<\eta<1.57$. 
It is a radially compact Cherenkov detector, and its PID performance is largely independent of the DIRC barrel radius and the bar length, making it ideal for a compact detector such as CORE, where the small radius ($\sim$0.5~m) not only reduces weight and cost, but also improves the acceptance for low-$p_T$ particles for which PID is required. 
The latter is true for all solenoid field settings, but is particularly important at 3~T. 
In the CORE geometry, the DIRC can thus cover the full momentum range, from low to high $p_T$.

The high-performance DIRC (hpDIRC) concept was developed as part of the generic R$\&$D program performed by the EIC PID collaboration (eRD14) with the focus on extending the momentum coverage of the DIRC
to provide more than $3\sigma$ separation for $\pi/K$ up to 6~GeV/$c$, $K/p$ up to 10~GeV/$c$, and to contribute to low-momentum electron identification in the barrel where pion backgrounds are the largest. 
The DIRC also provides an excellent timing resolution, especially in the vicinity of electron endcap (where the DIRC readout is located), making it synergetic with the LGAD TOF detectors located in the endcap.

\begin{figure}[htb]
\centerline{%

\includegraphics[width=1.\textwidth]{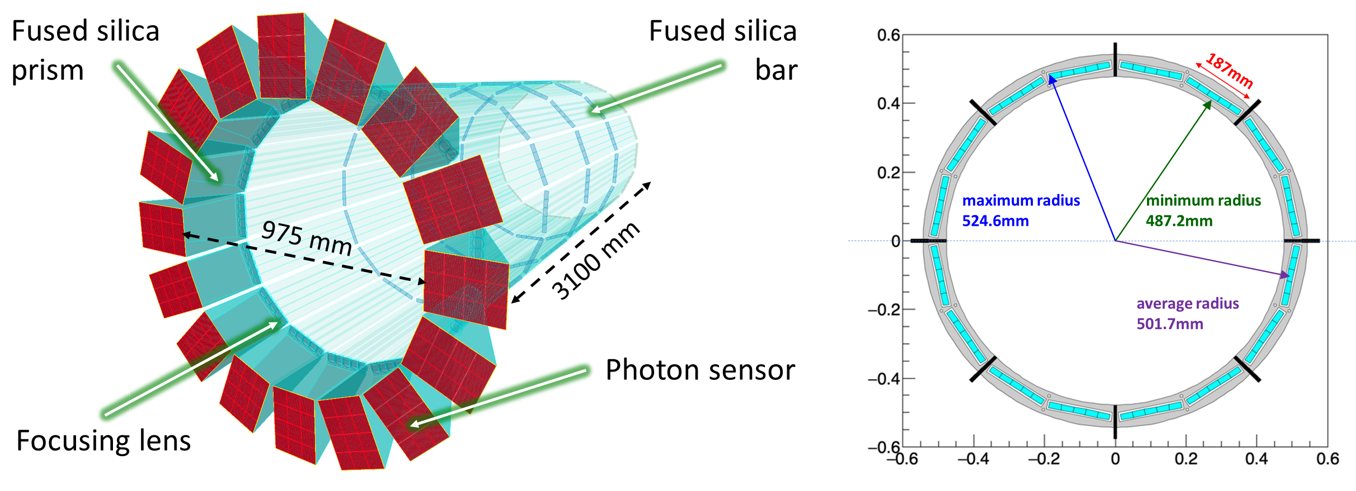}
}%
\caption{Geometry of the CORE high-performance DIRC detector in Geant4 simulation (left). Cross section of the bar boxes with support structure with the radial bar geometry indicated (right)}
\label{fig:DIRCDesign}
\end{figure}

The baseline design of the CORE hpDIRC detector was implemented in a detailed Geant4 simulation and is shown in Fig.~\ref{fig:DIRCDesign}a. 
It is divided into sixteen optically isolated sectors that comprise a bar box and a readout box, surrounding the beam line in a 16-sided polygonal barrel with an inner radius of about 50~cm, as shown in Fig.~\ref{fig:DIRCDesign}b.
Each bar box includes a set of five radiator bars made of synthetic fused silica, each 2.8~m long, with a cross section of 17~mm $\times$ 35~mm. 
The bars are placed side-by-side, separated by small air gaps, inside the light-tight bar box.
Mirrors are attached to one end of each bar to reflect Cherenkov photons towards the readout end, where they exit the bar and are focused by a 3-layer spherical lens on the back surface of the prism that serves as an expansion volume. 
The prism has a $32^{\circ}$ opening angle and dimensions of 237~mm $\times$ 175~mm $\times$ 300~mm. 
The detector plane of each prism is covered by an array of $3 \times 4$ MCP-PMTs with 3~mm $\times$ 3~mm pixels giving a total of about 49k channels to record the position and arrival time of the Cherenkov photons.
The pixel size was selected based on a Geant4 study as a compromise between the cost and performance for the hpDIRC design on one hand and technologically feasible sensor properties.

The conservative baseline design of the hpDIRC uses commercial microchannel plate PMTs (MCP-PMTs), available from Photek or Photonis. The hpDIRC simulation is based on Photonis Planacon XP85122 MCP-PMTs with realistic sensor parameters, including photon timing, collection, and quantum efficiency.

The mechanical design of bar boxes, readout sections, and support structure will be adapted from the PANDA Barrel DIRC detector as its key dimensions are almost identical to the CORE DIRC design. 

\begin{figure}[htb]
\centerline{%
\includegraphics[width=1.\textwidth]{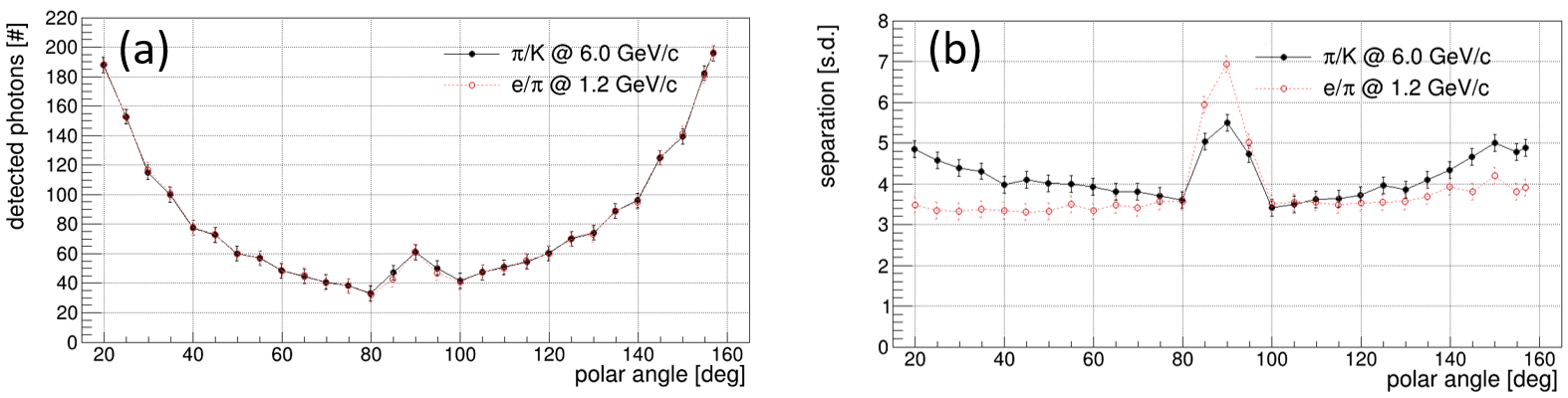}
}%
    \caption{Expected performance of the hpDIRC: Photon yield (a) and separation power (b) as a function of the particle polar angle.
    Results are based on the standalone Geant4 simulation and shown for $e/\pi$ at 1.2~GeV/$c$ and $\pi/K$ up to 6~GeV/$c$.
    }
\label{fig:DIRCPerformance}                                                 \end{figure}

Fig.~\ref{fig:DIRCPerformance} shows the expected performance from the standalone Geant4 simulation studies for two particular cases. 
The black points show the photon yield and separation power for charged pions and kaons as a function of the polar angle at a momentum of 6~GeV/$c$ while the red points show the same quantities for charged pions and electrons at 1.2~GeV/$c$.
The number of measured Cherenkov photons per particle (a) ranges from 35 to 200, depending on the polar angle.
The separation powers for both combinations of particle types (b) are derived from Gaussian fits to the log-likelihood difference for pairs of particle hypotheses and calculated as the difference of the mean values of the two Gaussians, divided by their average width. 
The angular resolution of the CORE tracking system at the hpDIRC radius is conservatively taken to be 0.5~mrad at 6~GeV/$c$ and 2.2~mrad at 1.2~GeV/$c$. 
The expected particle identification performance of the hpDIRC exceeds the CORE PID goal of $3\sigma$ (s.d.) separation power $\pi/K$ up to 6~GeV/$c$ for the entire polar angle range. It also provides good $e/\pi$ separation, and thus a significant suppression factor for pion backgrounds in the 0.7 - 1.4 GeV range, which is cumulative with the suppression provided by the EMcal and kinematic constraints. As shown in Fig.~\ref{fig:e_pi_ratio}, this is the most critical range for $e/\pi$ ID in the barrel region.

Cherenkov detectors, in addition to the nominal ring imaging operation, which provides positive ID of the particle type, can perform $\pi/K$ separation below the kaon threshold, in a so-called ``threshold mode" or ``veto mode". 
If a significant number of Cherenkov photons is detected by the hpDIRC sensors (typically, more than 10) for a particle with a momentum below the kaon threshold (at 0.5 GeV/$c$), the veto mode excludes the kaon and proton hypotheses and identifies the particle as a lower-mass particle, such as a pion or electron.
The veto mode will extend the CORE hpDIRC PID reach to lower momentum (the corresponding $p_T$ is lower still except at $\eta$=0) and allow $\pi/K$ separation above 0.2~GeV/$c$ for all values of $\eta$ except for a narrow range very close to zero, and above 0.3~GeV/$c$ for full angular range covered by the hpDIRC.

However, one should note that the $\pi/K/p$ ratios change at very low momenta. For the 10x275 GeV Pythia simulation shown in Fig.~\ref{fig:kaons_10x275_5x100}, in the $\eta$ range covered by the hpDIRC and for momenta between 0.2 and 0.5 GeV/$c$, the $\pi/K$ ratio is 20 and the $K/p$ ratio is 5.6. At mid rapidity, low-momentum hadrons are thus mostly pions that can be identified with high efficiency. The kaon / proton ambiguity for low momenta has no impact on inclusive measurements and only a marginal one on exclusive ones since most exclusive kaons have higher momenta and non-resonant $p\bar{p}$ production is a very small background for $\phi \rightarrow K^+K-$ (the primary background for the $\phi$ is two-pion production, \textit{e.g.}, $\rho^0 \rightarrow \pi^+\pi^-)$.

\begin{figure}[htb]
\centerline{
\includegraphics[width=0.55\textwidth]{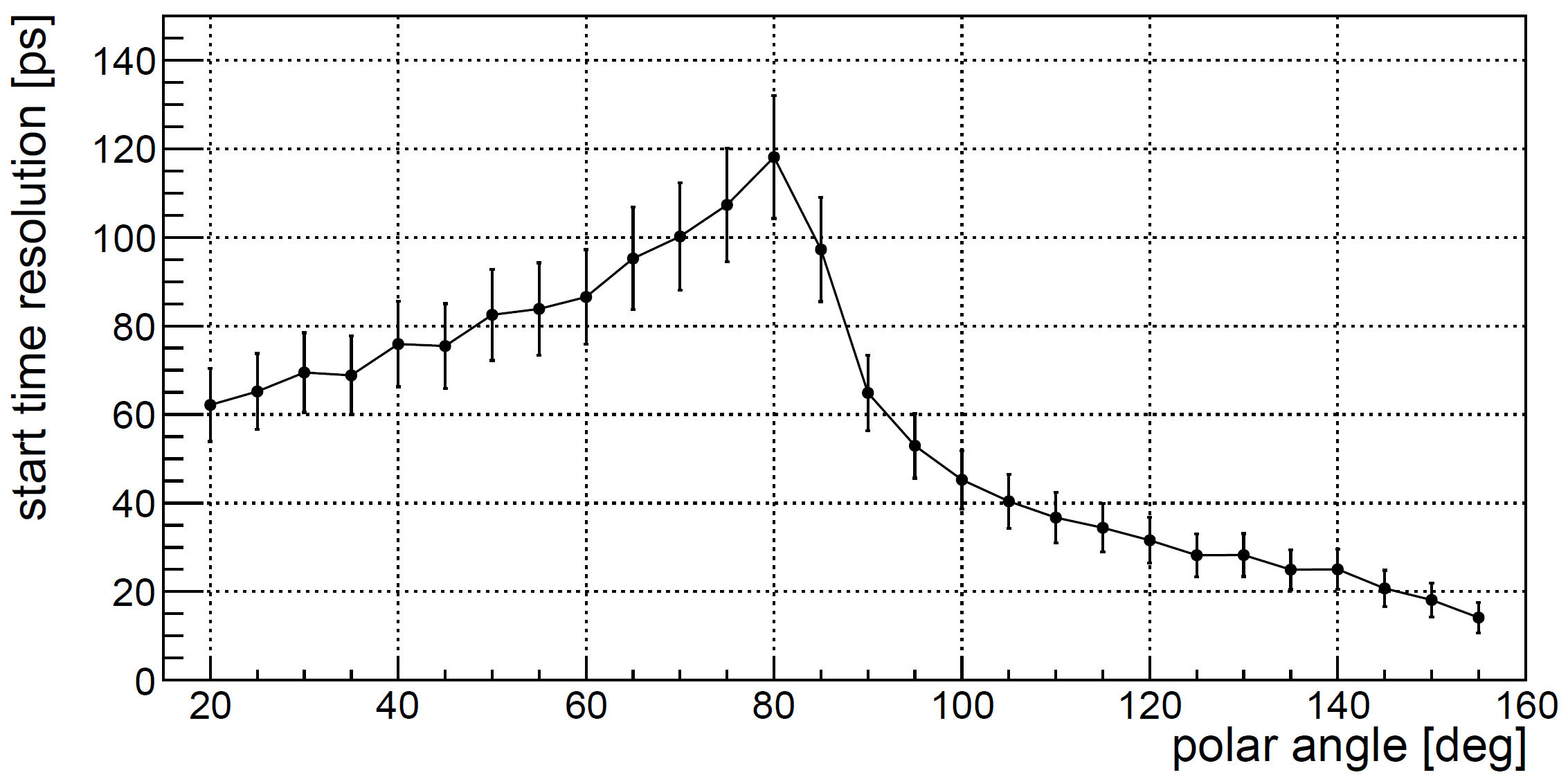}
}%
\caption{Resolution of the reconstructed time (``start time'') of the particle traversing the hpDIRC bar, as a function of the particle polar angle ($180^\circ$ signifies the electron beam direction, and hence the direction of the electron endcap).
Results are based on a standalone Geant4 simulation and shown for a single charged kaon with a momentum of 6~GeV/$c$.
}
\label{fig:DIRC_timing}
\end{figure}

In addition to providing hadronic PID, the hpDIRC is able to provide a useful contribution to the event time determination.
By calculating the average of the difference between the measured an expected arrival time for all detected photons in a sector, the time of the particle traversing the radiator bar can be determined with a resolution of 20--120~ps, depending on the polar angle, as shown in Fig.~\ref{fig:DIRC_timing}.

\begin{figure}[htb]
\centerline{%
    \includegraphics[width=1.\textwidth]{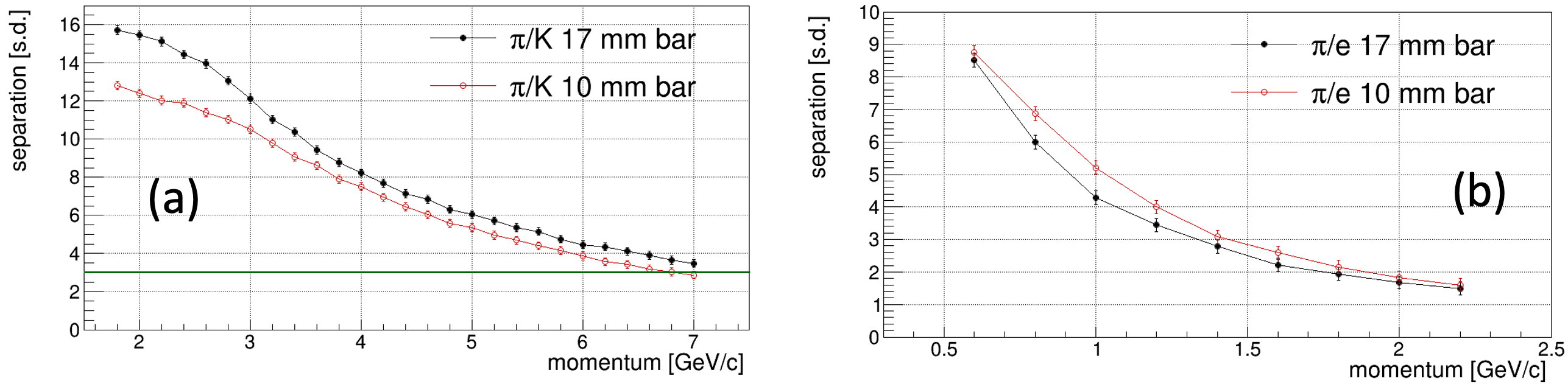}
}%
    \caption{Expected performance of the hpDIRC: Separation power ($n\sigma$) for $\pi/K$ (a) and $e/\pi$ (b) for two different bar thickness values as a function of the particle momentum.
    Results are based on a standalone Geant4 simulation and shown for a polar angle of 30$^\circ$.
    }
\label{fig:DIRCPerformance-thickness}
\end{figure}

The baseline design for the CORE hpDIRC assumes that the bars from the decommissioned BaBar DIRC are reused to reduce the system cost.
However, the small size of the CORE DIRC makes it affordable to build new bars, which could then be made thinner.

A Geant4 simulation study was performed in which all parameters of the baseline hpDIRC were kept unchanged, except for the bar thickness, which was reduced from 17~mm to 10~mm.
Fig.~\ref{fig:DIRCPerformance-thickness} shows the comparison of the expected performance for the baseline hpDIRC and the low-mass hpDIRC with thinner bars.
The key advantage of the thin hpDIRC is a significant reduction of the multiple scattering at lower momenta, improving the $e/\pi$ separation performance where pion backgrounds are the largest, while maintaining the 3~s.d. performance goal for $K/\pi$ separation.
Furthermore, the 41\% reduction in radiator material thickness would provide a significant benefit to the performance of the barrel EmCal and the lower weight of the bar boxes would simplify the mechanical support structure.

Thinner bars would emphasize the purity in the reconstruction of the scattered electron, which is essential for parity-violating DIS, and would create a natural complementarity between two detectors.

\subsubsection{LGAD Time of Flight}
\label{LGAD}
The AC-LGAD TOF is the primary hadron ID system for the electron endcap, covering $-3.5<\eta<-1.65$. The distance from the collision point is 1.2~m and the timing resolution is 25~ps. It covers the full endcap area of 0.6 $m^2$. Together with the DIRC, the LGAD also provides a measurement of the event start time ($t_0$).
A proven way of determining the latter in precision electron-scattering experiments, is the approach used by the CLAS detector at Jefferson Lab for over a decade during the 6 GeV era. This method was so successful that it was retained for the recently upgraded CLAS12 detector \cite{Burkert2020163419}.
It relies on measuring the time of arrival of an identified electron (a particle with $\beta$=1), and precisely reconstructing its path back to the vertex. In comparison with CLAS, the short flight path in combination with the position resolution of the all-silicon tracker will generally make the path uncertainty very small (CLAS12 has 80 ps resolution, but is much larger and uses drift chambers).
As shown in Fig.~\ref{fig:DIRC_timing}, the combination of AC-LGADs and DIRC can provide a resolution at the 25~ps level for $\eta < -0.55$, and at the 40~ps level up to $\eta < -0.35$, which is sufficient for determining the vertex time ($t_0$) with high precision. This $t_0$ can then be used for PID for hadrons that hit the electron endcap.
There are two limitations to this method of determining $t_0$. First, since the vertex z-resolution for low-momentum electrons would begin to deteriorate for $\eta < -2$ (as shown in Fig.~\ref{fig:tracker_resolution}, the TOF resolution would be affected unless there is another particle with either higher momentum or larger $\eta$. However, since this is almost always the case, this limitation has a very small impact on physics. The other limitation is for photoproduction ($Q^2$=0) events, where the scattered electron goes down the beam pipe to the low-$Q^2$ tagger instead of hitting the endcap detectors. However, photoproduction at the EIC is mostly relevant for heavy quark systems (\textit{e.g.}, $J/\psi$), which typically decay into electrons. These electrons can then be used for determining $t_0$.
In the rare cases when there is no identified electron, a ``multi-hadron'' method can be used for obtaining $t_0$ by a fit to all tracks. This approach is generally inferior to the electron method unless the number of hadrons is large, but in CORE the latter would only be relevant under the very particular circumstances described above.
The current CORE design is compatible with an extended LGAD coverage. An AC-LGAD layer could be added in-between the last MAPS disk of the tracker
and the aperture of the dRICH. This is not part of the CORE baseline, but could be a future upgrade.

The LGAD technology is similar to one used at Phase-2 upgrades of CMS and ATLAS at the HL-LHC. It is designed to provide a time resolution of 25~ps and a position resolution of 30 microns per hit, and will serve as both a TOF and the last tracking layer in the electron direction.
R\&D on adaptation of this technology for the EIC was undertaken by eRD29 and is described in section 14.3.7 (p.686) of the Yellow Report \cite{AbdulKhalek:2021gbh}. It has recently transitioned into the eRD112 targeted R\&D program, which focuses on developing the AC-LGADs technology to provide both excellent position and timing resolutions.

The standard LGADs used by CMS and ATLAS have a limitation on the spatial resolution. There are also dead areas  at the edges of the pixels and in-between the pixels, so that large-pitch pixels are necessary to avoid a low fill-factor. The AC-coupled LGAD technology essentially eliminates the dead area by utilizing resistive readouts, and can achieve a position resolution of a few microns by taking advantage of charge sharing among adjacent readout channels. The ``off-the-shelf'' resistive LGADs nevertheless provide a fallback option.

The AC-LGAD based TOF has the advantage of being highly compact, high magnetic field tolerant, and radiation hard (although this is not a primary concern at EIC). Readout and services are designed to be similar to the timing layers at CMS and ATLAS. Timing and amplitude of the signal will be read out and digitized by dedicated ASIC chips bump bonded to the LGADs sensors. Data will be aggregated by a service hybrids, which supplies low voltage power and consists of a readout board to transmit the data to the back-end electronic system. Detector modules are mounted on an aluminum support structure with liquid cooling.

\subsection{Electromagnetic Calorimetry}
High performance electromagnetic calorimetry (EMCal) is a key feature of the CORE design.
This is implemented with two technologies:
Lead Tungstate (PbWO$_4$) crystals, and a Tungsten-scintillator multilayer ``shashlyk''  \cite{Woody:2011bja}, which will be projective in the barrel and non-projective in the endcaps.
The EMCal is hermetic over the range $-3.5<\eta<3.5$,
with some loss in efficiency and degradation in resolution at the transition points around $\eta=-1.7$ and 1.4 (relative to $z$-axis).  The coverage at large negative pseudo-rapidity can likely be extended with
a small `far-backward' calorimeter before the first downstream electron quadrupole.  This can only be designed once there is a near-final design of the IP8 beamline and beampipe.

\begin{table}
\caption{\baselineskip 13pt
Physical properties of EMCal designs.   Decay time constants, 
and light yield thermal coefficient of PbWO$_4$ are taken from \cite{Follin:2021kgn}, measured at $20^\circ$ C.
For the W-Shaslik design, the absorber plates are
made of an 80\% W 20\% Cu alloy, which is readily machinable. ``Cell''
parameters are for the complete W-shashlyk module, not including effects
of the wavelength shifting fibers on a $1.5$ cm spacing square grid.}
\begin{tabular}{ccccccc}
\hline\hline
 Material & Density & $X_0$ & $R_M$  & Decay Time(s) & Thermal Coef \\
  & g/cm$^3$ & cm & cm &  nsec  & \% per $^\circ$C\\
  \hline
PbWO$_4$ & 8.30 & 0.89 & 2.0  & 2(58\%) \& 6(42\%) & -2.5\\
W & 19.3 & 0.35 & 0.94 & \\
\hline \hline
W-shashlyk & &\\  
 Material & Density & $X_0$ & $R_M$  & Layer  
  & \\ 
    & g/cm$^3$ & cm & cm &  mm &  \\
 \hline
 WCu & 17.2 & 0.41 & 1.02 & 1.25 &  \\
 Scint & 1.0 & 41. & 9.4 & 2.0 &  \\
Cell & 7.2 & 1.05  & 2.3 & 3.25& \\ 
\hline\hline
\end{tabular}
\label{tbl:EMCal}
\end{table}

\subsubsection{Lead Tungstate EM Calorimeters}
Lead Tungstate (PbWO$_4$) is a high-density material producing a large light yield in a short time interval
($<10$ ns), making it ideal for measuring electrons and photons in the energy range of interest at the EIC (1-20 GeV). Some basic parameters are shown in Table~\ref{tbl:EMCal}.
The electron Endcap EMCal design consists of 1432 rectangular
PbWO$_4$ crystals, each $2.0\times 2.0\times 20.0 \text{ cm}^3$. An outer wrapping of reflective and optical isolation layer is allocated a thickness of 0.25 mm (density $\approx 1 \text{ g/cm}^3 \ Z_\text{eff}\approx 6$), for a net spacing of 0.50 mm between crystals. The crystals
are stacked in 43 rows and 43 columns.
An inner square hole of $7\times 7$ crystals, or $(14.35\text{ cm})^2$ allows access to the first
flange of the beam pipe. The rows and columns are truncated to  approximately form a circular disk,
of maximal radius $45.8$ cm.  

Given the modest angular range of this calorimeter, the crystals will be flat and non-projective. This simplifies the manufacture and assembly, and allows for reuse of existing PbWO$_4$ crystals.
To avoid line-of-sight gaps, the horizontal and vertical detector
midplanes pass through the centers of a crystal row and column, respectively (hence the odd number of rows and columns of a detector stack with 4-fold symmetry).

To minimize any deficiency in resolution or efficiency at the transition from the endcap to the barrel, the outer 8 rings of crystals in the endcap are arranged in a semi-projective stepped pattern in $z$, as illustrated in Fig.~\ref{fig:CORE-2D}.

The crystals must be held to a constant, and preferably a uniform, temperature of $20^\circ$ C.
The endcap crystals are supported and temperature stabilized by a Cu ring of outer radius 47 cm, which is divided into
8 octants, with small gaps to allow mechanical flexibility. Each Cu octant in independently cooled by embedded channels. The Cu octants are wrapped in
a carbon fiber belt to form an integral mechanical unit. The Cu octants connect to a front cooling plate
and a rear cooling grid.  The Cu octants extend 10 cm behind the crystal array
to a mechanical backplane that supports both the crystals and digitizing electronics. This backplane is in turn
cantilevered from a support extending to the barrel flux return at $z = -2.5$ m.

The barrel region (outside the radius of the DIRC) for $\eta<0$ is also instrumented with
a PbWO$_4$ calorimeter \footnote{\baselineskip 13 pt
In fact, the PbWO$_4$ barrel calorimeter starts at slightly negative $\eta$, to allow the last W-shashlyk module to overlap $\eta=0$.}. This calorimeter is 2D projective in
$\phi$ and $\eta$, with the projection slightly offset from the IP. Each crystal is a 20 cm long trapezoidal prizm, with mid-length cross sections approximately a uniform
$2\times 2\text{ cm}^2$.
There are 53 rings of 204 crystals each in the barrel, for a total of 10,812 new crystals.
The total number of PbWO$_4$ crystals required for CORE is $200\cdot 53+1432=12224$.

We have modeled the performance of both $\eta<0$
PbWO$_4$ calorimeters with an energy resolution for
electrons, positrons, and gammas of
\begin{equation}
\left.\frac{\sigma(E)}{E} \right|_\text{PBWO$_4$}
= 1\% \oplus \frac{2\%}{\sqrt{E/\text{GeV}}} \oplus
\frac{1\%}{E/\text{GeV}}.
\end{equation}

\subsubsection{W-shashlyk EM Calorimeters}

The electromagnetic calorimetry in the range 
$\eta\in [0,3.5]$ is implemented with
tungsten-shashlyk (W-shashlyk) modules
which are projective in the barrel and non-projective in the endcap.
The use of tungsten rather than Pb increases the density, making the calorimeter more compact, and reduces the Moli\`ere radius, potentially improving both the
spatial resolution and two-photon separation power.
The precise ratio of W to plastic scintillator
can be tuned.  Our baseline inTable~\ref{tbl:EMCal} is based on input from the EIC Yellow Report 
(EIC-YR~\cite{AbdulKhalek:2021gbh}, Table 11.35 and Figures
11.80-11.81), 
eRD1 reports, particularly from January and July 2020\cite{eRD}, and \cite{Kuleshov:1528}.

In the barrel, individual W-shashlyk towers are trapezoidal prizms, with a $6\times 6\text{ cm}^2$  cross section (mid-length)
and a $4\times 4$ array of wavelength shifting fibers (WLS) running the length of the module. Based on the experience and simulations quoted in the EIC-YR we expect this geometry to have
an energy resolution for electrons and gammas of 
\begin{align}
\left.\frac{\sigma(E)}{E}\right|_\text{shashlyk} &= \left( 2\% \oplus \frac{6.3\%}{\sqrt {E/(1\text{ GeV})}}
\oplus \frac{2\%}{E/\text{GeV}}\right).
\end{align}
In the barrel region: $0\le \eta\le 1.4$ the modules are 25 radiation lengths in depth, to ensure full capture of the
EM showers up to the kinematic limit.
In the ion endcap region $1.33\le \eta\le 3.3$ the W-shashlyk 
depth is $20 X_0$.  The ion endcap HCal, directly behind the
W-shashlik EMCal will capture any residual energy of the showers. Notice there is substantial geometric overlap
between the barrel and endcap W-shashlyk acceptance, to capture
shower leakage from either EMCal.

The high-$\eta$ EMCal coverage is extended by a short EMCal after the
trackers in the
B0 magnet and a high resolution EMCal in front of the Zero-Degree Calorimeter (ZDC).
We define $\eta_i$ as the pseudo-rapidity relative to the ion beam line (as in Fig.~\ref{fig:CORE-2D}). 
There is no official design yet for
the IP8 B0 dipole. Based on the IP6 parameters, the B0 EMCal
can span a range $[7.5,20] $ mrad, or $4.6<\eta_i<5.6$.
In the horizontal plane at $\phi=0$, this is a range $3.6<\eta<3.8$
(relative to $\hat z$). The ZDC acceptance is a cone of $\pm 5$ mrad, or $\eta_i > 6.0$.

\subsection{Hadronic Calorimetry and Muon ID}
At the EIC, jets are in general best reconstructed from individual particles. However, for $\eta>2.5$ the momentum resolution of the tracker starts to deteriorate. The lab-frame energy of jets moving in the direction of the hadron beam is also higher then elsewhere due to simple kinematic effects. Thus, in the forward region the HCal is crucial for high-resolution jet measurements, and for hadronic methods of kinematics reconstruction discussed in section \ref{inclusive}.

For $\eta<1.2$, where jets will be reconstructed from individual identified particles, the momenta of which will be determined by the tracker and EMcal. However, there will be some energy leakage from neutral hadrons (mostly $K_L$'s) in about 1/3 of the jets. Since these neutral hadrons typically have low momenta, an energy measurement using a barrel HCal would not significantly reduce the uncertainty on the energy and direction of those jets. But knowing if one or more neutral hadrons were present within the jet cone can help in the analysis. Thus, as the baseline solution for CORE we plan to use a neutral hadron and muon ID detector, based on the Belle II ``KLM'', covering the barrel and electron endcap.
In addition to measuring neutral hadrons, the KLM is also an excellent detector for muon ID, which is important for production of heavy quarkonia (muon distributions from $J/\psi$ and $\Upsilon$ are discussed in section \ref{exclusive}), and processes such as timelike DVCS.

\subsubsection{Forward Hadronic Calorimetry}
In the hadron endcap, covering $1.2<\eta<3.5$, CORE will have a non-projective Fe/Sci HCal developed within eRD1 and also used for the STAR FCS upgrade. The anticipated resolution is $60\%/\sqrt{E}$ with a constant term of 6\%. Each module covers an area of 10x10 cm$^2$ and consists of interleaved layers of 20 mm Fe and 3 mm scintillator. In the STAR FCS there are 520 modules, each with 36 double layers. The STAR FCS is discussed in section 11.4.4 (p.515) of the Yellow Report \cite{AbdulKhalek:2021gbh}.
These modules are a good geometric fit for measuring particles hitting the outer part of the CORE endcap, and
will be re-used as-is in this position. The inner 1,602 modules will be made slightly longer (44 double layers), and be located directly behind the non-projective endcap EMcal. In the original UCLA design this was a W/SciFi, but W-shashlyk should also work equally well \cite{Oleg_private}.
The HCal will use a readout scheme similar to the STAR FCS, with additional independent readout of the three last scintillation tiles in the HCal towers with WLS fibers as a tail catchers. This was studied during Generic EIC detector R\&D.

Finally, we note that there is ongoing targeted R\&D (eRD107) to create a longitudinal segmentation within each module by having four sections with scintillators with alternating long and short time constants, making it possible to introduce software compensation. If the R\&D is successful, this approach could be applied to the new CORE Hcal modules (although it would require six SiPMs per module instead of one).

\subsubsection{Neutral Hadron and Muon Detection}
\label{KLM}

The Neutral Hadron and Muon system for CORE is based on the Belle KLM ($K_L - \mu$) detector
and its subsequent upgrades at Belle II \cite{Aushev:2014spa}.
It comprises layers of orthogonal scintillator strips with embedded WLS fibers and a single-end SiPM readout, interleaved with plates of the solenoid flux return steel.
The original Belle KLM used Resistive Plate Chambers (RPCs), but these are replaced by scintillator strips in the Belle II upgrade, which is what we also propose for the EIC.

\begin{figure}[htb!]
\begin{center}
\includegraphics[width=0.7\textwidth]{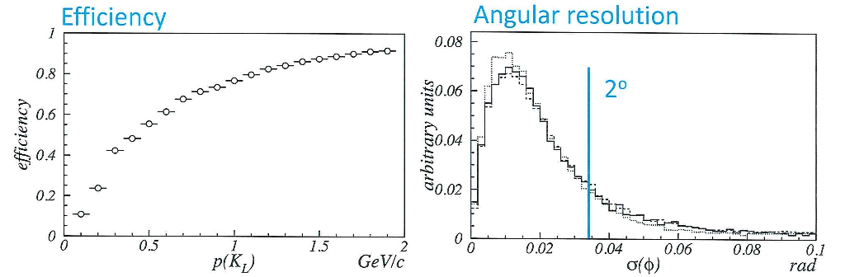}
\includegraphics[width=0.6\textwidth]{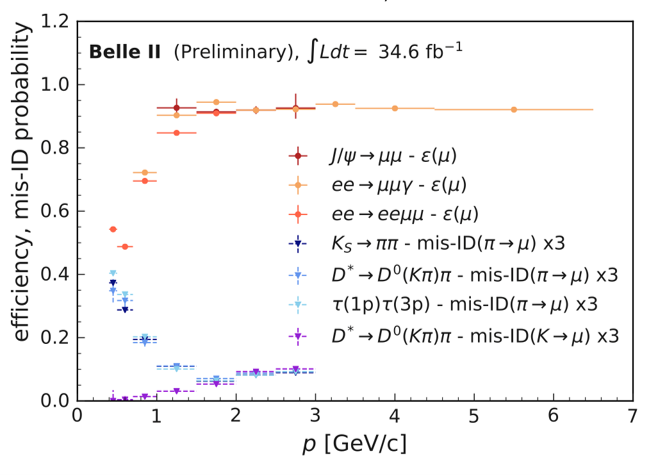}%
\hfill%
\caption[KLM performance at Belle]
{Top: $K_L$ efficiency (left) and angular resolution (right) from Belle data.
Bottom: muon ID and fake rate vs. layer number from early Belle II data.}
\label{fig:KLM_perform}
\end{center}
\end{figure}

In such a detector, muons are identified with high purity by
measuring their range
in the scintillator-steel detector stack. Neutral hadrons (mostly $K_L$s) are identified by the localized readout layer response following their interaction in a preceding steel plate.
Segmentation in the readout along the $z$ (beam axis) and $\phi$ (azimuth) directions, allows for a spacial coordinate measurement. Some aspects of the performance are shown in Fig.~\ref{fig:KLM_perform}.
The muon efficiency from early Belle II data in the lower panel, shows a steady rise from a turn on at $\sim 0.6$ GeV (determined by the material burden before the first readout layer) to a high-efficiency plateau; the mis-identification rate decreases with layer number out to $\sim 7$ layers. The muon momentum is determined by the inner tracking detectors.
In the upper plots, efficiency for $K_L$ detection is shown along with the angular resolution from tagged $K_L$'s using Belle data. Results from the Belle II scintillator upgrade are
not yet available,
but are expected to be as good or better. 

In the baseline CORE design, we will instrument the return steel of the entire barrel and electron-side endcap in this way for neutral hadron and muon detection. Besides the somewhat different geometry (a more elongated and compact barrel; smaller-radius endcap encircling the beam pipe), the main modifications in the application of the Belle design to CORE relate to a desire to shrink the radial extent of the readout gaps for overall compactness. 
Hence, the scintillator strip layer implementation generally follows that used at Belle, namely an octagonal steel plate structure of the barrel return iron accommodates the readout layers slid into air gaps. The barrel layer panels are rectangular, each panel comprising two orthogonal layers of scintillator strips glued onto a thin common substrate, enclosed by an aluminum frame and covered with additional support/protective sheaths. Two rectangular detector panels are placed in each layer of a barrel octant, one at the ion side and the other at the electron end, each inserted, respectively, where the barrel-endcap junction also provides service connections. For the endcap (on the electron-side), the scheme is similar, but with the matching endcap plated structure divided in halves (for easy removal to the side) and the active scintillator strip layers inserted at the outer radius in quadrant-shaped panels. We have chosen an insertion gap of 21.5 mm interleaved with 55.5 mm steel plates ($\sim $72\% steel) for the entire implementation. This provides a workable solution that also enables a sampling frequency similar to a standard Hcal - but with an individual layer response incorporating 3D positional readout.

A suitable scintillator strip geometry for the above has a cross sectional area of 7.5 mm $\times$ 30 mm, read out by a 1 mm diameter Wave Length Shifting (WLS) fiber. Use of a SiPM directly recording the WLS light from the fiber mirrored at the far end, gives a adequate number of photo electrons, even for 300 cm strip lengths, as encountered in the barrel $\phi$-readout strips \cite{Baldini_2017}. A 30 mm strip width gives an angular resolution of order $\sim $10 msr for the inner most layers of the barrel (comparable to the muon multiple scattering in \textit{e.g.}, Belle), and we adopt this width for both the barrel and endcap implementations, as a compromise between total channel number and lateral strip response.
The resulting nominal strip counts for 14 barrel readout 
layers are: ``$\phi$'' strips 36-64 (lengths 1.5-3m) and 
48-98 ``$z$'' strips (lengths 1.2-2m) per octant. Hence a 
full barrel total of $\sim 30$k strips. For the 12 readout 
layer endcap, the count is more uniform with 84 strips in 
each orthogonal plane per layer per octant (with lengths 
$x$ and $y$ up to 2.4m), and an endcap total of $\sim 8.1$k 
strips. Overall, this makes for $\sim 38$k channels of CORE muon 
and neutrals detector readout.

The CORE readout electronics for the Neutral Hadron and Muon detector subsystem similarly follow that envisioned for an all-scintillator upgrade to the Belle II KLM Barrel. Modest amplification and shaping is done with 32-channel amplifier ASICs, which then feed 64-channel transient waveform digitizer (HDSoC) ASICs. Individual hits from either one or two of these HDSoC ASICs per orthogonal coordinate, for both in a layer, and collected into a Readout Unit, which following the Belle II convention is also known as a System Control and ReadOut of Data (SCROD). A tabulation of the required number of these devices, given the channel segmentation described above is provided in Table~\ref{tbl:KLM}.

\begin{table}[htb!]
\caption{NHM/KLM readout summary.}
\begin{tabular}{cccc}
\hline\hline
 Item & Quantity & Channels serviced & Comment \\
  \hline
Pre-amp ASICs & $\sim$1.5k & 32 & U. Hawaii 180nm design \\ \hline
HDSoC ASICs & $\sim$720 & 64 & Nalu Scientific \\ \hline
SCROD & $\sim$270 & $\sim$38k & KLM ReadoutUnit (RU) \\ \hline
FELIX cards & 7 & all & CORE common, assuming no data concentrator \\ 
\hline\hline
\end{tabular}
\label{tbl:KLM}
\end{table}

\subsection{Near-beam Detectors}

The interaction region (IR) design for CORE is consistent with the accelerator design as detailed in the CDR. The central CORE detector can be used in either IR6 or IR8. Both IRs can be optimized for the shorter length of CORE. This would reduce $\beta^{max}$ and chromaticity, thereby making the IR less challenging for the accelerator. While CORE is a good option for either IR, it is particularly synergetic with the the secondary focus currently developed for IR8.

\subsubsection{Far-forward Detectors}

The concept of a forward detection layout with a large dispersion and a second focus reducing the size of the beam at the detection point, which now form the basis of IR8, was pioneered by member of CORE a decade ago. CORE members were also active in developing the current implementation of this interaction region (and are listed as part of the IR8 team in the reference material on the project FAQ page).
However, while this layout nicely demonstrates the very significant impact on the physics capabilities of the EIC made possible by the exceptional forward acceptance made possible by the $2^{nd}$ focus, we feel that there is still room for further improvements.
We were originally hoping to fold some of these ideas into the CORE proposal, but the guidance from the project has been that changes to IR8 should wait until a decision is taken on magnet technologies, which will happen after the submission of the detector proposals. Thus, while CORE members are interested in working to further improve IR8, this will have to happen on a parallel track.

However, regardless of which IR CORE would be placed in, its shorter length (4 m on the hadron side) would make it possible to move the accelerator magnets, including the first, small ``B0'' dipole and final-focus quads (FFQs), closer to the collision point. The impact on luminosity is discussed in section \ref{luminosity}. In addition, moving the magnets closer will result in a small improvement in the forward acceptance. The opposite happened when the detector space on the hadron side in IR6 was increased from 4.5 m to 5 m. Going to 4 m would be analogous, but reversed.

The main task for the proposal is thus to list and cost the detectors at the different locations. Here it is important to note that while the optics and acceptances for the two IRs are very different, the basic layout of the detector subsystems is quite similar, and so are the instrumentation requirements. Thus, from the latter perspective the discussion in section 11.6 (p.546) of the Yellow Report \cite{AbdulKhalek:2021gbh} generally applies to both IRs.
The main difference is that in order to fully exploit the potential of IR8 to detect and identify nuclear fragments, some additional instrumentation would be required.

\begin{figure}[htb!]
\begin{center}
	\includegraphics[width=0.8\textwidth]{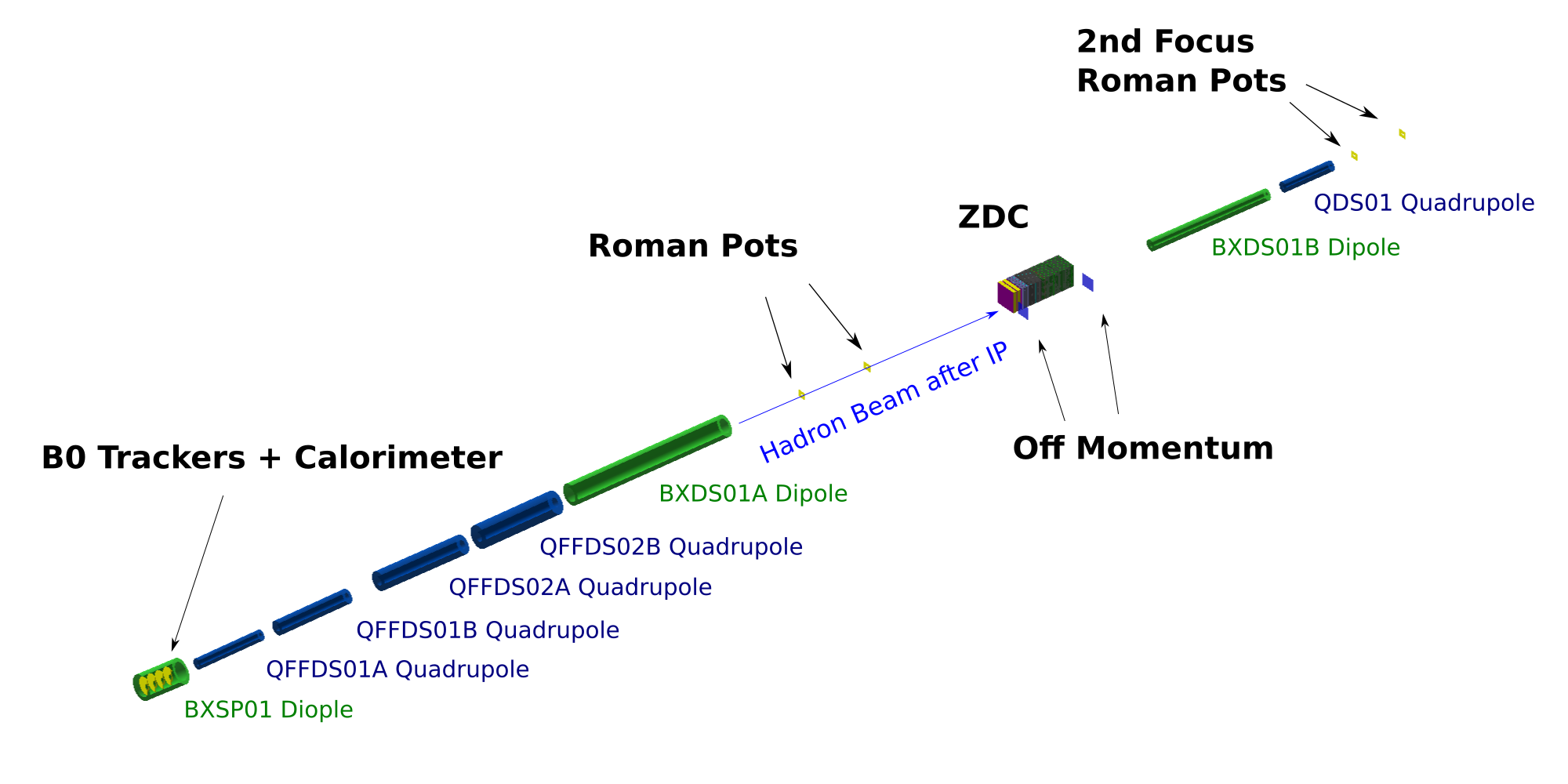}
	\caption{Overview of IR8.}
	\label{fig:IR8}
\end{center}
\end{figure}

The layout of IR6 is shown in the Yellow Report, and that of IR8 in Fig.~\ref{fig:IR8}. 
The instrumentation for detecting charged particles is grouped in three locations: the B0 dipole, the Roman pots (RPs) and off-momentum detectors (OMDs) after the first large dipole, and the RPs in the 4 m section around the $2^{nd}$ focus. In addition, there is a zero-degree calorimeter (ZDC) with an EMcal directly in front of it for detection of neutrals. A small EMcal is also located inside the B0.
Charged particles are tracked using silicon-pixel detectors. Due to the small Bdl of the B0 dipole, the position resolution requirements there are the most stringent, and space limitations demand a compact stack of trackers. The RPs and OMBs are grouped in pairs, except in the 4 m section at the $2^{nd}$ focus where there would be one RPs at the focus and one on each side. All stations would have tracking detectors, and the second one in each group also would benefit from a timing layer. This could be implemented using an AC-LGAD tracker, which would provide both position and time information (although a pixel readout would be required to handle the near-beam rates that limit how close to the beam one can operate the detector).
While IR6 may not have a $2^{nd}$ focus, it would still have a set of ``downstream'' RPs.

The unique capability provided by the second focus to detect the most of the nuclear fragments that are produced in e-A collisions creates many new physics opportunities. To help identify the individual isotopes, the last RP could contain a thin quartz bar acting as a radiator for a Cherenkov PID detector that would provide Z-information in addition to the rigidity ($\sim A/Z$) measurement in the forward spectrometer. A rare isotopes program at the EIC would also greatly benefit from high-resolution EMcal with a high photoelectron (p.e.) yield per MeV in front of the ZDC to enable gamma spectroscopy on short-lived isotopes. Although it is more expensive, LYSO would be a better choice than PbWO$_4$ due to its higher p.e. yield. While the latter is not important for the higher photon energies in the central detector (other aspects are more relevant there), the (boosted) nuclear photons typically have 10 to 100 MeV, where the p.e. yield does matter.

A high-resolution EMcal in front of the ZDC and additional PID for ion fragments would benefit the e-A program as a whole.
However, we believe that a rare-isotope program complementary to FRIB could bring additional funding opportunities for these two subsystems if feasibility is fully established. Thus, we do not include them in the baseline cost of IR8, and assume that the EMcal is of the simpler W/SciFi type.

\subsubsection{Far-Backward Detectors}

The far-backward detectors measure electrons scattered at very low-$Q^2$ (quasi-real photoproduction), and assist in luminosity determination through measurement of bremsstrahlung photons. Since the the electron beam geometry is similar in both IR6 and IR8, the layout of the far-backward detector is also going to be similar. Thus, although the discussion of these detectors in section 11.7 (p.572) of the Yellow Report \cite{AbdulKhalek:2021gbh} focuses on IR6, it is generally applicable to both IRs.

The low-$Q^2$ electron tagging is done by two near-beam stations following the first beamline dipole, and as well as the EMcal in the electron endcap. In CORE, the latter can be complemented by a small-angle EMcal in front of the endcap iron, which starts a $z$ = -3 m.

For the bremsstrahlung photon measurements, three detector subsystems will be used. One subsystem is a pair spectrometer (PS) that measures the energies of electron-positron pairs generated by pair production from bremsstrahlung photons. Downstream of the PS, a photon calorimeter (PHOT) placed in the photon beamline measures photon flux directly by absorbing the incident photon beam. In addition to the PS and PHOT detectors, the electron detectors will be used in special low-current runs in conjunction with the PS and PHOT to determine the detector acceptances and degree of bremsstrahlung-photon collimation.     
A preliminary design of the PS utilizes a 1.8 Tesla 18D36 dipole magnet to separate electrons and positrons arising from pair-converted bremsstrahlung photons. The conversion occurs upstream of the magnet within an aluminum converter placed in the incident photon beam. The magnetic field of the dipole magnet is parallel to the floor, causing the electron-positron pair to separate in a vertical plane.
Beyond the magnetic field produced by the dipole, the electron-positron pair continues an additional 3 meters within a vacuum chamber to a segmented hodoscope oriented perpendicularly to the floor; this hodoscope determines the trajectories, momenta, and opening angle of incident pair-produced electrons and positrons. The hodoscope consists of two segmented planes of detectors, with the planes separated in $z$ by half a meter. The first detector plane is located one meter downstream of the vacuum chamber. The first detector plane (most upstream) contains a finely segmented detector array PS-F, with a coarsely segmented array PS-C used to form a downstream coincidence. Each PS detector plane consists of an UP and a DOWN arm. Each PS-F arm has 260 scintillator tiles and each PS-C has 36 scintillator tiles per arm.      

\begin{figure}[htb!]
\begin{center}
	\includegraphics[width=0.5\textwidth]{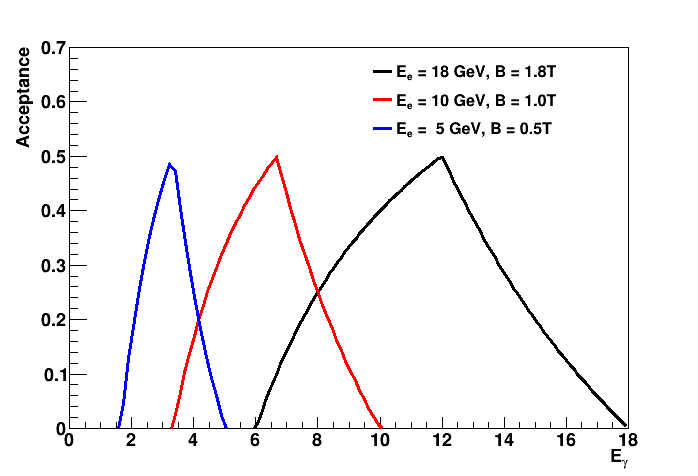}
	\caption{Acceptance of the pair spectrometer versus energy for bremsstrahlung photons. 
	The electron beam energy and pair spectrometer magnetic field setting are given on the legend of the plot.}
	\label{fig:psAcc}
\end{center}
\end{figure}

The acceptance of the pair spectrometer as a function of bremsstrahlung photon energy is shown in Fig.~\ref{fig:psAcc}. The solid lines represent different electron beam energies and pair-spectrometer magnetic-field settings, where black represents an 18 GeV electron beam with a 1.8 Tesla field, red represents 10 GeV electrons with 1.0 Tesla field and blue is for a 5 GeV electron beam and 0.5 Tesla magnetic field.

\subsection{Electronics \& Services}
\begin{figure}[htb!]
\begin{center}
	\includegraphics[width=0.8\textwidth]{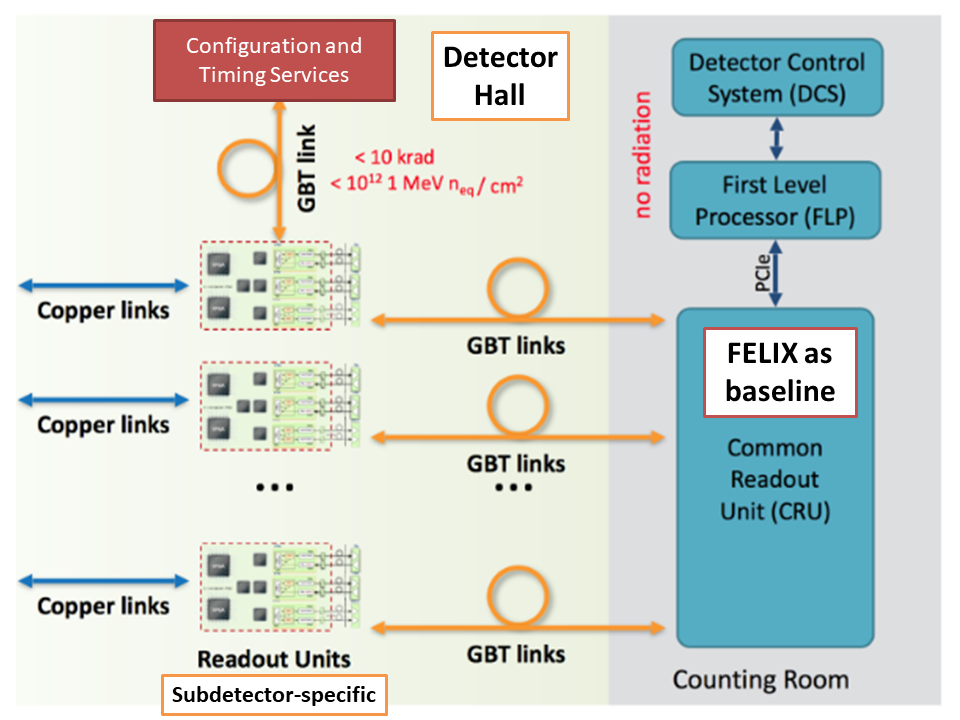}
	\caption{Modern readout overview with in-situ digitization and back-end data collection/even-building.}
	\label{fig:readout_overview}
\end{center}
\end{figure}

Control and readout of the various sub-detector systems is based upon a model that has become common at the LHC experiments \cite{Prieto:2020cac} and Belle II \cite{Zhou:2020qed}, and illustrated in Fig.~\ref{fig:readout_overview}. Various aspects of these key components are described briefly in the subsequent subsections.

\subsubsection{In-Situ Digitization}
Front end electronics are integrated as close to the detector as feasible, to reduce cabling and associated performance, cost, and space penalties. For higher channel density detectors, this is usually an ASIC optimized for the detector signal of interest, and where allowed elsewhere commercial ADCs are used. This is summarized in Table~\ref{tbl:CORE_readout}. The interface cards upon which these devices reside or interface will be physically different, but logically similar. Specifically, they will need to integrate seamlessly into the electrical and mechanical infrastructure, which includes communications, DAQ, slow-controls and power/cooling services.

\begin{table}[htb!]
\caption{CORE readout summary, broken down by subdetector. Readout Units (RU) are physically different for each subsystem, but logically and functionally similar, and readout is common and based upon the FELIX standard \cite{Paramonov:2021jpz}. }
\begin{tabular}{lllcccl}
\hline\hline
 Detector & Sensor & Digitization & \# of  & \# of  & \# of  &  \\
 Subsystem & Type & ASIC/ADC & Channels & RU &  FELIX & basis \\
  \hline
Si-Tracker & MAPS & ITS2, ITS3 & 30 G & $\sim$500 & 12 & eRD25 \\ \hline
hpDIRC & MCP-PMT & AARDVARC & 49 k & $\sim$120 & 4 & eRD14, Nalu \\ \hline
Dual RICH & SiPM & ALCOR & 65 k & $\sim$190 & 4 & eRD14, INFN \\ \hline
TOF (eEndcap) & DC-LGAD & ALTIROC & 2 M & $\sim$150 & 8 &  \\
\hline
EMCal: PBWO4 & SiPM & COTS ADC & 20 k & $\sim$150 & 2 &  \\ \hline
EMCal: W-shashlyk & SiPM & COTS ADC & 30 k & $\sim$150 & 3 &  \\ \hline
Ion HCal & SiPM & COTS ADC & 50 k & $\sim$150 & 4 & eRD25 \\ \hline
KLM (mu/Hcal) & SiPM & HDSoC & 38 k & $\sim$270 & 7 & Belle II, Nalu \\ \hline
Far-FWD/BWD & AC-LGAD & ALTIROC & TBD & TBD & 2 &  \\ 
\hline\hline
\end{tabular}
\label{tbl:CORE_readout}
\end{table}

\subsubsection{Communications, DAQ}
Common infrastructure connections provide requisite timing, programming, configuration, and housekeeping services to each of the subdetector-specific Readout Units (RU). Standards for these interfaces will be provided to all groups designing RUs so that they can be tested, installed and commissioned using CORE common modules. Gigabit fiber-optic transceivers on the RUs forward zero-suppressed data to the Common Readout Units (CRU) [FELIX] and server nodes which host them, which provide a uniform and scalable system of resources for event building.

\subsubsection{Slow-Controls and Services}
Detector Control Systems manage and monitor the interrelated Low Voltage, High Voltage, RU and CRU. It is foreseen that a message and error logging system, similar to those deployed in other large, running experiments will be employed.

\subsection{Integration and Maintenance}
The CORE detector is in compliance with all procedures developed by the EIC project. The endcaps can be detached and moved directly to the sides. Afterward the detector can be moved away from the beamline and rotated. To facilitate the removal of the hadron endcap, the endcap EMcal on the hadron side is supported from the barrel flux return iron. After removing the EMcal, the MPGD and dRICH (which has six sectors) can be moved out. On the electron side, the small and light endcap EMcal can be cantilevered in from behind using a support anchored in the barrel flux return iron. When the support and endcap EMcal are moved back, the DIRC readout (or even the full bar boxes) can be removed. The silicon tracker can in principle be removed through either endcap, but the simplest design would likely be a clam-shell that is removed through the hadron endcap after the dRICH is dismounted. Assembly  would occur in the reverse order, with the barrel EMcal first, followed by the DIRC and silicon tracker.

\section{Upgrade paths}
\label{upgrade_paths}
The baseline CORE detector presented and costed in this proposal has all the subsystems required to execute the full EIC physics program. In that situation we consider the possibilities offered by advancing technology as opportunities or options rather than Upgrades. In that context, we discuss below a series of options we see on the horizon.

\begin{itemize}
\item{One possible shorter-term upgrade could be to add a LGAD TOF layer behind the last MAPS disk on the hadron side, as discussed in section \ref{LGAD}.}
\item{Another possibility is discussed in section \ref{risk}. If the delivery of PbWO$_4$ crystals for the barrel EMcal would be slower than anticipated, the W-shashlyk section could be extended to negative $\eta$, creating a possibility for a future upgrade.}
\item{The silicon technology used for the tracker is evolving quickly, and it is likely that a significantly enhanced version could replace the initial one as part of a mid-life upgrade.}
\item{Photosensors for the DIRC and RICH detectors is another area where there is a lot of progress. In the future, this could mean that by replacing just the photosensors and readout electronics, the RMS timing resolution - and hence momentum reach - could improve for the DIRC.}
\item{There are also more speculative upgrade paths. For instance, a breakthrough in the use optical metamaterials as Cherenkov radiators could make it possible to replace the dRICH with a much more compact system that would enable moving the accelerator magnets even closer to the IP, which would be straightforward to implement with the CORE layout.}
\end{itemize}

While is difficult to speculate on future technologies, the flexible design of the CORE detector, with a compact inner core enclosed in a spacious flux return, creates many opportunities well aligned with technologies that could be available in the 2035-2040 time frame.






\section{Construction}
\label{construction}

The CORE proto-collaboration membership and structure enjoys sufficient expertise and support to have pulled together this proposal, including the cost and schedule at a conceptual level.

We believe that our proposal is innovative and contains numerous opportunities for participation in providing the major sub-systems of the experiment. It will benefit from the natural support from major laboratories, particularly for major procurements, such as the solenoid magnet. We expect this support to emerge once the experiment has received encouragement from the EIC Experiment Advisory Committee.

Table \ref{table:construction} shows the participation and anticipated evolution of commitments to the detector.

\begin{center}
\begin{table}[htb!]
\resizebox{\textwidth}{!}{%
\begin{tabular}{ |c|c|c|c| } 
 \hline
 Sub-system & Primary Responsibility & Key Participants & Potential Source of Support \\ 
 \hline
 \hline
 Solenoid & P. Brindza (ODU/JLab) &   & JLab/BNL Magnet Groups \\
 \hline
 Silicon tracker & S. Bueltmann (ODU) &   & EIC Silicon consortium \\
 \hline
 MPGD & M. Hohlmann (FIT) & K. Gnanvo (JLab) & JLab Detector Group \\
 \hline
 DIRC & G. Kalicy (CUA) & J. Schwiening (GSI) & PANDA DIRC Group (GSI) \\
 \hline
 dRICH & C. Joo (UConn) &  & CFNS @ Stony Brook, PID consortium \\
 \hline
 LGAD &  &  & LGAD consortium \\
 \hline
 PbWO$_4$ EMcal & C. Mu\~noz Camacho (IJCLab) & C. Hyde (ODU)  & Electron Endcap EMcal consortium\\
 \hline
 W-shashlyk EMcal &  & LLC Uniplast &  \\
 \hline
 Forward Hcal &  &  & Calorimeter consortium \\
 \hline
 KLM & W. Jacobs (IU) & A. Vossen (Duke) &  \\
 \hline
 Pair spectrometer & M. Dugger (ASU) &  &  \\
 \hline
 Low-$Q^2$ tagger & L. Guo (FIU) &  &  \\
 \hline
 ZDC & M. Murray (KU) & &  \\ \hline
 Forward Tracking & M. Murray (KU) & & \\ \hline
 Electronics & G. Varner (UH) & I. Mostafanezhad (Nalu) & JLab/BNL Electronics Groups \\
 \hline
\end{tabular}}
\caption{Participation and preliminary commitments to construction. Potential Source of Support are not committed.}
\label{table:construction}
\end{table}
\end{center}

\subsection{Timelines}
\label{timelines}
In what follows, we have assumed the Critical Decisions in the project schedule as laid out on page 3 of the presentation made by the Project Director, Jim Yeck, to the Meeting of the EIC Users Group on October 28, 2021. The call for proposals specifies that for Detector 1, the completion of detector construction must be achieved by CD-4A, July 2031. We assume that funding would start at CD3, July 2024 providing 7 years for construction. 

For CORE a number of detector sub-systems fit into a construction period of order 2 years. These include the MPGD, LGAD, and the W-shashlyk EMcal. For the latter a prospective vendor (Uniplast) has indicated that it could build all the modules needed for the barrel and hadron endcaps in half a year.

A second category comprises those sub-systems supported by the ongoing targeted R\&D. For those, the respective R\&D groups have prepared detailed construction timelines. These include the silicon tracker, the dRICH, the DIRC, and the Hcal in the hadron endcap for which four-year construction and installation schedule estimates are typical. The CORE versions of these systems are smaller and will thus require fewer hours to build than the examples provided by the respective groups, for example the silicon consortium. The Hcal is modular in nature, the time required to assemble the 1,602 modules will largely depend on the size of the workforce, but again fits comfortably into a four-year construction period. For all these systems a schedule showing completion before CD-4A (Detector 1) would enjoy two years (50\%) schedule contingency.

The third category are the major CORE-specific systems. These include the solenoid, the neutral hadron and muon system (KLM), and the PbWO4 crystals for the barrel EMcal. Of these the solenoid and the lead tungstate crystals are expected to be the schedule drivers. For the magnet, we have developed a conceptual schedule which including engineering design, procurement, which is of order 4-5 years followed by about 1 year of installation.  This assumes that we can support some preliminary engineering and design during the development of CD2 and CD3. This schedule would leave some contingency before CD4A, however, a prudent plan would advance the magnet for early procurement as part of a CD3A package. The procurement of the PbWO$_4$ crystals will depend on the production rates that can be achieved at Crytur and SICCAS, but our currently timeline for delivery of all the required crystals is compatible with operation in 2033. To meet the CD4A date, we would negotiate with the vendors to increase their capacity, which does appear possible, and advance the Lead Tungstate as a candidate for CD3A, early acquisition.
In conclusion, with the caveat associated with the lead tungstate, CORE construction and installation, with a duration of 7 years, would be complete by CD4A.

If selected as Detector 2, the construction and installation duration required would be the same as that for the Detector 1 situation. The challenge would be to establish a funding stream, ``hors project'' which would support a construction start by July 2026. Only if that can be achieved, could the schedule meet a CD4 completion. In turn, early completion of the selection process would be important.

\section{Collaboration}
\label{collaboration}

CORE is a proto-collaboration in the true sense of the word. As such, we tried to balance between two goals. On one hand, we wanted to demonstrate that the proto-collaboration already enjoys sufficient expertise and support to not only put together this proposal, but also that its current members can take the lead in designing and building key components of the detector. However, at the same time, we expect the membership to grow and we wanted to ensure that all members could participate in setting up the formal structures of the full collaboration. This aspect would be particularly important in a Detector 2 scenario, for which one would need to build the broadest possible collaboration. 

We believe that our proposal is innovative and contains numerous opportunities for participation, and we expect a broader support to emerge once the experiment has received encouragement from the EIC Experiment Advisory Committee. We also expect that the collaboration will benefit from the natural support from major laboratories, particularly for major procurements, such as the solenoid magnet and silicon tracker.  

Thus, the CORE proto-collaboration developed the proposal as a group of interested parties, with a direct participation by all its members, without (as of yet) establishing a formal collaboration structure. However, an important role the proto-collaboration was also to establish close links with the various technology-focused consortia that the EIC community has self-organized into.
These include the Silicon Consortium, the Hadronic and Electromagnetic Calorimetry Consortia, the PID consortium (which includes RICH and DIRC groups), and the Low-Gain Avalanche Detector (LGAD) Consortium, and the MPGD/tracking consortium.
All of these consortia have membership and collaborations beyond the EIC project, notably with PANDA/FAIR and LHC upgrade projects. The CORE membership includes participants in these consortia, and we have benefited from extensive discussions with people across the EIC Users Group membership.

We do not claim that the signatories of this proposal at this time have the resources to alone build the CORE detector, but the proto-collaboration forms a strong basis for a full collaboration. Since the number of detector proposals exceeds the number of possible detectors (which are limited to IP6 and IP8), we expect a consolidation to occur, and that many more users who are attracted by the unique capabilities of CORE and the physics opportunities that they bring will join the collaboration.

\subsection{Inclusion and Diversity}
CORE, as a collaboration and detector project, cannot be successful without a meaningful commitment to inclusion and diversity. We must recruit enthusiastic collaborators. In particular, we must mentor and support our junior members.

As a first step, we adhere to the APS Code of Conduct for Meetings
(\url{https://www.aps.org/meetings/policies/code-conduct.cfm}), edited for context:
\begin{quotation}
All participants will conduct themselves in a professional manner that is welcoming to all participants and free from any form of discrimination, harassment, or retaliation. Participants will treat each other with respect and consideration to create a collegial, inclusive, and professional environment within the CORE consortium. Creating a supportive environment to enable scientific discourse  is the responsibility of all participants.

Participants will avoid any inappropriate actions or statements based on individual characteristics such as age, race, ethnicity, sexual orientation, gender identity, gender expression, marital status, nationality, political affiliation, ability status, educational background, or any other characteristic protected by law. Disruptive or harassing behavior of any kind will not be tolerated. Harassment includes but is not limited to inappropriate or intimidating behavior and language, unwelcome jokes or comments, unwanted touching or attention, offensive images, photography without permission, and stalking.
\end{quotation}
It is not sufficient to refrain from overt discrimination. We must also acknowledge that the culture of physics, with an emphasis on pursuit of excellence and tough critique of ideas, methodology, and results can all too easily spill over into micro-aggression and behavior bordering on bullying.  Two recent articles from Physics Today bring this in stark relief. A June 2021 article from the online ``Careers \& Education'' section poses the question: ``Why does biophysics attract a disproportionate number of women?'' \cite{BioPhysicsWomen:2021}.
The blunt answer is given by a prominent biophysics professor:
\begin{quotation}
``It is entirely because there are fewer \textit{jerks}\footnote{Euphemism} in biophysics.''
\end{quotation}
More broadly, in the October edition of Physics Today, the article
``Lessons from 35 years in industry'' \cite{PhysicsTodayIndustry:2021}
provides the following advice:
\begin{quotation}
As for behavior, I have shocking news: Physicists can be arrogant. If asked how you know something, answers such as “it’s trivial” or “conservation of momentum” may be amusing to other physicists but are highly offensive to engineers and incomprehensible to management.
\end{quotation}
With these examples, and doubtless numerous others, we pledge to dedicate
ourselves to a welcoming and nurturing environment, that is not just for people like ourselves, but for everyone with the interest and opportunity to join.

\clearpage

\clearpage

\section{Acknowledgements}

We would like to acknowledge the many people within the EIC user community who have offered us advice, critiques, software support, physics insight, and general support.  These include the members of the various EIC detector technology consortia mentioned in the text. In particular, we thank
Evaristo Cisbani (INFN),
Abhay Desphande (CFNS/SBU),
Doug Higinbotham (JLab),
Kolja Kauder (BNL),
Edward Kistenev (BNL),
Wei Li (Rice),
Dimitri Romanov (JLab), 
Ernst Sichtermann (LBNL),
Oleg Tsai (UCLA),
and Craig Woody (BNL).

\vskip 2em
\centerline{\textbf{REFERENCES}}
\bibliography{references}
\end{document}